\documentclass[journal]{IEEEtran}

\usepackage{amsmath,amssymb,amsfonts} % Combines AMS packages
\usepackage{algorithmic}
\usepackage{algorithm}
\usepackage{array}
\usepackage[caption=false,font=normalsize,labelfont=sf,textfont=sf]{subfig}
\usepackage{textcomp}
\usepackage{stfloats}
\usepackage{url}
\usepackage{verbatim}
\usepackage{graphicx}
\usepackage[export]{adjustbox} % Included only once
\usepackage[footnote]{acronym} % Included only once
\usepackage[dvipsnames]{xcolor} % Included only once

\usepackage{multirow}
\usepackage{hhline}
\makeatletter
\newcommand{\thickhline}{%
    \noalign {\ifnum 0=`}\fi \hrule height 1.1pt
    \futurelet \reserved@a \@xhline
}
\newcolumntype{"}{@{\hskip\tabcolsep\vrule width 1.1pt\hskip\tabcolsep}}
\makeatother

\usepackage{tikz}

\usepackage[hidelinks]{hyperref}
\hypersetup{hidelinks}

\usepackage{diagbox}
\usepackage{bm}
\usepackage{orcidlink}
\usepackage{ifthen}
\usepackage{cancel}
\usepackage{bbm}
\usepackage{siunitx}
\usepackage{ulem}

\usepackage[
    abbreviations,
    automake,
    nonumberlist,
    nogroupskip,
    postdot,
    style=alttree,
    section=subsection,
    numberedsection=autolabel
]{glossaries-extra}

\newglossary*{optics}{Optics and Instruments}
\newglossary*{algorithms}{Algorithms and Processing}

\makeglossaries

\newabbreviation[type=algorithms]{rr}{RR}{ridge regression}
\newabbreviation[type=algorithms]{pinv}{PINV}{pseudo-inverse}
\newabbreviation[type=algorithms]{tsvd}{TSVD}{truncated singular value decomposition}
\newabbreviation[type=algorithms]{lv}{LV}{Loris-Verhoeven}
\newabbreviation[type=algorithms]{svd}{SVD}{singular value decomposition}
\newabbreviation[type=algorithms]{dct}{DCT}{discrete cosine transform}
\newabbreviation[type=algorithms]{dft}{DFT}{discrete Fourier transform}
\newabbreviation[type=algorithms]{idft}{IDFT}{inverse discrete Fourier transform}
\newabbreviation[type=algorithms]{idct}{IDCT}{inverse discrete cosine transform}
\newabbreviation[type=algorithms]{irca}{IRCA}{interferometer response characterization algorithm}
\newabbreviation[type=algorithms]{ista}{ISTA}{iterative shrinkage-thresholding algorithm}
\newabbreviation[type=algorithms]{fista}{FISTA}{fast iterative shrinkage-thresholding algorithm}

\newabbreviation[type=optics]{tbi}{TBI}{two-beam interference}
\newabbreviation[type=optics]{mbi}{MBI}{multiple-beam interference}
\newabbreviation[type=optics]{fp}{FP}{Fabry-Perot}
\newabbreviation[type=optics]{fpi}{FPI}{Fabry-Perot interferometer}
\newabbreviation[type=optics]{mi}{MI}{Michelson interferometer}
\newabbreviation[type=optics]{fts}{FTS}{Fourier-transform spectroscopy}
\newabbreviation[type=optics]{opl}{OPL}{optical path length}
\newabbreviation[type=optics]{opd}{OPD}{optical path difference}
\newabbreviation[type=optics]{imspoc}{ImSPOC}{Imaging SPectrometer On Chip}
\newabbreviation[type=optics]{cc}{CC}{Classic ColorChecker}

\newabbreviation[type=algorithms]{rmse}{RMSE}{root mean squared error}
\newabbreviation[type=algorithms]{mcw}{MCW}{matching central wavenumbers}

\newcommand{\argmin}[1]{\mathop{\mathrm{argmin}}_{#1}}
\newcommand{\argmax}[1]{\mathop{\mathrm{argmax}}_{#1}}

\newcommand{\vect}[1]{\mathbf{#1}}
\makeatletter
\newcommand{\diag}[1]{\mathop{\operator@font diag}\{#1\}} 
\makeatother
\newcommand{\innerproduct}[2]{\langle #1, #2 \rangle}

\newcommand{\matr}[1]{\mathbf{#1}}
\makeatletter
\newcommand{\Diag}[1]{\mathop{\operator@font diag}\{#1\}} 
\makeatother

\newcommand{\T}{{\sf T}}

\DeclareMathOperator{\dct}{DCT}
\DeclareMathOperator{\idct}{IDCT}
\newcommand{\hadamprod}{\odot}    % Hadamard product
\newcommand{\hadamdiv}{\oslash}    % Hadamard product
\newcommand{\R}[1]{\mathbb{R}^{#1}}
\newcommand{\RR}[2]{\mathbb{R}^{#1 \times #2}}

\definecolor{purple}{rgb}{0.7,0,0.5}

\begin{document}

% \title{Reconstruction of Spectra from Interferograms\\ in Multiple-beam Interference Spectroscopy}

% \title{Spectrum Reconstruction and Numerical Analysis\\ in Multiple-beam Interference Spectroscopy}

\title{Multiple-beam Interference Spectroscopy: Instrument Analysis and
Spectrum Reconstruction}

\author{
	Mohamad Jouni\orcidlink{0000-0002-7258-0935},~\IEEEmembership{Member,~IEEE,}
	Daniele Picone\orcidlink{0000-0002-0226-8399},~\IEEEmembership{Member,~IEEE,}
	Mauro Dalla~Mura\orcidlink{0000-0002-9656-9087},~\IEEEmembership{Senior Member,~IEEE}
	\thanks{
        % Manuscript received DD Month 202Y; revised DD Month 202Y and DD Month 202Y; accepted DD Month 202Y. Date of publication DD Month 202Y; date of current version DD Month 202Y.
        This work was supported partly by the AuRA region and FEDER under the project ImSPOC-UV (convention FEDER n. RA0022348) by Grant ANR FuMultiSPOC (ANR-20-ASTR-0006), and under project ``Pack Ambition International 2021'' by Grant 21-007356-01FONC and Grant 21-007356-02INV.
        \textit{(Corresponding author: Mauro Dalla~Mura.)}
	}
	\thanks{
		Mohamad Jouni and Daniele Picone are with the Centre National de la Recherche Scientifique (CNRS), Grenoble Institute of Technology (Grenoble INP), Grenoble Images Speech Signals and Automatics Laboratory (GIPSA-Lab), Université Grenoble Alpes, 38000 Grenoble, France (e-mail: mohamad.jouni@grenoble-inp.fr; daniele.picone@grenoble-inp.fr).
	}
	\thanks{
		Mauro Dalla Mura is with CNRS, Grenoble INP, GIPSA-Lab, Université Grenoble Alpes, 38000 Grenoble, France, and also with the Institut Universitaire de France (IUF), 75231 Paris, France (e-mail: mauro.dalla-mura@grenoble-inp.fr).
	}
	%
	% \thanks{Digital Object Identifier XXXX/YYYY}
}

\markboth{
	IEEE Journal Name,~Vol.~V, No.~N, Month~202Y
}%
{
	JOUNI \MakeLowercase{\textit{et al.}}: ReSpect-MBI
}

%\IEEEpubid{
%	\begin{minipage}{\textwidth}
%		\centering
%		0000--0000/00\$00.00~\copyright~202Y IEEE. Personal use is permitted, but republication/redistribution requires IEEE permission.
%		\\
%		See https://www.ieee.org/publications/rights/index.html for more information.
%	\end{minipage}
%}
% \IEEEpubid{
% 		0000--0000/00\$00.00~\copyright~202Y IEEE. Personal use is permitted, but republication/redistribution requires IEEE permission.
%		\\
%		See https://www.ieee.org/publications/rights/index.html for more information.
% }
% Remember, if you use this you must call \IEEEpubidadjcol in the second column for its text to clear the IEEEpubid mark.

\maketitle

% In this work, we describe a unified framework to encompass the majority of previous works...
% Unified categorization of previous works / methods
% Extend the numerical analysis

\begin{abstract}
Hyperspectral imaging systems based on \gls{mbi}, such as Fabry-Perot interferometry, are attracting interest due to their compact design, high throughput, and fine resolution.
Unlike dispersive devices, which measure spectra directly, the desired spectra in interferometric systems are reconstructed from measured interferograms.
Although the response of \gls{mbi} devices is modeled by the Airy function, existing reconstruction techniques are often limited to Fourier-transform spectroscopy, which is tailored for \gls{tbi}.
These methods impose limitations for \gls{mbi} and are susceptible to non-idealities like irregular sampling and noise, highlighting the need for an in-depth numerical framework.
To fill this gap, we propose a rigorous taxonomy of the TBI and MBI instrument description and propose a unified Bayesian formulation which both embeds the description of existing literature works and adds some of the real-world non-idealities of the acquisition process.
Under this framework, we provide a comprehensive review of spectroscopy forward and inverse models.
In the forward model, we propose a thorough analysis of the discretization of the continuous model and the ill-posedness of the problem.
In the inverse model, we extend the range of existing solutions for spectrum reconstruction, framing them as an optimization problem. Specifically, we provide a progressive comparative analysis of reconstruction methods from more specific to more general scenarios, up to employing the proposed Bayesian framework with prior knowledge, such as sparsity constraints.
Experiments on simulated and real data demonstrate the framework’s flexibility and noise robustness.
%
% An implementation of
The code is available at
https://github.com/mhmdjouni/inverspyctrometry.
\end{abstract}


\glsresetall

\section{Introduction}

\IEEEPARstart{M}{easuring} the \textit{spectrum} of a light source in a scene is at the core of imaging spectroscopy and has deep implications in various fields, such as geology, gas detection, security, remote sensing, disaster prevention, and more~\cite{Eism12, Mano16, zhu2023spectrum, pan2021sparse}.
In recent times, both the scientific community and industrial venues have shown interest in image spectrometers based on the principle of \textit{interferometry}
\cite{BornW13:book, Hari10, Hech16, PicoGDFL23:oe},
which is a fundamental tool in domains such as in
optics \cite{zhu2022review, blasi2023pattern, fang2020full},
radar \cite{gao2019point},
radio astronomy \cite{holler20122},
among other
measurement systems \cite{khan2008millimeter, zhang2022gas, hungund2023real}.
Compared to dispersive spectrometers, they allow for improved SNR \cite{fellgett2006nature}, compact instruments with reduced cost \cite{GuerLFD18:imspoc}, and acquisitions with finer spectral resolution \cite{cook1995multiplex}.
%\cite{okada-1989-infor-proces-spect}

\def \x {0.92}

\begin{figure*}[t]
    \centering
    
    \includegraphics[width=\x\linewidth]{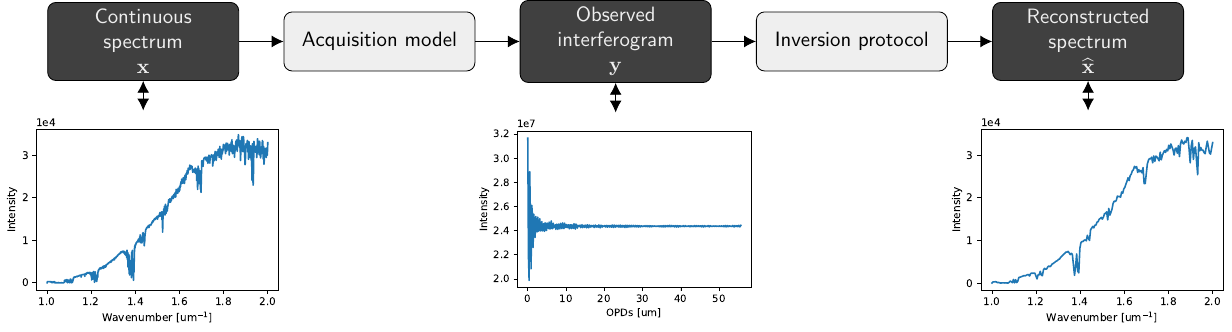}
    
    \caption{
    	The acquisition and inversion pipelines of the reconstruction of spectra from interferograms. It portrays both the role of the instrument for capturing the interferogram, which is modeled by a mathematical transformation of the input, and the necessity of an inversion pipeline for spectral reconstruction.
%    	$\delta$ and $\sigma$ denote the \glspl{opd} and wavenumbers respectively.
     }
    \label{fig:inversion_model}
\end{figure*}

% \IEEEpubidadjcol

Interferometers are devices that measure the intensity of the interference of superimposed coherent light beams after traveling different optical paths, whose difference is known as the \gls{opd}. As such, the spectrum of a point source is collected in a transformed domain, across different \glspl{opd}, resulting in the so-called \textit{interferogram}.
Since the desired spectra are not immediately intelligible to the final user, the acquired interferograms need to be processed \cite{fuhrmann2004spectrum, Dona19}.
\figurename~\ref{fig:inversion_model} illustrates this concept.
The two main phenomena of interference are the \textit{\gls{tbi}} and the \textit{\gls{mbi}} \cite{BornW13:book, Hari10}.

Devices based on \gls{tbi}, such as the \glsfmtlong{mi}, are widely employed in spectroscopy.
%\textcolor{red}{I feel there are too many abbreviations. Probably just better to remove some, such as MBI, TBI, MI, FPI, and remove the nomenclature altogether.}
This class of devices belongs to \gls{fts} \cite{Bell73:book}, where the
%theoretical
acquisition model at a given \gls{opd} is expressed as the Fourier transform of the incident spectrum, so the retrieval of the spectrum from an interferogram boils down to a Fourier inversion, i.e. an \gls{idft}.

For devices based on \gls{mbi}, the acquisition is described through a superimposition of a potentially infinite number of coherent light waves \cite{BornW13:book, Hari10}.
The response function of such devices at a given \gls{opd} is modeled as an \textit{Airy function}, and it can easily be shown that its Fourier series expansion is composed by harmonics that decay exponentially according to the reflectivity of the interferometer \cite{snell1995multiplex}.
Recently, \gls{mbi} devices, especially based on the \gls{fpi}, have become preferable for imaging spectroscopy as they are easier to construct, provide finer spectral resolution with more compact designs \cite{yang-2019-compac-ultrah}, are less influenced by environmental disturbances \cite{wang2010comparison}, and offer higher optical throughput \cite{jacquinot1954luminosity}.
\figurename~\ref{fig:interferometers} illustrates the operating principle of a \glsfmtlong{mi} and a \gls{fpi}.

% \subsection{Related Works and Limitations}
\subsection{Background}
\label{subsec:related_works}

Traditionally, the properties of \gls{mbi} interferometers have been
% widely
used for detecting emission lines and filtering absorption lines in incident light \cite{GousPLRVCFD22:igarss}. More recently, some imaging designs and prototypes based on the \gls{fpi} have been proposed to operate as spectrometers and aim for the full spectrum reconstruction of the incident light. Such devices typically operate in very low reflectivity regimes in order to emulate the operating principle of the standard \gls{fts}
\cite{chen-2006-simpl-terah}.
% \cite{jovanov-2011-trans-fourier}
For example, different scanning-based devices have been proposed that obtain recordings at different \glspl{opd} by adjusting the thickness of the \gls{fpi}. Among these,
some operate in the ultraviolet to near-infrared domains
\cite{PisaZ09:oe, ZuccCEP15:tim},
% \cite{zucco-2014-fabry-perot, pisani-2017-high}
%
other ones in the thermal range
\cite{wright-2022-hyti},
% \cite{wright2016tircis, lucey-2014-fourier-fabry, wright-2023-hyti}
%
and others with hybrid designs
% and high resolution
\cite{xu2018ultra, yang-2020-first-order}.
% \cite{yang-2019-compac-ultrah, yang-2020-theor-analy}
%
In \cite{GuerLFD18:imspoc}, a snapshot device has been proposed, featuring a multi-aperture design. The device consists of an array of \glspl{fpi} with different thicknesses arranged in a staircase pattern, enabling the simultaneous acquisition of recordings at different \glspl{opd}.

For most of these devices, the companion techniques proposed for spectrum reconstruction are based on the \gls{idft}. 
This strategy has however several limitations in terms of accuracy.

In particular, this solution is only valid for \gls{mbi}-based devices when the intensity of the
%higher-order
harmonics is actually negligible, which is not true in most practical scenarios.
An often overlooked side-effect of harmonics is their spectral overlapping within the spectral range of interest. \Gls{idft}-based techniques are particularly sensitive to this issue,  causing important distortions in the reconstruction samples.
An analytical solution to address this issue has been proposed in \cite{al-saeed-2016-fourier-trans}, where the \gls{dft} of the interferogram is expanded using a Haar function. However, this solution is only valid if the reflectivity of the instrument is assumed constant over the spectral range.
This challenge becomes even more relevant with multi-aperture snapshot devices \cite{GuerLFD18:imspoc} where the \glspl{fpi} may exhibit slightly or very different physical or manufacturing characteristics (e.g., gain, reflectivity).

Moreover, the spectral resolution of \gls{dft}-based solutions is strictly limited by the Nyquist-Shannon sampling theorem \cite{jerri1977shannon} (hereafter referred to as the \textit{sampling theorem}) and ignore the potential to exploit the contribution of the aliased non-zero spectral samples due to the harmonics that can be potentially collected outside of the spectral range of the instrument. 

% This is relevant in snapshot devices where the domain of \glspl{opd} is physically fixed.
Finally, the acquired interferogram in most practical applications is irregularly sampled. This is due to the \glspl{opd} being associated to physical components of the device, which are in turn sensitive to manufacturing defects. As a result, some distortions occur in the reconstruction, as \gls{dft}-based techniques assume that the interferogram is regularly sampled. 

% \gls{dft}-based techniques are not easily applied in case as irregular sampling of the \glspl{opd} and the filtered spectral range, and do not account to noise on the measurements.

\subsection{Aims and Contributions}

This work addresses the need for a common formulation and reconstruction procedure of \gls{mbi} spectroscopy currently missing in the respective literature.
We provide a comprehensive review that both embeds the description of existing works and adds some of the real-world non-idealities of the acquisition process.
We describe a unified framework based on a Bayesian formulation \cite{gurel2020compressive, wang2022combination} of the problem with a more accurate representation of the physical acquisition model.
Specifically, a \textit{transfer matrix} is generated from an Airy function that embeds the physical design conditions of a given \gls{mbi}-based device, accommodating the challenging case of multi-aperture snapshot devices, and the reconstruction is carried out as a linear inverse problem, incorporating prior information.
However, since the interferometric model exhibits a domain transform that is related to the Fourier transform, we provide an analysis of the discretization of the continuous model \cite{gupta2018continuous, bohra2020continuous} and the well-posedness of the problem in terms of Hadamard \cite{Idie13:book}.
%
% With that goal in mind, the novelty lies in setting a framework for the analysis and resolution of spectrum reconstruction from interferometric measurements, as detailed below.

This work acts as an extension to our preliminary results in \cite{JounPD23:icassp, picone2024spectro}.
Within the comprehensive review, the novel contributions are summarized as follows:
\begin{enumerate}
	\item
	For the acquisition model analysis, we extend the findings of the \gls{mbi} theoretical model in \cite{cook1995multiplex} with a numerical analysis on its discretized version, i.e., the transfer matrix. In particular, we investigate the conditions of well-posedness of the problem in terms of Hadamard \cite{Idie13:book} under the textbook formulation of both \gls{tbi} and \gls{mbi} acquisition systems.
	First, we summarize the conditions for the proper discretization of the continuous interferometric transformation, and formulate the two regimes in terms of the \gls{dct}.
	Then, we formalize the limitations of spectrum reconstruction in terms of spectral resolution, recoverability, and Hadamard well-posedness.
	
	\item
	For the reconstruction of spectra, first, we represent the model with a transfer matrix that embeds information of real acquisition scenarios such as irregular sampling, variable reflectivity, harmonic contribution, and measurement noise.
	Then, by using the matrix within a Bayesian framework, we extend the techniques of spectrum reconstruction to linear inverse problems and incorporate prior knowledge such as LASSO \cite{gurel2020compressive}.
	For that, we provide a pedagogical progressive analysis of the ability of these techniques to tackle the non idealities from more specific to more general scenarios.
\end{enumerate}
With this analysis, our aim is to establish a novel unified framework that accommodates both \gls{tbi} and \gls{mbi} systems. Under this umbrella, we are able to describe instruments with different operating principles, reflectivity regimes, physical conditions, and prior information.

The remainder is organized as follows.
Section \ref{sec:instrumental_model} presents a physical background of \gls{tbi} and \gls{mbi}.
Section \ref{sec:transfer_matrix} presents the analysis of the system.
Section \ref{sec:inversion_algorithms} talks about spectrum reconstruction.
Sections \ref{sec:results} and \ref{sec:real_data_experiments} present the experiments and results.
Finally, a conclusion is drawn in Section \ref{sec:conclusions}.

%While machine learning-based approaches may provide a way to intrinsically regularize this formulation, their applicability, at least in their pure form, is limited as reference data is typically unavailable. Moreover, hybrid (model- and learning-based) approaches such as algorithm unrolling \cite{MongLE21:spm} and deep priors \cite{ZhanLZZVT21:tpami} rely heavily on a good understanding of the system, which we aim to address here as preliminary step and solid basis for the \textit{interpretability} of such systems.

%%%%%%%%%%%%%%%%%%%%%%%%%%%%%%%%%%%%%%%%%%
\section{Acquisition (Instrumental) Model}
\label{sec:instrumental_model}

\def \x {0.31}
\def \y {0.36}

\begin{figure*}[t]
	\centering
	
	\subfloat[Michelson (\gls{tbi})]{
		\centering
		\includegraphics[width=\x\textwidth]{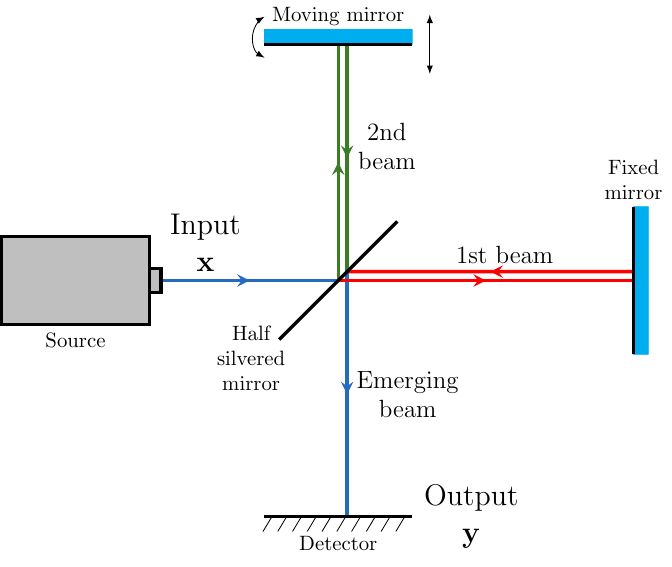}
		\label{fig:michelson_interferometer}
 	}
  \hspace{2.5cm}
	\subfloat[\glsxtrlong{fp} (\gls{mbi})]{
		\centering
		\includegraphics[width=\y\textwidth]{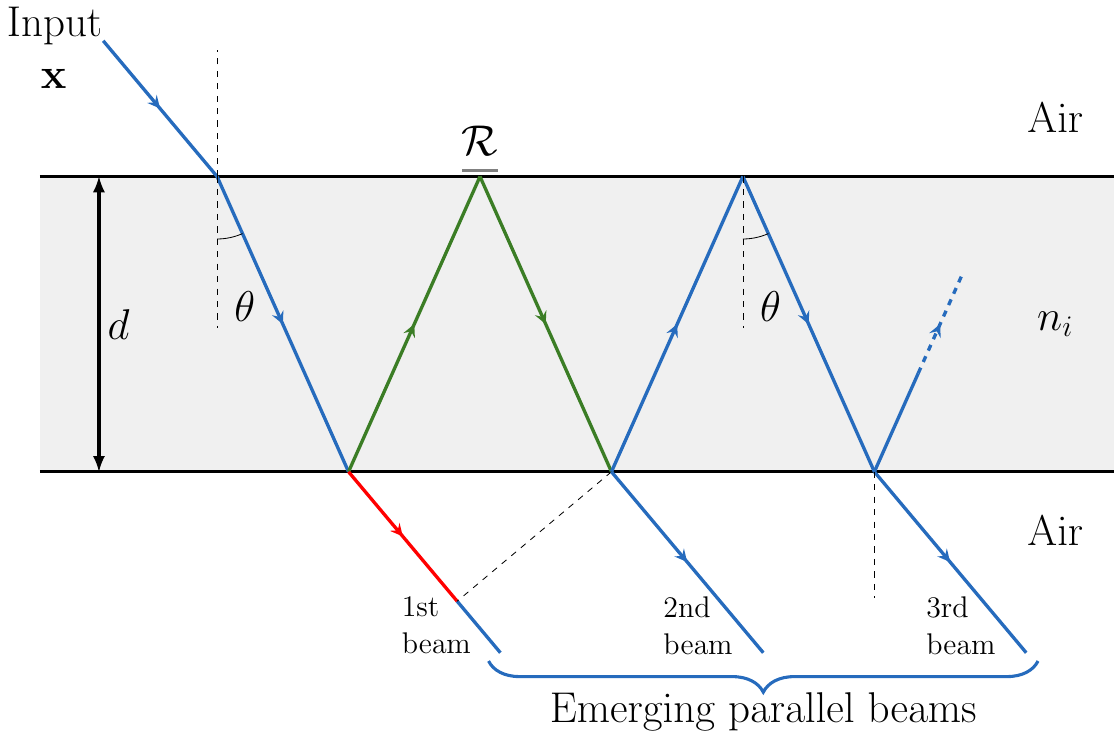}
		\label{fig:fabry_perot_interferometer}
	}
	
	\caption{Illustration of the operation principle of a Michelson interferometer (left) and a \glsxtrlong{fpi} (right).}
	\label{fig:interferometers}
\end{figure*}

%\begin{figure}[t]
%	\centering
%	
%	\begin{subfigure}[t]{\x\linewidth}
	%		\centering
	%		\includegraphics[width=\columnwidth]{figures/interferometers/interferometer_michelson/full/main.pdf}
	%		\caption{Michelson interferometer}
	%		\label{fig:michelson_interferometer}
	%	\end{subfigure}
%	\begin{subfigure}[t]{\y\linewidth}
	%		\centering
	%		\includegraphics[width=\columnwidth]{figures/interferometers/interferometer_fabry_perot/full/main.pdf}
	%		\caption{\glsxtrlong{fp} étalon}
	%		\label{fig:fabry_perot_interferometer}
	%	\end{subfigure}
%
%	\caption{Illustration of different interferometers.}
%
%\end{figure}

In this section, we recall the physical principles behind \gls{tbi} and \gls{mbi}. The goal is to describe the acquisition model of the two phenomena throughout the mathematical expression which describes the transformation from the incident light (optical part) to the measured signal (observation part).
In the following, the input is described as the continuous spectral radiance of the incident light, while the observation is defined by the resulting discrete interferogram \cite{BornW13:book, Hari10}.

\subsection{Interferogram Formation}
\label{subsec:interferometric_transmittance_response}

Let us denote the interferogram signal $y(\delta)$ as a function in the domain of the \glspl{opd} $\delta$, and the spectrum $x(\sigma)$ as a function in the domain of the wavenumbers $\sigma$.
Throughout this paper, we use the wavenumbers, defined as the reciprocal of the wavelengths, as they are more convenient and more widespread in the literature of \gls{fts} and \gls{mbi}.
For consistency throughout this paper, we express the \glspl{opd} $\delta$ and wavenumbers $\sigma$ in $\si{\um}$ and $\si{\um}^{-1}$, respectively. 
%Some authors prefer to express the spectrum in the domain of the wavelengths, that is the reciprocal of the wavenumbers, but our formalism is more common in the literature of \gls{fts}.

In the following, we assume that the device operates in the arbitrary spectral support $\Omega = [\sigma_{\min}, \sigma_{\max}]$ (where $0<\sigma_{\min}<\sigma_{\max}$).
We also assume that the device
%(or part of it)
is capable of generating a certain fixed \gls{opd} $\delta$. The specific mechanisms for generating this \gls{opd} will be detailed when relevant.

With this formalism, the interferogram $y(\delta)$ can be modeled by the following equation \cite{BornW13:book, Hari10}:
\begin{equation}
	y(\delta) =
	\int_{\Omega}^{}
	A(\delta, \sigma) x(\sigma) d\sigma
	\label{eq:interferogram_formation}
	\,,
\end{equation}
where
%$l \in \{0, \dots, L-1\}$ such that $L$ is the number of samples in the interferogram,
%$\delta$ is the \gls{opd} of the interfering beams,
%corresponding to the measurement of $y(\delta)$,
%and
$A(\delta, \sigma)$ defines the \textit{response function} of the device. In the context of interferometry, one can interpret $A(\delta, \sigma)$ as a domain transfer function from the wavenumbers
%$\sigma$
to the \glspl{opd}.
%$\delta$.

%Interferometric instruments measure multiple samples of the interferogram, either sequentially using a single interferometer and varying its \glspl{opd}, or simultaneously using several interferometers at the same time. \textcolor{red}{If you want to cut, this has been explained already.}
%Ideally, one would like to collect infinitely fine intervals, which would generate a continuous interferogram, but we are limited by the characteristics of the instrument.

%\subsection{Interferometric Response Function}

The interferometric response function $A(\delta, \sigma)$ represents the intensity ratio of light interference as a function of $\sigma$ for each $\delta$ \cite{BornW13:book, Hari10}. 
In the following sections, we present the expressions of the response functions for the \gls{tbi} and \gls{mbi}. These expressions are described in the context of their most common applications, namely the \glsfmtlong{mi} and \gls{fpi} respectively. However, the reader can consider the described results without loss of generality, as they could be extended without much effort to other use cases as well. 
% in some simplified scenarios. However, their expression is given without loss of generality and can be easily extended to other models or combination.

\subsection{Two-beam Interference (TBI)}
% \textcolor{red}{I dislike subsubsections. Just do subsection?}

In a \glsfmtlong{mi}, the incident light is split into two beams, which then travel different optical paths $d_1$ and $d_2$ until they get reflected by their respective mirrors, before meeting at one point at the detector. Their \gls{opd} $\delta = d_1 - d_2$ causes a phase difference $\Delta\phi = 2 \pi \delta \sigma$ among the two beams. By adjusting the relative position of one of the mirrors, the \gls{opd} can be varied, which allows to study the interference patterns for different phase differences \cite{BornW13:book, Hari10}.

The response function $A(\delta, \sigma)$ is then defined as the ratio between the captured optical intensity (given by the two beams) and the input intensity $x(\sigma)$, yielding \cite{BornW13:book, Hari10}:
\begin{equation}
	\begin{gathered}
		A(\delta, \sigma)
%		=
%		\frac{I_o(\delta, \sigma)}{x(\sigma)}
		=
		2 \mathcal{T}(\sigma) (1 + \cos(2 \pi \delta \sigma))
		\,,
	\end{gathered}
	\label{eq:transmittance_response_michelson_nonmonochromatic}
\end{equation}
where $\mathcal{T}(\sigma) \in [0, 1]$ denotes the transmittance of the \glsfmtlong{mi}.
The multiplicative factor $2$ appears in order to compensate the optical energy in the special case where the beams are evenly split.

For simplicity, by setting $\mathcal{Q}(\sigma) = 2 \mathcal{T}(\sigma)$, the observed interferogram $y(\delta)$ is obtained from \eqref{eq:interferogram_formation} as follows:
\begin{equation}
	\begin{aligned}
		y(\delta)
		&=
		\int_{\Omega} \mathcal{Q}(\sigma) x(\sigma) d\sigma
		+
		\int_{\Omega} \mathcal{Q}(\sigma) x(\sigma) \cos(2 \pi \delta \sigma) d\sigma
		\\
		&=
		\mathcal{F}_c\{\mathcal{Q} x\}(0) + \mathcal{F}_c\{\mathcal{Q} x\}(\delta)
		\,,
	\end{aligned}
	\label{eq:interferogram_formation_michelson_nonmonochromatic}
\end{equation}
%\textcolor{red}{I prefer the more standard expression $\mathcal{F}_c\{\mathcal{Q}(\sigma) x(\sigma)\}(\delta)$ for Fourier cosine transform, or at most $\mathcal{F}_c\{\mathcal{Q}x\}(\delta)$ if you really really want to save space (Regardless, I would avoid $\mathcal{Q}\cdot x$). I also prefer to express $\innerproduct{\mathcal{Q}}{x}$ as $\mathcal{F}_c\{\mathcal{Q}(\sigma) x(\sigma)\}(0)$ or $\mathcal{F}_c\{\mathcal{Q}x\}(0)$, for a more consistent notation throughout (as in, you need to explain (\sigma)just one thing).}
%
where $\mathcal{F}_c\{\mathcal{Q} x\}(\delta)$ represents the \textit{Fourier cosine transform} of the spectral distribution $\mathcal{Q}(\sigma) x(\sigma)$ evaluated in $\delta$. Its expression for $\delta=0$, namely $\mathcal{F}_c\{\mathcal{Q} x\}(0) = \frac{1}{2} y(0)$, can be expressed as a function of the detected intensity $y(0)$ in the case the beams travel equal paths. Eq.~\eqref{eq:interferogram_formation_michelson_nonmonochromatic} serves as the historical foundation of the term \glsxtrlong{fts}.
%The spectrum $x(\sigma)$ can ideally be obtained through a simple \gls{idft} \cite{BornW13:book, Hari10}. \textcolor{red}{Would not put this remark here; this just makes it confusing since the DFT is a discrete to discrete, while you are in a context of continuous to discrete (check continuous time Fourier transforms (CTFTs) for more details). Also why suddenly IDFT from a cosine transform? Just remove this sentence?}
%such that :
%\begin{equation}
%	\begin{aligned}
%		x(\sigma)
%		=
%		\frac{4}{\mathcal{Q}(\sigma)}
%		\int_{0}^{\infty}
%		\left(y(\delta) - \frac{1}{2} y(0)\right)
%		\cos(2 \pi \sigma \delta)
%		d\delta.
%	\end{aligned}
%	\label{eq:spectrum_reconstruction_michelson_nonmonochromatic}
%\end{equation}

\subsection{Multiple-beam Interference (MBI)}
\label{subsubsec:multiple_beam_interference}

The design of an \gls{fpi} is typically manufactured using interferometric cavities. Geometrically, a cavity is described as two parallel surfaces separated by a thickness $d$, which encloses a homogeneous optical material with a refractive index $n_i$. 
The overall cavity exhibits a transmittance $\mathcal{T}(\sigma)$ and reflectivity $\mathcal{R}(\sigma) \in [0, 1]$.
The incident light is transmitted through the first surface at an angle $\theta$ into the interior of the plate, where it gets reflected back and forth between the two surfaces, and attenuated by the reflectivity $\mathcal{R}(\sigma)$ every time it bounces over the cavity surface.
After each round trip within the cavity, a beam emerges out of the \gls{fpi}.
These potentially infinite emerging parallel beams are collected at a detector, exhibiting an \gls{opd} $\delta = 2 n_i d \cos(\theta)$ and a phase difference $\Delta\phi = 2 \pi \delta \sigma$ between two consecutive beams \cite{BornW13:book, Hari10}.
%The most common way to obtain an interferogram is
%An interferogram can then be obtained by allowing the thickness $d$ of the \gls{fpi} to be varied. \textcolor{red}{Honestly just remove this sentence, as the aim of this section is to have the expression of $A$, not to tell how to obtain an interferogram. This will be done in detail in section 3?}

%The expression of the response function function $A(\cdot, \cdot)$, expressed as the ratio between the output and input intensities, can be modeled through the contribution of a series of $N \in \mathbb{N}$ emerging (transmitted) waves from the interferometer as \cite{BornW13:book, Hari10}:
%\begin{equation}
%	A(\delta, \sigma)
%	=
%	\left|
%	\mathcal{T}(\sigma)
%	\sum_{n=0}^{N-1}
%		\mathcal{R}^n(\sigma)
%		e^{- j \, n \, (2 \pi \delta \sigma)}
%	\right|^2
%	\label{eq:transmittance_response_fabry_perot_N_waves}
%\end{equation}
%which boils down to a geometric series of ratio $\mathcal{R}(\sigma)
%e^{j (2 \pi \delta \sigma)}$.
% The response function $A(\cdot, \cdot)$, representing the interference of the emerging (transmitted) waves from the interferometer, 
It can be shown \cite{BornW13:book, Hari10} that the response function of the \gls{fpi} can be expressed as the following closed form:
\begin{equation}
	A(\delta, \sigma) =
	\frac {
		\mathcal{T}^2(\sigma)\
	} {
		1 + \mathcal{R}^2(\sigma)
		-
		2\mathcal{R}(\sigma)
		\cos(2 \pi \delta \sigma)
	}
\,,
	\label{eq:transmittance_response_fabry_perot_infty}
\end{equation}
which is widely known as the Airy function.
%and in some cases as the $\infty$-wave model \cite{PicoGDFL23:oe}.
%Starting with the Poisson kernel, which is given by \cite{Cars21:book}:
%\begin{equation}
%	\begin{gathered}
%		\mathcal{P}_r(\theta)
%		= \sum_{n=-\infty}^{\infty}
%		r^{|n|}
%		\,
%		e^{in\theta}
%		=
%		\frac{1-r^2}{1-2r\cos(\theta)+r^2}
%		\\
%		\textrm{where}
%		\;\;
%		0 \leq r < 1
%		\label{eq:poisson_formula}
%	\end{gathered}
%\end{equation}
%\mdm{define $\theta$...}
%%
%the $\infty$-wave model of expression \eqref{eq:transmittance_response_fabry_perot_infty} can be written as follows (we omit the expression of $\sigma$ in $\mathcal{T}(\sigma)$ and $\mathcal{R}(\sigma)$ for the sake of readability):
%\begin{equation}
%	\begin{aligned}
%		A(\delta, \sigma)
%		&=
%		\frac{
%			\mathcal{T}^2
%		}{
%			1 + \mathcal{R}^2
%			-
%			2\mathcal{R}
%			\cos(2 \pi \delta \sigma)
%		}
%		\\
%		&=
%		\frac{\mathcal{T}^2}{1-\mathcal{R}^2}
%		\times
%		\frac{
%			1-\mathcal{R}^2
%		}{
%			1 + \mathcal{R}^2
%			-
%			2\mathcal{R}
%			\cos(2 \pi \delta \sigma)
%		}
%	\end{aligned}
%	\label{eq:infty_wave_model_poisson_kernel_appendix_1}
%\end{equation}
%
Equivalently, imposing $\mathcal{Q}(\sigma) = \frac{\mathcal{T}^2(\sigma)}{1-\mathcal{R}^2(\sigma)}$ for simplicity, eq. \eqref{eq:transmittance_response_fabry_perot_infty} can be expanded as a Fourier series \cite{cook1995multiplex}:
\begin{equation}
	\begin{aligned}
		A(\delta, \sigma)
		&=
		\mathcal{Q}(\sigma)
		\left[
		1
		+
		2
		\sum_{n=1}^{\infty}
		\mathcal{R}^{n}(\sigma)
		\cos(2\pi \, n\delta \, \sigma)
		\right]
		\\
		&=
		C_0 (\sigma)
		+
		\sum_{n=1}^{\infty}
		C_n (\sigma)
		\cos(2\pi \, n\delta \, \sigma)
		\,,
	\end{aligned}
	\label{eq:transmittance_response_fabry_perot_fourier_series}
\end{equation}
where $C_n(\sigma)$ $\forall n \in \mathbb{N}$ are the coefficients such that:
\begin{equation}
%    \begin{aligned}
	C_0(\sigma) = \mathcal{Q}(\sigma)\,
	;
	\;\;\;
	C_n(\sigma) = 2 \mathcal{Q}(\sigma) \mathcal{R}^{n}(\sigma)\,
	,
	\;\;\;
	\forall n\geq 1\,.
%    \end{aligned}
	\label{eq:transmittance_response_fabry_perot_fourier_coefficients}
\end{equation}

Compared to \gls{tbi}, for each \gls{opd} $\delta$, there is a potential harmonic contribution coming from the oscillations $n\delta$ $\forall n \geq 2$.
Since the harmonics decay exponentially as $n \to \infty$, then the model can be approximated as:
\begin{equation}
	\begin{aligned}
		A(\delta, \sigma)
		=
		\sum_{n=0}^{N-1}
		C_n(\sigma)
		\cos(2\pi \, n\delta \, \sigma)
		+
		%\cancelto{\scriptstyle{0}}
        {\mathcal{O}}(N)
		\,,
	\end{aligned}
	\label{eq:transmittance_response_fabry_perot_fourier_series_N_waves}
\end{equation}
where ${\mathcal{O}}(N) \approx 0$ includes all the terms associated to more than $N$ reflections, which are negligible as the coefficients $C_n(\sigma)$ become negligible for $n \geq N$.
The corresponding interferogram is expressed as follows:
\begin{equation}
	\begin{gathered}
		y(\delta)
%		=
%		\int_{\Omega}
%		\sum_{n=0}^{N-1}
%		C_n(\sigma)
%		\cos(2 \pi n\delta \sigma)
%		\,
%		x(\sigma)
%		d\sigma
%		\\
%		=
%		\int_{\Omega}
%		\mathcal{Q}(\sigma) x(\sigma)
%		d\sigma
%		\innerproduct{\mathcal{Q}}{x}
%		+
%		2
%		\sum_{n=1}^{N-1}
%		\int_{\Omega}
%		\mathcal{Q}(\sigma) \mathcal{R}^n(\sigma) x(\sigma)
%		\cos(2 \pi n\delta \sigma)
%		d\sigma
%		\\
		=
		\mathcal{F}_c\{\mathcal{Q} x\}(0)
		+
		2
		\sum_{n=1}^{N-1}
		\mathcal{F}_c\{\mathcal{Q} \mathcal{R}^n x\}(n\delta)
%		\mathcal{F}_c(\mathcal{Q} \cdot \mathcal{R}^n \cdot x)(n\delta)
%		+
%		\cancelto{\scriptstyle{0}}{\mathcal{O}}(N).
		\,.
	\end{gathered}
	\label{eq:interferogram_formation_fabry_perot}
\end{equation}
Here, each record $y(\delta)$ contains the contribution of not only the fundamental term, but also that of its harmonics.

% \subsubsection{\glsfmtshort{fts} as a special case of \glsfmtshort{mbi}}
%\label{subsubsec:multiple_beam_interference_2_waves}

When the reflectivity $\mathcal{R}(\sigma)$ is too low such that $C_{2}(\sigma)$ becomes negligible, eq. \eqref{eq:interferogram_formation_fabry_perot} boils down to the fundamental terms as follows:
\begin{equation}
	\begin{aligned}
		y(\delta)
		=
		\mathcal{F}_c\{\mathcal{Q} x\}(0)
		+
		2
		\mathcal{F}_c\{\mathcal{Q} \mathcal{R} x\}(\delta)
%		+
%		\cancelto{\scriptstyle{0}}{\mathcal{O}}(2)
		\,.
	\end{aligned}
	\label{eq:interferogram_formation_fabry_perot_2_wave}
\end{equation}
%Due to the factor $\mathcal{R}(\sigma)$ in the Fourier term, an analytical solution may not be exact, but
Since $\mathcal{R}(\sigma)$ is assumed very small, one could approximate the term $\mathcal{F}_c\{\mathcal{Q} x\}(0)$ to the mean value of the interferogram,
%that is, as a function of the observation,
and eq.~\eqref{eq:interferogram_formation_fabry_perot_2_wave} becomes roughly similar to eq. \eqref{eq:interferogram_formation_michelson_nonmonochromatic} up to the multiplicative factor $\mathcal{R}(\sigma)$.
%\textcolor{red}{It is still not the same formula as the one for Michelson (as you used the formalism of the book of Borne, which has no intensity degradation $\mathcal{R}$ for Michelson interferometric replica); either just say that it is an expression that is roughly in the same form of eq. (3),  or add $\mathcal{R}$ in eq.(3). $\mathcal{R}\approx 1$ in Michelson anyway, allowing to make the equation symmetric, but not to have the bias term expressed as function of $y(0)$. You can't have all I guess, as Michelson can measure $y(0)$ very easily, while it is almost impossible for Fabry-Perot interferometers.}

%\textcolor{red}{Where is the $\mathcal{R}$ term in eq.(3)?}
%As such, the spectrum can be reconstructed similarly to \eqref{eq:spectrum_reconstruction_michelson_nonmonochromatic} as:
%\begin{equation}
%	x(\sigma)
%	=
%	\frac{4}{2\mathcal{Q}\mathcal{R}}
%	\int_{0}^{\infty}
%	\left(y(\delta) - \bar{y}\right)
%	\cos(2 \pi \sigma \delta)
%	d\delta
%	\label{eq:interferogram_formation_fabry_perot_2_wave_variant_r}
%\end{equation}
%
%On a side note, if $\mathcal{R}$ is assumed independent of $\sigma$, then
%$
%\innerproduct{\mathcal{Q}}{x}
%=
%\frac{y(0)}{1 + 2\mathcal{R}}
%$
%and the reconstruction is exact:
%\small
%\begin{equation}
%	x(\sigma)
%	=
%	\frac{4}{2\mathcal{Q}\mathcal{R}}
%	\int_{0}^{\infty}
%	\left(y(\delta) - \frac{1}{1 + 2\mathcal{R}} y(0)\right)
%	\cos(2 \pi \sigma \delta)
%	d\delta
%	\label{eq:interferogram_formation_fabry_perot_2_wave_fixed_r}
%\end{equation}
%\normalsize

\section{Transfer Matrix Numerical Analysis}
\label{sec:transfer_matrix}

In this section, we aim to derive a proper discrete version of the direct model that describes the acquisition system.
The motivation for this is twofold: first, in the \gls{opd} domain, we must ensure that the number of acquisitions is finite; second, in the wavenumber domain, we need to sample the continuous range to allow for numerical computations. 
To this end, we discretize the response function $A(\delta, \sigma)$,
defining its sampled version as the transfer matrix in Section \ref{subsec:transfer_matrix_representation}.

Additionally, when treated as a linear inverse problem, spectrum reconstruction can be ill-posed or ill-conditioned in terms of Hadamard \cite{Idie13:book}, leading to solutions that are either not unique or unstable.
To formalize the nature of the inverse problem at hand, we perform a numerical analysis on the characteristics of the transfer matrix.

First, we start with a brief sampling analysis of the response function in Section \ref{subsec:sampling_analysis} \cite{cook1995multiplex, jerri1977shannon}.
In Section \ref{subsec:dct} we propose a formalization of the textbook models of \gls{tbi} and \gls{mbi} in terms of the \gls{dct}-II.
A naive reconstruction procedure involves manipulating this result, which ultimately requires the use of the \gls{idct}.
However, such procedure exhibits some limitations for the \gls{mbi} case in terms of spectral resolution.
In Section \ref{subsec:spectral_resolution}, we give a summary of such limitations \cite{cook1995multiplex}.

Alternatively, one may also approach the reconstruction problem as \textit{least square} solution, which instead involves a Moore-Penrose inversion of the transfer matrix. In Section \ref{subsec:conditioning}, we propose to examine this approach in terms of the condition number of the transfer matrix. We specifically highlight how the system's physical parameters affect this condition number and emphasize the need for regularization even in the ideal textbook \gls{mbi} scenario.

While sections \ref{subsec:transfer_matrix_representation}, \ref{subsec:sampling_analysis}, and \ref{subsec:spectral_resolution} are derived from the literature, we streamline the findings with practical insights on the algebraic characteristics of the corresponding transfer matrix. These conclusions are utilised to derive our proposed extended analysis in sections \ref{subsec:dct} and \ref{subsec:conditioning}.

\subsection{Transfer Matrix Representation}
\label{subsec:transfer_matrix_representation}

The discrete representation of the response function is obtained by sampling the wavenumbers domain into $K$ points $\bm{\sigma} \in \{\sigma_0, \dots, \sigma_{K-1}\}$. Additionally, we consider devices capable of generating a set of \glspl{opd} $\bm{\delta} \in \{\delta_0, \dots, \delta_{L-1}\}$. The corresponding captured optical intensities $y(\delta_l)$ for $l\in\{0, ..., L-1\}$ effectively sample the interferogram into $L$ points.

We can then define the \textit{transfer matrix} $\matr{A}\in\mathbb{R}^{L\times K}$, whose elements $a_{lk}= \matr{A}_{[l, k]} = A(\delta_l, \sigma_k)$ are obtained by sampling the response function.
This transfer matrix $\mathbf{A}$ defines the transformation from the discrete support of wavenumbers $\bm{\sigma}$ to that of \glspl{opd} $\bm{\delta}$, and each row of $\mathbf{A}$ can be interpreted as an interferometer's response at a given \gls{opd}.

If the transfer matrix is used for system inversion, having fewer observations than unknowns (i.e., $L \leq K$) results in an \textit{underdetermined} linear system. Conversely, if there are more observations than unknowns, the system is \textit{overdetermined}.

In the analysis of the following sections, we assume that the coefficients of $\mathbf{A}$ strictly follow the models described in eq. \eqref{eq:transmittance_response_michelson_nonmonochromatic} and \eqref{eq:transmittance_response_fabry_perot_infty} for the \gls{tbi} and \gls{mbi}, respectively.

\def \xx {0.215}
\def \yy {13}

\begin{figure*}
	\centering
	\begin{tabular}{c|c|c|c|c}
		&
		\textbf{\gls{tbi} (Michelson)}
		&
		\multicolumn{3}{c}{\textbf{\gls{mbi} (\glsfmtlong{fp})}}
		\\
		\hline
		&
		\;\;\;\;$\mathcal{T} = 0.5$
		&
		\;\;\;\;$\mathcal{R} = 0.2$
		&
		\;\;\;\;$\mathcal{R} = 0.5$
		&
		\;\;\;\;$\mathcal{R} = 0.8$
		\\
		\hline
		\raisebox{\yy mm} {\rotatebox[origin=c] {90} {\textbf{Transfer Matrix}}}
		&
		\includegraphics[width=\xx\textwidth]{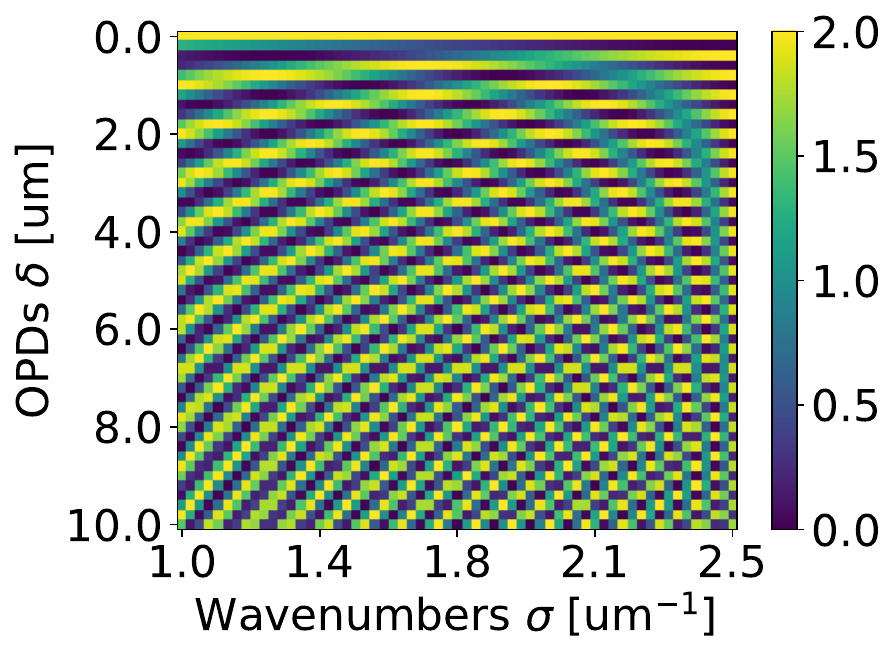}
		&
		\includegraphics[width=\xx\textwidth]{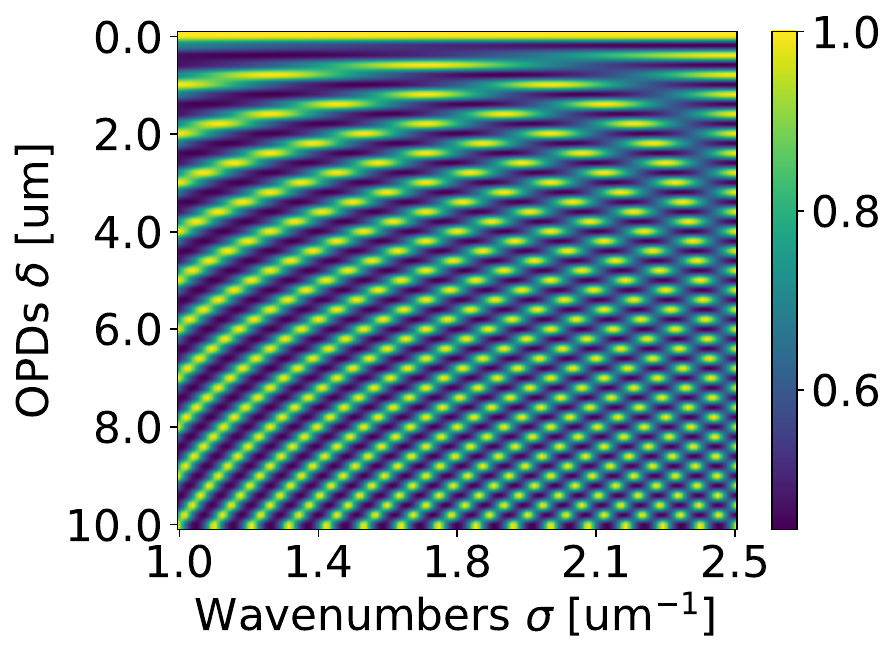}
		&
		\includegraphics[width=\xx\textwidth]{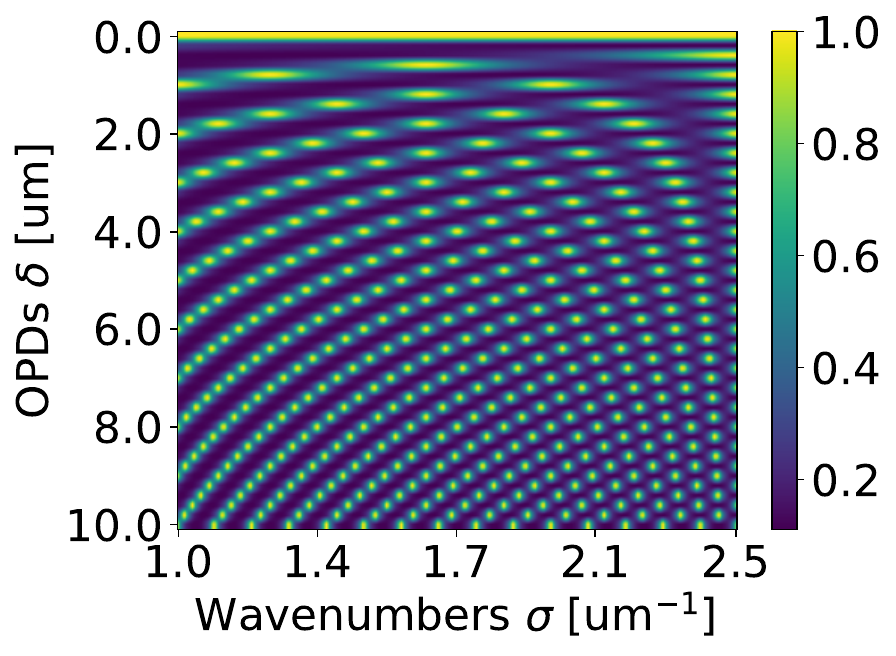}
		&
		\includegraphics[width=\xx\textwidth]{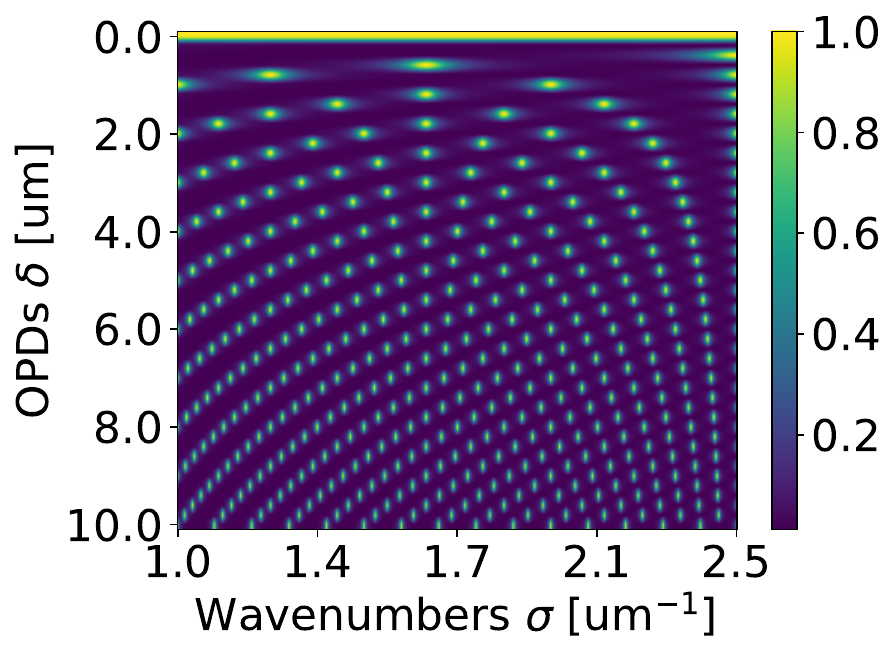}
		\\
		\raisebox{\yy mm} {\rotatebox[origin=c] {90} {\textbf{Singular Values}}}
		&
		\includegraphics[width=\xx\textwidth]{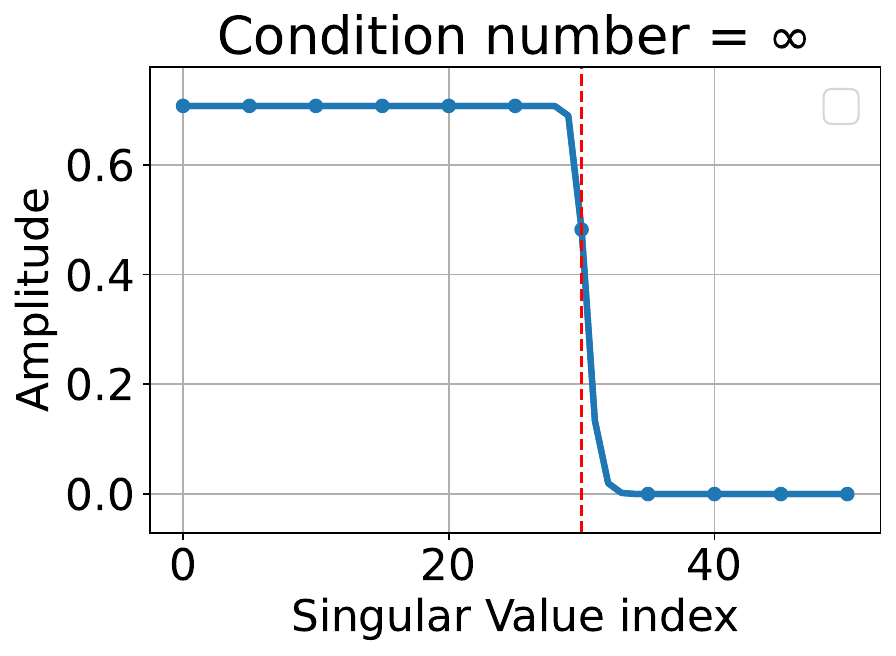}
		&
		\includegraphics[width=\xx\textwidth]{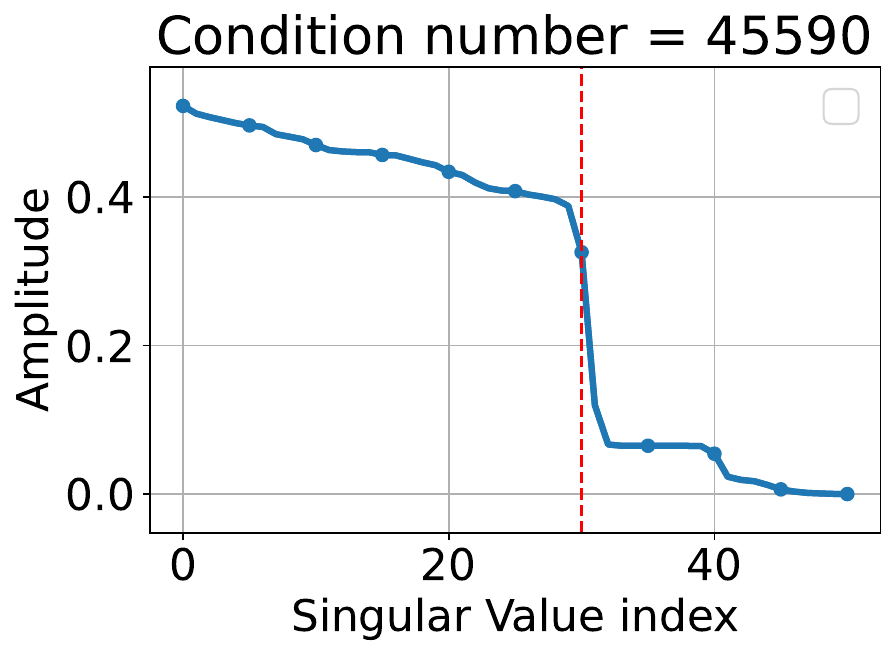}
		&
		\includegraphics[width=\xx\textwidth]{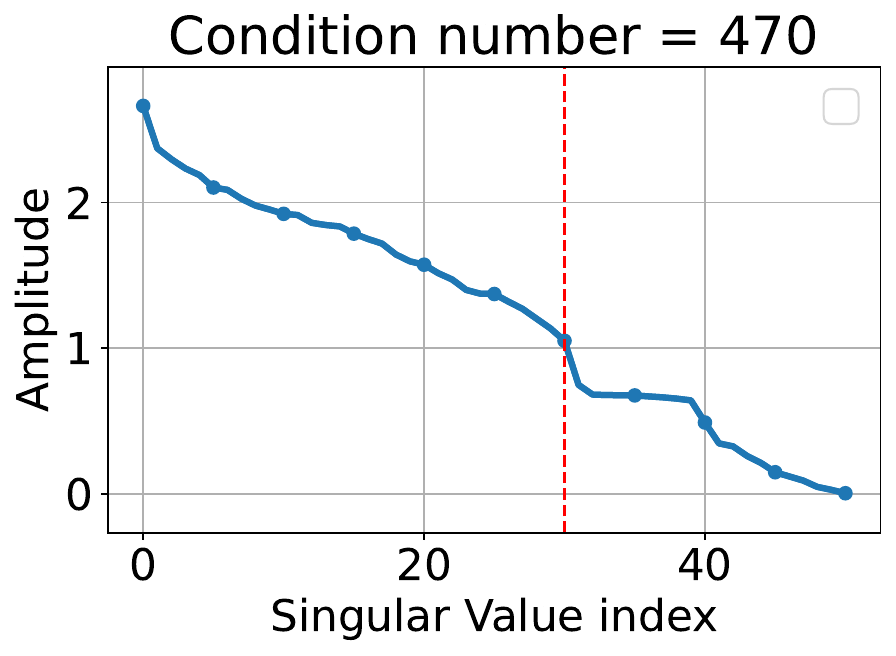}
		&
		\includegraphics[width=\xx\textwidth]{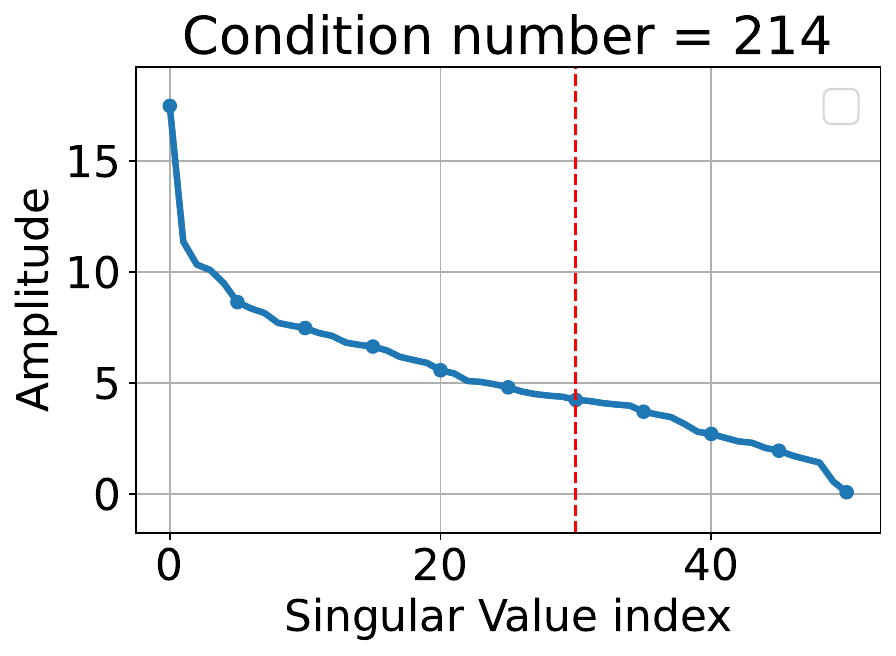}
		\\
		\raisebox{\yy mm} {\rotatebox[origin=c] {90} {\textbf{Interf. Response}}}
		&
		\includegraphics[width=\xx\textwidth]{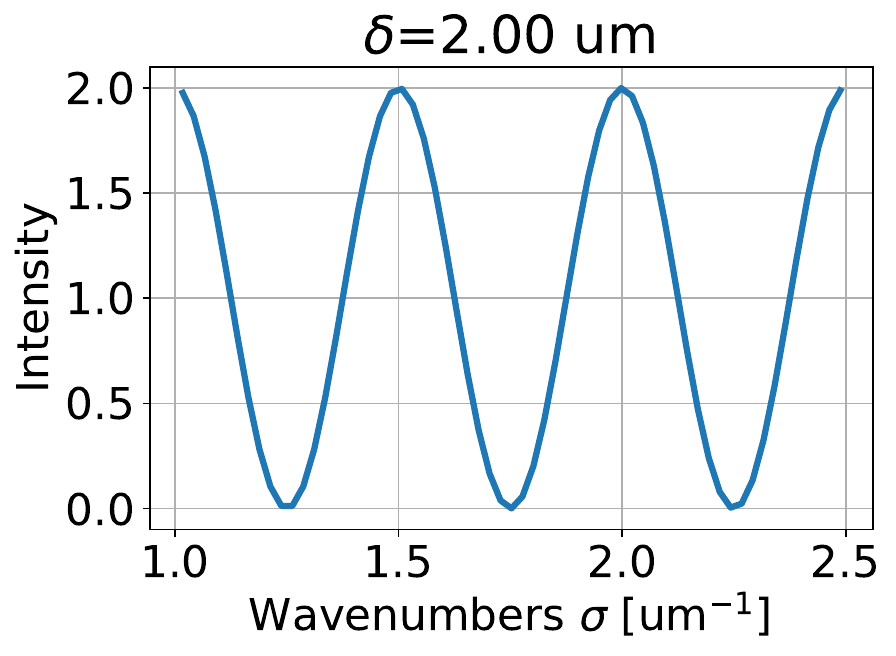}
		&
		\includegraphics[width=\xx\textwidth]{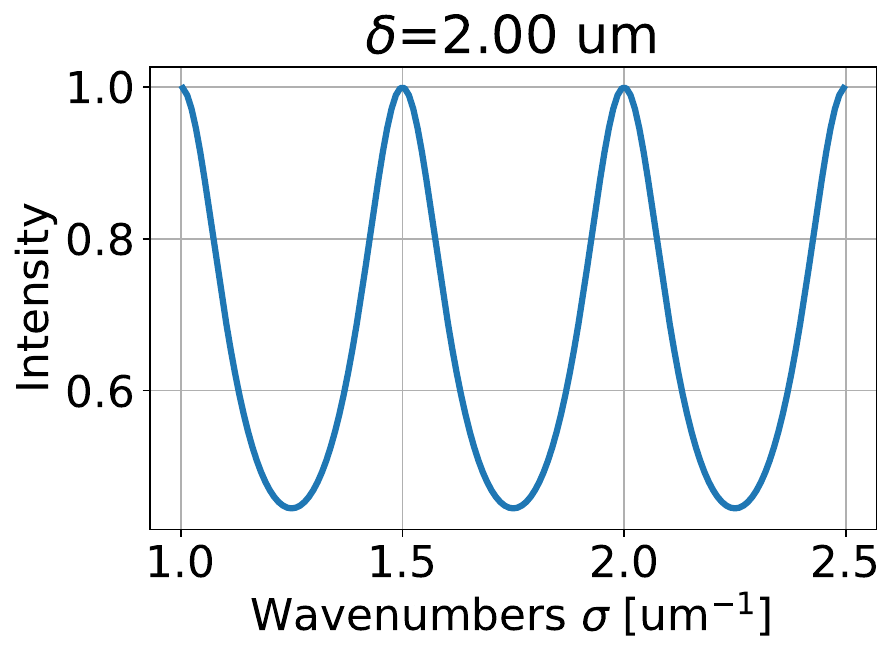}
		&
		\includegraphics[width=\xx\textwidth]{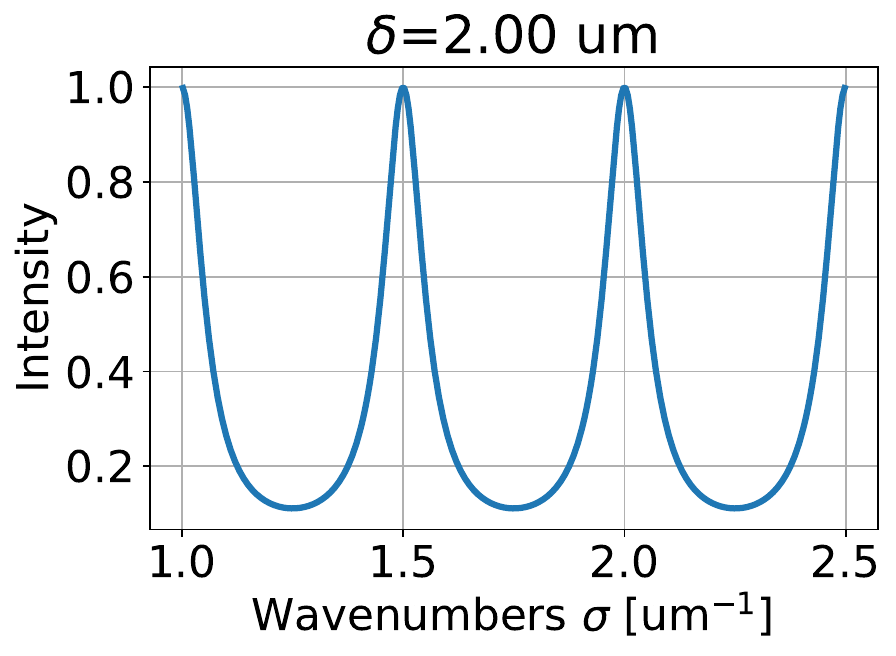}
		&
		\includegraphics[width=\xx\textwidth]{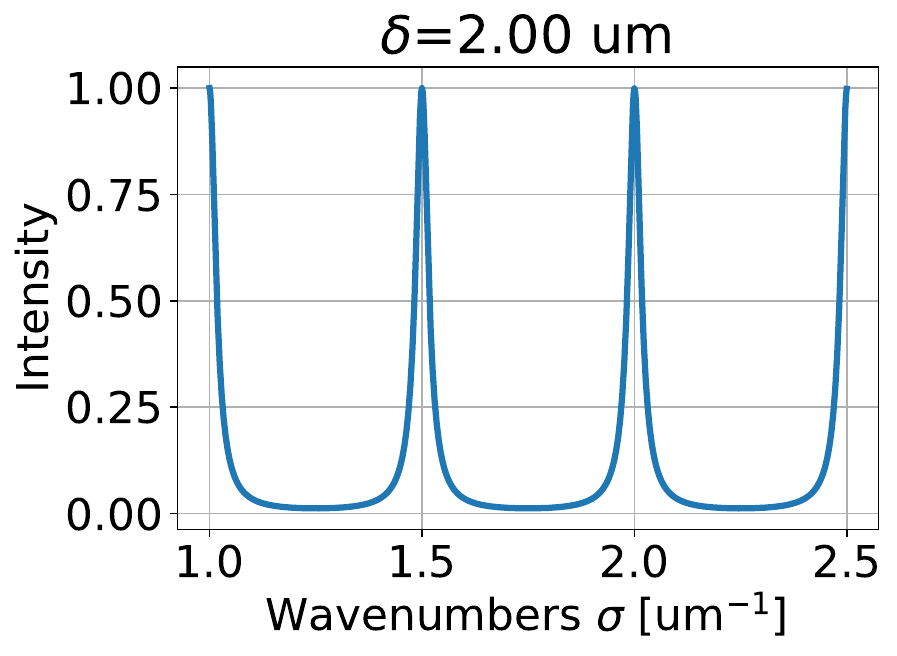}
		% \\
		% \raisebox{\yy mm} {\rotatebox[origin=c] {90} {\textbf{Fourier Transform}}}
		% &
		% \includegraphics[width=\xx\textwidth]{figures/direct_model/mich_oversampled/transfer_matrices/opd_dct.pdf}
		% &
		% \includegraphics[width=\xx\textwidth]{figures/direct_model/fp_0_low_r/transfer_matrices/opd_dct.pdf}
		% &
		% \includegraphics[width=\xx\textwidth]{figures/direct_model/fp_0_med_r/transfer_matrices/opd_dct.pdf}
		% &
		% \includegraphics[width=\xx\textwidth]{figures/direct_model/fp_0_high_r/transfer_matrices/opd_dct.pdf}
	\end{tabular}
        \caption{
            Transfer matrices of some \gls{tbi} (Michelson) and \gls{mbi} (\gls{fp}) instruments. Each row in the transfer matrices represents the interferometer response at a given \gls{opd}.
            The spectral support of the instruments is $\Omega \in [1, 2.5] \, \si{\um}^{-1}$, falling in the near-infrared and visible domains.
            The range of \glspl{opd} is arbitrarily chosen with $L=51$ samples and a step of $\Delta\delta = 0.2 \, \si{\um}$, giving a Nyquist wavenumber $\sigma_{\textrm{Nyq}} = \sigma_{\max}$.
            As such, $\Omega$ covers a ratio $\alpha=0.6$ of the Nyquist range $[0, \sigma_{\textrm{Nyq}}]$.
            The wavenumbers in \gls{tbi} are oversampled such that $K > L$ in order to observe the rankness and the sampling limit via the singular values, while those of \gls{mbi} follow eq. \eqref{eq:sampling_analysis_wavenumbers_mbi}.
            % \textcolor{red}{Why do you plot the OPD response in term of $\sigma$ instead of $\phi = 2\pi\sigma\delta$?}
        }
	\label{fig:transmittance_responses}
\end{figure*}

\subsection{Sampling Analysis}
\label{subsec:sampling_analysis}

For a proper representation of the response function $A(\delta, \sigma)$, it is necessary that the transfer matrix $\mathbf{A}$ verifies the sampling theorem \cite{jerri1977shannon, cook1995multiplex}. This is required in order to properly represent the physical phenomena described by the continuous model.
In particular, two conditions can be identified: one in the \gls{opd} domain and one in the wavenumber domain.
% wavenumber domain is required to properly represent the oscillations of the filtering effect of the interferometers, while the one in the domain in the \glspl{opd}, is required to assure that the interferogram is properly sampled.

\subsubsection{\glsfmtshort{opd} domain}

The \gls{opd} domain condition defines the main guidelines for the manufacturing of the device in order to accurately sample the observed interferogram $y(\delta)$. To this end we assume that the instrument samples in an \gls{opd} range $\bm{\delta} \in \{0, \dots, \delta_{\max}\}$ with a fixed step size $\Delta\delta$.

Using the sampling theorem, the maximum wavenumber representable by the discrete interferogram is $\sigma_{\text{Nyq}}=1/(2\Delta\delta)$. 
Given the spectral support of the device $\Omega=[\sigma_{\min},\,\sigma_{\max}]$, we need to impose $\sigma_{\text{Nyq}}\geq \sigma_{\max}$, obtaining \cite{cook1995multiplex}:
\begin{equation}
    \Delta\delta
    \leq
%    1/(2 \sigma_{\max})\,.
    \frac{1}{2 \sigma_{\max}}\,.
    \label{eq:sampling_analysis_opd_step_tbi}
\end{equation}

\subsubsection{Wavenumber domain}
The condition of the wavenumber domain is required to properly represent the oscillations of the filtering effect of the interferometer by sampling it without aliasing. In other words, $\matr{A}$ must be able to represent the full spectral support from the responses shown in eq. \eqref{eq:transmittance_response_michelson_nonmonochromatic} and \eqref{eq:transmittance_response_fabry_perot_fourier_series_N_waves} in the \gls{tbi} and \gls{mbi}, respectively.
This is equivalent to representing the highest cosine oscillation, yielding the following conditions for the sampling rate $\Delta\sigma$ of the wavenumbers \cite{jerri1977shannon}:
%\begin{subequations}
%    \begin{align}
%        \Delta\sigma
%        &\leq
%        1/(2 \delta_{\max})
%%        \frac{1}{2 \delta_{\max}}
%		\,,
%        &\text{(\gls{tbi}),}
%        \label{eq:sampling_analysis_wavenumbers_tbi}
%        \\
%        \Delta\sigma
%        &\leq
%        1/(2 \, (N-1) \, \delta_{\max})
%%        \frac{1}{2 \, (N-1) \, \delta_{\max}}
%		\,,
%        &\text{(\gls{mbi}).}
%        \label{eq:sampling_analysis_wavenumbers_mbi}
%    \end{align}
%\end{subequations}
\begin{equation}
	\Delta\sigma
	\leq
%	1/(2 \, (N-1) \, \delta_{\max})
	    \frac{1}{2 \, (N-1) \, \delta_{\max}}
	\,,
	\label{eq:sampling_analysis_wavenumbers_mbi}
\end{equation}
whereas in the \gls{tbi}, as $N=2$, this leads to $\Delta\sigma \leq 1/(2 \delta_{\max})$.

If eq. \eqref{eq:sampling_analysis_wavenumbers_mbi} is not verified, the transfer matrix $\matr{A}$ does not properly represent the continuous system, as some of the interferometer responses are aliased.
%
% On the other hand, as $\Delta\sigma$ gets further reduced than the limit, $\matr{A}$ remains a proper discretization of the forward continuous system. 

% However, in the context of inversion, while $K$ augments within a given wavenumber range $\Omega$:
% \begin{itemize}
% 	\item There is a limit to how informative this improvement can get.
% 	This is related to the recoverable spectral resolution given a $\delta_{\max}$, which is discussed in Section \ref{subsec:spectral_resolution}.
	
% 	\item One could end up with less observations than unknowns, i.e., $L \leq K$, defining an underdetermined linear system, or, otherwise, an overdetermined one. 
 % In either cases, a straightfoward solution would to apply a ``minimum norm least squares solution''.
	% This is discussed in Section \ref{subsec:conditioning} with respect to the matrix rank and condition number.
% \end{itemize}

\subsection{Proposed formulation in terms of the \glsfmtshort{dct}}
\label{subsec:dct}

Here, we formalize the conditions under which the physical models of the \gls{tbi} and \gls{mbi} can be expressed in terms of the \gls{dct}.
Namely, the \textit{Fourier cosine transforms} in eq. \eqref{eq:interferogram_formation_michelson_nonmonochromatic} and \eqref{eq:interferogram_formation_fabry_perot} become analogous to the \gls{dct}-II.
% \cite{AhmeNR74:tc}.

For that, considering the full Nyquist bandwidth $\Omega = [0, \sigma_{\text{Nyq}}]$, the \glspl{opd} and wavenumbers have to be regularly sampled into an equal number of samples $K = L$, with steps $\Delta\delta$ and $\Delta\sigma$, respectively, such that $\forall \, l, k \in \{0, \dots, K-1\}$:
\begin{equation}
	\delta_l = l \Delta\delta
	\,
	,
	\;\;\;
	\sigma_k = \left(k+\frac{1}{2}\right) \Delta\sigma
	\,
	,
	\;\;\;
	\Delta\sigma \Delta\delta = \frac{1}{2 K}
    \,,
	\label{eq:dct_conditions}
\end{equation}
Then, we can write the cosines in the models \eqref{eq:transmittance_response_michelson_nonmonochromatic} and \eqref{eq:transmittance_response_fabry_perot_infty} as:
\begin{equation}
	\cos \left[ 2 \pi \, \delta_l \, \sigma_k \right]
	=
	\cos\left[\frac{\pi}{K} \left(k + \frac{1}{2}\right)l\right]
	.
	\label{eq:cosine_dct}
\end{equation}

The above equation shows that the \gls{dct}-II matrix representation is a special case of the transfer matrix of \gls{tbi}, and that of \gls{mbi} without the harmonics, when the conditions in \eqref{eq:dct_conditions} are met.

Applying eq.~\eqref{eq:cosine_dct} to \eqref{eq:interferogram_formation_michelson_nonmonochromatic} and \eqref{eq:interferogram_formation_fabry_perot}, respectively, we obtain:
\begin{subequations}
    \def \sumn {\sum_{n=1}^{N - 1}}
    \begin{align}
        y_l
        &=
        \frac{1}{2}y_0
        +
        \dct_l(\vect{q} \hadamprod \vect{x})
        \,,
        &\text{(TBI)}\,,
        \label{eq:interferogram_formation_michelson_nonmonochromatic_discrete}
        \\
        y_l
        &=
        \innerproduct{\vect{q}}{\vect{x}}
        +
        2
        \sumn
        \dct^{(n)}_l(\vect{q} \hadamprod          \vect{r}^n \hadamprod \vect{x})
        \,,
        &\text{(MBI)}\,,
        \label{eq:interferogram_formation_fabry_perot_2_wave_discrete}
    \end{align}
\end{subequations}
where we denote by $\vect{q} \in \R{K}$ and $\vect{r} \in \R{K}$ as the discretized versions of $\mathcal{Q}(\sigma)$ and $\mathcal{R}(\sigma)$ respectively, and by $\hadamprod$ and $\hadamdiv$ as the Hadamard (element-wise) product and division respectively. Additionally, $\dct_l(\cdot)$ denotes the $l$-th element of the \gls{dct}, while $\dct^{(n)}(\cdot)$ denotes the \gls{dct} carried out with the set of oscillations $\{n\delta_l\}$ $\forall n \geq 2$, i.e., the harmonics.
Note that imposing $K=L$ in eq. \eqref{eq:interferogram_formation_fabry_perot_2_wave_discrete}
% following eq. \eqref{eq:dct_conditions},
causes aliasing in the presence of non-negligible harmonics as it violates condition \eqref{eq:sampling_analysis_wavenumbers_mbi}.
% as eq. \eqref{eq:sampling_analysis_wavenumbers_mbi} implies that $K \geq (N-1) L$.

%\input{sections/old/spectral_resolution.tex}

\subsection{Spectral Resolution}
\label{subsec:spectral_resolution}

In this section, we discuss some of the implications related to inverting the interferogram with approaches that are customarily employed in the \gls{tbi} regime.
Given the representation in terms of the \gls{dct}, it is often assumed to be a natural fit to apply approaches that operate in the Fourier domain, such as the one that will be discussed in Section \ref{subsec:inverse_discrete_cosine_transform}. We carry out this analysis in terms of the spectral resolution of the reconstructed spectrum.

Having an \gls{opd} support within $[0, \delta_{\max}]$ is equivalent to windowing the interferogram signal by $\delta_{\max}$. In the wavenumber domain, this is equivalent to a convolution by a cardinal sine function of width $1/(2 \delta_{\max})$.
Roughly, that is equal to the spectral resolution \cite{cook1995multiplex}.
% In \gls{mbi}, by a factor of  $n\in\{2, \dots, N-1\}$.

In \gls{mbi}, applying the \gls{idct} recovers not just the spectrum in $[\sigma_{\min}, \sigma_{\max}]$, but also some replicas placed at $[n\sigma_{\min}, n\sigma_{\max}]$, i.e. broadened by a factor of $n$, with $n\in\{2, \dots, N-1\}$. These replicas exhibit an exponential decay due to the harmonics.
One could think of recovering the signal directly from those replicas, which enhances the spectral resolution roughly to $1/(2n\delta_{max})$. While these broadened replicas provide finer detail, the robustness to noise is reduced because the replicas' intensities are attenuated by a factor of $\mathcal{R}^n$.

Regardless, this technique is often unfeasible, due to overlaps between these replicas. 
To avoid such overlaps, we must meet the condition: 
\begin{equation}
    \frac{\sigma_{\max}-\sigma_{\min}}{ \sigma_{\min}} \leq \frac{1}{N-2}\,.
    \label{eq:overlap}
\end{equation}
This is normally not a problem for \gls{tbi} system, as this equation is automatically verified.
Additionally, the situation worsens for larger $\Delta\delta$, as potential overlaps can arise through aliasing.  Specifically, to represent all replicas without aliasing, we must be able to represent the spectrum up to a maximum wavelength $(N-1)\sigma_{\max}$. This imposes a stricter constraint than eq. \eqref{eq:sampling_analysis_opd_step_tbi}, formulated as follows:
\begin{equation}
    \Delta\delta \leq \frac{1}{2(N-1) \sigma_{\max}}\,.
    \label{eq:sampling_analysis_opd_step_mbi}
\end{equation}

In most practical scenarios, the above condition is overly restrictive for manufacturing real devices, especially multi-aperture ones. Therefore, performing such inversion for the \gls{mbi} with the same basic techniques that are applied for \gls{tbi} instruments is often inefficient. It is instead preferable to conduct a more ad-hoc analysis, which we discuss in the following section.

Another manufacturing constraint is given by the need to increase of $\delta_{\max}$ to improve the spectral resolution. This requirement limits the miniaturizing capability of the instrument, particularly for multi-aperture devices \cite{GuerLFD18:imspoc}.

\subsection{Proposed System Analysis and Condition Number}
\label{subsec:conditioning}

%In the discrete representation, when treated as an inverse problem based on the transfer matrix, it is easy to see in eq. \eqref{eq:sampling_analysis_wavenumbers_mbi} that the multiplicative factor $N$ can lead to more unknowns than observations, i.e., $K > L$.
%In the context of spectral reconstruction via the pseudo-inverse of the matrix, which will be detailed in Section~\ref{subsec:pinv}, this defines an underdetermined system.
%Moreover, inverse problems are typically ill-posed or ill-conditioned.
%We discuss the conditions of this inversion via the condition number and the matrix rank in each case.

When the spectrum reconstruction is obtained as a the least square solution, the approach involves a Moore-Penrose inversion
% \cite{penrose1955generalized}
of the transfer matrix, as detailed in Section \ref{subsec:pinv}.
In this context, it is useful to characterize the inverse problem in terms of Hadamard, to verify if there is a need for regularization.
We discuss this problem in terms of the condition number and of the matrix rank, both for the \gls{tbi} and for the \gls{mbi}.
In the case of \gls{tbi}, the system is often overdetermined and orthogonal, while in \gls{mbi}, it is typically ill-conditioned, but could be overdetermined or underdetermined. % as discussed in Section \ref{subsec:sampling_analysis}.

We denote by $R_{\matr{A}} \leq \min(L, K)$ the rank of $\matr{A}$ and by $\{\psi_r\}_{r \in [1, ..., R_{\mathbf{A}}]}$ the set of singular values in descending order. The condition number of $\mathbf{A}$ is defined as \cite{Idie13:book}:
\begin{equation}
	c=\psi_1 / \psi_{R_{\mathbf{A}}} \geq 1,
	\label{eq:condition_number}
\end{equation}
where if $c=1$, the singular values are equal and the problem is well-conditioned.
If $c$ is large or $\infty$, the problem becomes ill-conditioned or ill-posed, respectively.

\figurename\,\ref{fig:transmittance_responses} visualizes different cases of transfer matrices of \gls{tbi} and \gls{mbi} models.
We consider one case of \gls{tbi} based on the Michelson model, as well as three cases of \gls{mbi} based on the Fabry-Perot model with increasing reflectivity. In \gls{tbi} the wavenumbers are oversampled (beyond the Nyquist limit), while in \gls{mbi} the sampling rate follows eq. \eqref{eq:sampling_analysis_wavenumbers_mbi}.
In each case, we show the plots of singular values with the computed condition number. For the sake of demonstration, we plot an arbitrary interferometer responses (that is, a row of the matrix) at a given \gls{opd}.
% Finally, we plot the corresponding Fourier transform to show the harmonic Dirac terms.
In the following, we use \figurename\,\ref{fig:transmittance_responses} as support to give some numeric examples, and we denote by $0 < \alpha \leq 1$ the portion that $\Omega$ occupies of the full Nyquist bandwidth $[0, \sigma_{\textrm{Nyq}}]$.

%We assume that $\Delta\delta$ operates in the following regime:
%\begin{equation}
%	\sigma_{\max} \leq \frac{1}{2\Delta\delta} \leq (N-1) \sigma_{\max}
%\end{equation}
%or in other words, we verify the Nyquist condition \eqref{eq:sampling_analysis_opd_step_tbi}, but not eq. \eqref{eq:sampling_analysis_opd_step_mbi} if $N\geq 3$.

Here, we are not limited to the condition of no overlap of eq. \eqref{eq:overlap}, contrary to the case of \gls{idct}. The signal can be fully recoverable as long as the continuous acquisition system is properly sampled and the condition number is low.
To avoid aliasing we would however still require the sampling condition to hold (with the $N-1$ term), but we assume a more realistic case where $\sigma_{\max} < \sigma_{\text{Nyq}} < (N-1)\sigma_{\max}$. Specifically \figurename \ref{fig:transmittance_responses} shows $\sigma_{\text{Nyq}} = \sigma_{\max}$.
Technically, this allows for the same spectral resolution of Section \ref{subsec:spectral_resolution} but without the overlap condition, yet the condition number has to be verified.

\subsubsection{TBI}
The matrix is generated from fundamental cosine functions with equal magnitudes whose oscillations are equal to the \glspl{opd}, $\{\cos(2 \pi \delta \sigma)\}$. If $\alpha=1$ so that $\Omega = [0, \sigma_{\textrm{Nyq}}]$, we have an orthonormal basis. If $\Omega$ is sampled at the Nyquist limit following the conditions of Section \ref{subsec:dct}, $\matr{A}$ becomes square and orthogonal. Then, $R_{\matr{A}} = K = L$ and $c=1$.

In practical scenarios, where $\alpha < 1$ so that $\Omega \subset [0, \sigma_{\textrm{Nyq}}]$, we discuss that $R_{\matr{A}} \approx \alpha L$. For instance:
\begin{itemize}
    \item If $\Delta\sigma = 1 / (2\delta_{\max})$, i.e., at the Nyquist limit, we have $K \approx \alpha L$ samples in $\Omega$ (more rows than columns). Only the columns are linearly independent, which makes $\matr{A}$ semi-orthogonal.
    The system is said to be overdetermined (more observations than unknowns), with $R_{\matr{A}} = K < L$. Here, the problem is well-posed since $c=1$.
    
    \item If $\Delta\sigma < 1 / (2\delta_{\max})$, i.e., oversampled, we have a linear dependency in both the columns and rows as $\Omega$ covers only a bandpass interval whose limit is at $\alpha L$ samples.
    $\matr{A}$ is rank-deficient with $R_{\matr{A}} = \alpha L < \min{(K, L)}$. The singular values after the $(\alpha L)$-th index drop to $0$, leading to an ill-posed problem as $c \rightarrow \infty$.
\end{itemize}
% \sout{
% $\matr{A}$ being rank-deficient means that there is a loss of information and a low spectral resolution that is limited only to the choice of $\delta_{\max}$.
% }
% \textcolor{red}{There is no loss of information. It's just that you are trying to recover the spectra with a smaller step $\Delta\sigma$ than it is allowed by the spectral resolution.}

To illustrate this more clearly, let us consider a numerical example, referring to the Michelson case of \figurename~\ref{fig:transmittance_responses}. In this example, we have $L=51$, $\delta_{\max} = 10 \si{\um}$, and $\Omega=[1, 2.5]\,\si{\um}^{-1}$ with $\alpha=0.6$. The wavenumbers are oversampled such that $K > L$.
The interferometer response shows a perfect cosine form.
The singular values drop to $0$ after the $31$-st index, which means that $\matr{A}$ is rank-deficient where $R_{\matr{A}} = 31 \approx \alpha L$ and $c \rightarrow \infty$.
% \sout{
%     In this case, the problem is ill-posed.
% }
% \textcolor{red}{It is, but just because you made it so yourself, so that could be criticized. ``You can choose whatever $\Delta\sigma$ you want, so why choosing it to be ill-posed?'' You can write something like:}
% \textcolor{blue}{
Therefore, for the \gls{tbi}, assuming eq.~\eqref{eq:sampling_analysis_opd_step_tbi}, one can easily avoid the ill-posedness of the problem by properly choosing the sampling step $\Delta\sigma\geq 1 / (2\delta_{max})$, that is the minimum in terms of spectral resolution.
% }

\subsubsection{MBI}
We assume here to be in the case in which either eq. \eqref{eq:sampling_analysis_opd_step_mbi} is not necessarily verified, or in other words some aliasing may appear due to the harmonics.
%
% (forgive the fast wording.) \\
% As the reflectivity increases, even though the forward model requires finer sampling rate in the wavenumbers to properly represent the continuous one, following \eqref{eq:sampling_analysis_wavenumbers_mbi}
%
In such case, the inversion of the matrix (e.g., the Moore-Penrose inverse) imposes the physical constraint that the \glspl{opd} are undersampled, causing the harmonics to mirror around $\sigma_{\textrm{Nyq}} = 1 / (2\Delta\delta$) and overlap.

\def \x {0.7}

\begin{figure}[t]
	\centering
	\includegraphics[width=\x\linewidth]{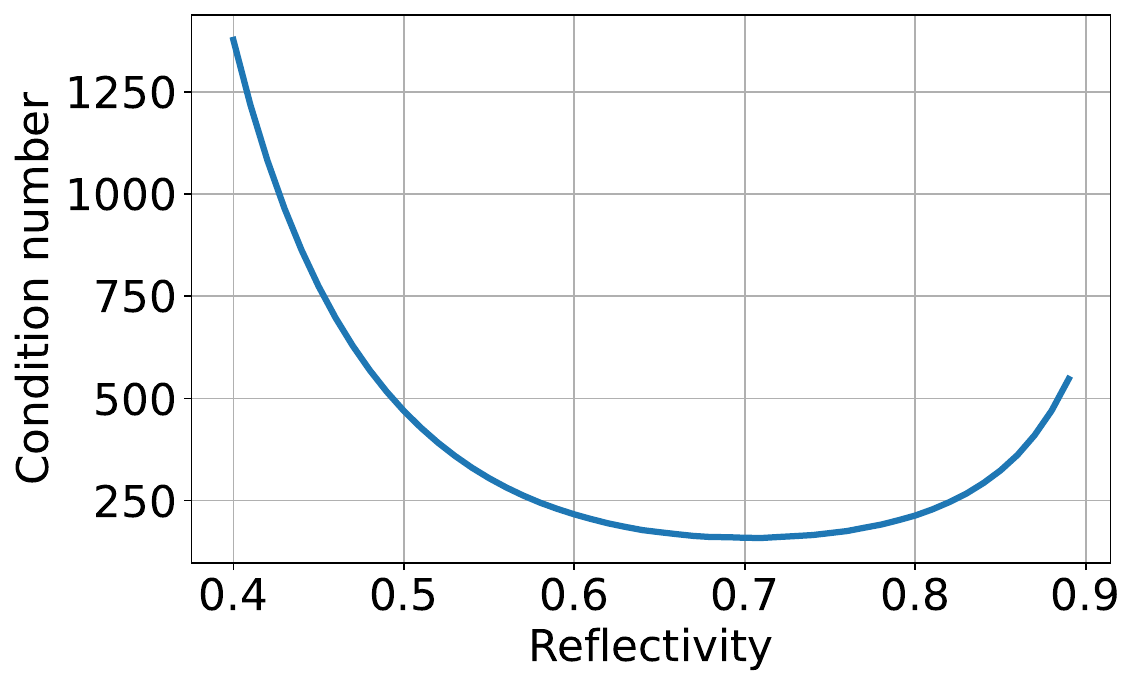}
	\caption{
		The change of the condition number of the transfer matrix of a \gls{fpi} with respect to reflectivity. The range of \glspl{opd} and wavenumbers is the same as that of \figurename~\ref{fig:transmittance_responses}. The condition number shows a minimum at $\mathcal{R}=0.7$.
	}
	\label{fig:condition_number_reflectivity}
\end{figure}

To investigate this effect, coming from eq. \eqref{eq:transmittance_response_fabry_perot_fourier_series_N_waves}, we can decompose the transfer matrix $\matr{A}$ into the sum of components $\matr{A}^{(n)}$, each generated from weighted cosines of the $n$-th harmonic, as follows:
\begin{equation}
        \matr{A}_{[l, k]}
        =
        \sum_{n=0}^{N-1}
        \matr{C}_{[k]}^{(n)}
        % \cos(2\pi \, n\delta_l \, \sigma_k)
        \cos\left[\frac{\pi}{K} n l \left(k + \frac{1}{2}\right)\right]
        =
        \sum_{n=0}^{N-1}
        \matr{A}_{[l, k]}^{(n)}
        \,,
\end{equation}
where $\matr{C}_{[k]}^{(n)}$ is the discretized version of $C_n(\sigma)$ from eq. \eqref{eq:transmittance_response_fabry_perot_fourier_coefficients}.
% When the reflectivity is very small, the components for $n \geq 2$ are exponentially attenuated compared to the fundamental component. The singular values then show a steep curve. As the reflectivity increases, the harmonics gain more weight and the steepness smoothes out.

% progressively disrupting the orthogonality of the fundamental terms that were observed in the case of \gls{tbi}.
% \textcolor{red}{What is the definition of "disrupting the orthogonality" exactly?}
%
% As the reflectivity increases, the harmonic order $N$ progressively increases, and the spectral support $\Omega$ requires finer sampling rate $\Delta\sigma$ following \eqref{eq:sampling_analysis_wavenumbers_mbi}. This leads to a multiplicative increase in the number of samples $K$.
% \textcolor{red}{I do not get the one above. In theory, the issue is not on the wavenumbers, but on the \glspl{opd}? With the wavenumbers we can always choose the one that allows for the best available resolution, assuming we find it.}

% \textcolor{red}{I would talk about the progression here.}
% \textcolor{blue}{

These harmonics (i.e., $n \delta_l$ $\forall n\geq2$) result in cosines at aliased \gls{opd} with exponentially smaller weights, which overlap with existing fundamental cosines.

If $\mathcal{R}$ is very small, the overlaps involve very fast decaying cosines, which barely impact the recoverability in the range of interest $\Omega$. However, the harmonic coefficients that fall outside of that support are too much attenuated to allow for extra information to recovering the signal, making them indistinguishable to noise.
If $\mathcal{R}$ is very large, then the terms that fall outside the support $\Omega$ are more informative, while the information within $\Omega$ is mostly disrupted.
Therefore, a balance has to be found. We can roughly investigate this effect by analyzing the singular values of $\matr{A}$.
We analyze the condition number for $\alpha=0.6$ in \figurename~\ref{fig:transmittance_responses}.
% }

% \sout{
% On one hand, we have more unknowns than observations such that $K > L$, leading to an underdetermined system. Moreover, $\matr{A}$ is no more orthogonal as the singular values start to show a decreasing slope, and we have an ill-conditioned problem as $c>1$.
% On another hand, compared to \gls{tbi}, the previously observed $0$-singular values become progressively non-zero as new independent harmonic terms are formed, increasing the rankness of $\matr{A}$ up to full-rankness $R_{\matr{A}} = \min(K, L)$ and relaxing the dependency on $\delta_{\max}$.
% %
% However, this improvement is limited as observed through the progression of the condition number with respect to the reflectivity. The more the problem is ill-conditioned, the poorer the resolution. This can be due to potential noise amplification \cite{cook1995multiplex}.
% }
%
Let us compare the \glspl{svd} of the \gls{mbi} models.
As the reflectivity increases, the steepness at the $31$-st index is smoothed out as the matrix becomes full-rank.
Furthermore, in \figurename~\ref{fig:condition_number_reflectivity}, while we observe an improvement in the condition number as the reflectivity increases, it reaches a minimum at $\mathcal{R}=0.7$ for the given set of \glspl{opd}.

The condition number shows that there is effectively a need for penalization or regularization even in the ideal textbook formalism.
% Spectrum reconstruction in \gls{mbi} is often an ill-conditioned and underdetermined problem.
%
% With real world non-idealities such as irregular sampling and variable reflectivity, this analysis becomes even more complex. However, the transfer matrix can still embed this kind of knowledge
%
With real-world non-idealities like irregular sampling and variable reflectivity, this analysis becomes more complex. Nevertheless, the transfer matrix can still incorporate and represent this information.
In the following section, we apply the same formalism but go beyond the textbook representation of the transfer matrix. We progressively introduce a more general formalism that is not limited to a specific form of the transfer matrix, gradually building towards the more comprehensive Bayesian framework.

%Instead, we describe an increasingly more generic framework 

% However, we now also try to tackle the non idealities as we showcase an increasingly more generic framework for reconstruction towards a Bayesian framework.

\section{Spectrum Reconstruction}
\label{sec:inversion_algorithms}

In this section, we describe a series of techniques for spectrum inversion, ending with our proposed solution.
Specifically, the reconstruction problem is formalized as finding an estimation $\hat{\vect{x}}\in\R{K}$ of an unknown spectrum $\vect{x} \in \R{K}$ from the observed interferogram $\vect{y} \in \R{L}$ given knowledge of the forward sensing model $\matr{A} \in  \RR{L}{K}$.
Generally speaking, $\matr{A}$ is generated from the acquisition model of a given device following the framework of Section \ref{sec:transfer_matrix}.
To streamline the exposition, these techniques are organized in an order that reflects their progression of applicability from more specific to more general scenarios. This increase in generality comes however with a decrease in computational speed.

In particular, we start with the \gls{idct} in Section \ref{subsec:inverse_discrete_cosine_transform}, the Moore-Penrose inverse in Section \ref{subsec:pinv}, the penalized inversion matrix in Section \ref{subsec:svd_based_approaches}, concluding with the Bayesian formulation in Section \ref{subsec:bayesian_framework}.
Finally, we specialize this framework for our proposed solution in Section \ref{subsec:loris_verhoeven}.

\subsection{Inversion from the Fourier Domain}
\label{subsec:inverse_discrete_cosine_transform}

Under the specific conditions outlined in Section \ref{sec:transfer_matrix}, where the transfer matrix $\mathbf{A}$ strictly adheres to the model defined for the \gls{tbi} or \gls{mbi}, a straightforward reconstruction strategy for the spectrum entails an inversion from its expression given in the Fourier domain by the interferogram.

We begin by examining the \gls{tbi} expression that we derived in eq. \eqref{eq:interferogram_formation_michelson_nonmonochromatic_discrete}. This equation shows that the sampled interferogram $\mathbf{y}$ involves a \gls{dct}-II of the input spectrum. To invert this expression, a natural procedure involves employing the \gls{idct}.
As the \gls{dct} is nearly orthonormal, its inverse operation entails a multiplication by a matrix whose coefficients mirror those of the \gls{dct}. The multiplication by this inverse matrix is formally known as \gls{dct}-III in the literature and denoted by $\idct(\cdot)$ in the following.

% \textcolor{red}{I would call this section something like "Inversion from the Fourier domain". Then I would state which are the conditions to apply these methods, and mention that for the TBI the method is equivalent to the IDCT.}

% We explore the mathematical effect of \gls{idct} to reconstruct the spectrum in each of \gls{tbi} and \gls{mbi}.
%%
%Generally speaking, the advantage of inversion with \gls{idct} is that it is simple, quick, and easy to implement on hardware.
%%
%However, it requires hard conditions mentioned in \eqref{eq:dct_conditions}, is generally not an accurate solution for \gls{mbi} spectroscopy, is limited in terms of spectral resolution even in high reflectivity regimes, and does not account to the presence of noise in the acquisitions.
%
% First, we recall the definition of the \gls{idct}, i.e., the inverse of \eqref{eq:dct_ii_elementwise}. Assuming a discrete signal $\vect{u} \in \R{M}$ and $\vect{v} = \dct(\vect{u}) \in \R{M}$, then $\forall k = \{0, \dots, M-1\}$:
% \begin{equation}
% 	\begin{gathered}
% 		u_k
% 		=
% 		\sum_{l=0}^{N-1} v_l \cos\left[\frac{\pi}{N} l\left(k + \frac{1}{2}\right)\right] - \frac{1}{2} v_0
% 	\end{gathered}
% \end{equation}
% Equivalently, we can write:
% \begin{equation}
% 	\begin{gathered}
% 		\vect{u}
% 		=
% 		\idct(\vect{v})
% 		=
% 		\matr{W}^{\T}
% 		\vect{v}
% 		-
% 		\frac{1}{2}
% 		v_0
% 	\end{gathered}
% \end{equation}
% where $\matr{W}^{\T}$ is the transpose of $\matr{W}$ ($\matr{W}$ is unitary so the inverse is just given by the transpose).

Specifically, by manipulating eq. \eqref{eq:interferogram_formation_michelson_nonmonochromatic_discrete}, the spectrum can be reconstructed using the following procedure:
\begin{equation}
	\begin{aligned}
            \idct\left(\vect{y}- \frac{1}{2} y_0\right)
		= 
		\idct
		\left(
		\dct(\vect{q} \hadamprod \vect{x})
		\right)
		\\
		\iff
		\;
		\hat{\vect{x}}
		=
		\idct\left(\vect{y} - \frac{1}{2} y_0\right)
		\hadamdiv
		\vect{q}\,.
	\end{aligned}
        \label{eq:idct}
\end{equation}

For the \glsfmtshort{mbi}, applying the same methodology presents certain challenges. In fact, performing a similar manipulation on eq. \eqref{eq:interferogram_formation_fabry_perot_2_wave_discrete} yields:
\begin{equation}
    \begin{aligned}
	\idct(\mathbf{y} - \innerproduct{\vect{q}}{\vect{x}})
	&=
	\idct\left(
		2 \mathcal{Y}^{(1)}
		+
		2 \sum_{n=2}^{\infty} \mathcal{Y}^{(n)}
	\right)
	\\
	&=
	2 \, \vect{q} \hadamprod \vect{r} \hadamprod \vect{x}
	+
	\bm{\xi}\,,
    \end{aligned}
\end{equation}
where we have defined the harmonic terms $\mathcal{Y}^{(n)}$ and the corresponding ``residuals'' $\bm{\xi}\in\R{K}$ from the inversion as:
\begin{subequations}
    \begin{align}
        \mathcal{Y}^{(n)} &= \dct^{(n)} (\vect{q} \hadamprod \vect{r}^{n} \hadamprod \vect{x})\,,
        \\
        \bm{\xi} &= \idct \left(2\sum_{n=2}^{\infty} \mathcal{Y}^{(n)}\right)
        \,.
    \end{align}
\end{subequations}

Essentially, the estimation of the spectrum would be accurate up to an unknown bias factor $\innerproduct{\vect{q}}{\vect{x}}$ on the interferogram $\mathbf{y}$, if not for the presence of the residuals $\bm{\xi}$.
This analysis mirrors that of Section \ref{subsec:spectral_resolution}, with the residuals $\bm{\xi}$ representing the replicas due to harmonics; as such, the sampling theorem imposes that eq. \eqref{eq:sampling_analysis_opd_step_mbi} must be met to avoid aliasing.

In \cite{al-saeed-2016-fourier-trans}, an analytical method for spectrum reconstruction has been proposed in order to compensate the harmonic overlap. This method uses a Haar function to expand the \gls{dft} of the interferogram and the unknown spectrum. Howver, the applicability of this method is still limited by reflectivity and transmittance of the acquisition system, which is assumed constant over the spectral range. Moreover, its spectral resolution is limited by the Nyquist bandwidth since it is based on the \gls{dft} of the interferogram.

\subsection{Moore-Penrose inversion (Pseudo-inversion)}
\label{subsec:pinv}

In practical scenario, the device can be characterized with a controlled measurement \cite{PicoGDFL23:oe}. This procedure can be employed to obtain coefficients of the transfer matrix $\mathbf{A}$ which allow for a more realistic representation of the system with respect to the theoretical models described in Section \ref{sec:instrumental_model}.
This includes some non-idealities that affect most real world acquisition systems (e.g., irregular sampling, variable reflectivity, harmonic contribution, measurement noise).
This necessity demands for more general approaches for inversion that can be applied to transfer matrices $\mathbf{A}$ of any sort.

The most naive approach in this direction is to estimate the spectrum through an inversion of $\mathbf{A}$, or more specifically a Moore-Penrose inversion
% \cite{penrose1955generalized}
(also known as the pseudo-inversion), as the matrix is not necessarily square, nor full rank. In other words, the estimation is given by:
\begin{equation}
    \hat{\mathbf{x}} = \mathbf{A}^{\dagger}\mathbf{y}\,,
    \label{eq:pinv}
\end{equation}
where $\mathbf{A}^\dagger\in\mathbb{R}^{L \times K}$ denotes the \glsfmtlong{pinv} of $\mathbf{A}$.
%If the \gls{svd} of $\mathbf{A}$ is given by $\mathbf{A} = \mathbf{U}\mathbf{\Sigma}\mathbf{V}^T$, then $\mathbf{A}^\dagger=\mathbf{V}\bm{\Sigma}^\dagger\mathbf{U}^T$, where $\bm{\Sigma}^\dagger$ is a diagonal matrix whose diagonal contains the reciprocal of all the non-zero singular values of $\mathbf{A}$, ordered in decreasing order.
As the non-zero singular values of $\mathbf{A}^{\dagger}$ are the reciprocal of those of $\mathbf{A}$, the two matrices exhibit the same condition number.
%However, this approach is only applicable if $\mathbf{A}$ has a small enough condition number.
%as it can be easily shown that $\mathbf{A}^\dagger$ and $\mathbf{A}$ have the same condition number.
%If this condition does not hold, the problem is not well-conditioned in the sense of Hadamard, leading to unstable solutions for $\hat{\mathbf{x}}$.
In this case, the analysis of Section \ref{subsec:conditioning} holds, and the approach is applicable when $\mathbf{A}$ has a small enough condition number.

In particular, the \gls{idct} procedure, analyzed in the previous section for the ideal \gls{tbi} expression, is a special case of the pseudo-inversion that operates under well-behaved conditions.

%\textcolor{red}{Here, I would start by stating that, in more general cases, the matrix $\mathbf{A}$ could assume any form, such as those obtained by characterizing the device under controlled measurement scenarios. Then you can state that for this scenarios, a naive approach for estimating the spectrum is through an inversion of $\mathbf{A}$, or more specifically a Moore-Penrose inversion (also known as pseudo-inversion) as the matrix is not necessarily square. However, this approach works properly only under the condition that the condition number of $\mathbf{A}$ is small enough (ideally 1 in the case $\mathbf{A}$ is orthogonal.)}

% A naive solution for \eqref{eq:direct_model} is the closed form $\hat{\mathbf{x}}=\mathbf{A}^{\dagger}\mathbf{y}$, where $\mathbf{A}^\dagger$ denotes the \glsfmtlong{pinv} of $\mathbf{A}$.
% However, as shown in \figurename\,\ref{fig:transmittance_responses}, some of the singular values have zero or very small amplitudes, and the singular values $\psi'_r=1/\psi_r$ of $\mathbf{A}^{\dagger}$ tend to become extremely large.
%Hence, the condition number of $\mathbf{A}^{\dagger}$ can either be $\infty$ or extremely large, showing that the inversion problem cannot be solved without some sort of regularization. 

% In the special case of \gls{tbi} with ideal conditions, $\matr{A}$ and $\mathbf{A}^{\dagger}$ are equivalent to the \gls{dct}-II and \gls{idct}-II respectively.

\subsection{Inversion with Penalized Singular Values}
\label{subsec:svd_based_approaches}

To overcome the limitations of the pseudo-inversion approach, a series of methods were developed in the literature to directly adjust the condition number of $\mathbf{A}^\dagger$, introducing some penalization on its singular values \cite{Idie13:book}.
For such methods, let us impose that the estimation $\widehat{\mathbf{x}}=\widetilde{\mathbf{A}}\mathbf{y}$ is carried out with a modified version $\widetilde{\mathbf{A}}$ of $\mathbf{A}^{\dagger}$. 

Given $r \in \{1, \dots, R_{\matr{A}}\}$, let us define $\psi_r$ and $\widetilde{\psi}_r$ as the singular values of $\mathbf{A}$ and $\widetilde{\mathbf{A}}$, respectively. 
Two of the most widespread techniques for singular value penalization are:

%\textcolor{red}{As the pseudo-inversion is limited to the case ..., a series of approaches were developed in the literature to adjust the condition number of the transfer matrix. In those approaches the singular values of $\mathbf{A}$ are penalized}

%Some simple strategies are available in the literature \cite{Idie13:book}, for which the estimation $\widehat{\mathbf{x}}=\widetilde{\mathbf{A}}\mathbf{y}$ is carried out with a modified version $\widetilde{\mathbf{A}}$ of $\mathbf{A}^{\dagger}$, with penalized singular values $\psi_r'$.
%Among such strategies the most co:
\subsubsection{\Glsentryfull{tsvd}} \cite{Hans90:jssc}, where a given percentage $0 < \lambda < 1$ of the singular values of $\mathbf{A}^\dagger$ is kept unmodified and the rest are set to zero, i.e.:
\begin{equation}
	\widetilde{\psi}_r =
	\begin{cases}
		1 / \psi_{r},
		&
		\text{if } r < \lambda \, R_{\matr{A}}\,.
		\\
		0,
		&
		\textrm{otherwise}\,.
	\end{cases}
\end{equation}

\subsubsection{\Glsentryfull{rr}} \cite{GoluHO99:jmaa}, where the singular values are dampened by a penalization parameter $\lambda>0$:
    	\begin{equation}
   		\widetilde{\psi}_r =
   		\frac{\psi_{r}}
   		{\psi_{r}^2 + \lambda^2}
   		\,,
   	\end{equation}
which can be easily be shown that it is equivalent to the closed-form solution of the following minimization problem:
    \begin{equation}
	\hat{\vect{x}} = \argmin{\vect{x}}{
		\frac{1}{2} \|\vect{y} - \matr{A} \vect{x}\|^2_2
		+
		\lambda^2 \, \|\vect{x}\|_2^2
	}\,.
	\label{eq:regularized_techniques_ridge_regression}
\end{equation}

%It can be easily verified that with a proper choice of the parameters $\lambda_{\textrm{\gls{tsvd}}}$ or $\lambda_{\textrm{\gls{rr}}}$, the condition number of $\widetilde{\matr{A}}$ is attenuated with respect to that of $\matr{A}^\dagger$, potentially leading to a well-conditioned problem.

\subsection{Bayesian Inversion}
\label{subsec:bayesian_framework}

In the Bayesian framework, the optical acquisition phenomenon is modeled as:
\begin{equation}
	\vect{y} = \matr{A} \vect{x} + \vect{e}\;,
	\label{eq:direct_model}
\end{equation}
where the observed interferogram $\vect{y}$ is expressed as the sum of a deterministic component $\matr{A}\vect{x}$ and a stochastic additive noise $\vect{e} \in \R{L}$.

In the case $\vect{e}$ is assumed to be Gaussian, the maximum a posteriori estimator $\hat{\vect{x}}$ of the latent variable $\vect{x}$ can be written in the form \cite{LoriV11:ip} :
\begin{equation}
    \begin{aligned}
        \hat{\vect{x}} & = \argmin{\vect{x}}{
            h(\vect{x}) +
            \lambda \, g(\mathbf{L}\vect{x})
        } \\
        & = \argmin{\vect{x}} \frac{1}{2} \|\matr{A} \vect{x} - \vect{y}\|^2_2 +
		\lambda \, g(\mathbf{L}\vect{x})
		\,,
    \end{aligned}
    \label{eq:bayes}
\end{equation}
where $h(\vect{x}) = \frac{1}{2} \|\matr{A} \vect{x} - \vect{y}\|^2_2$ is the data fidelity term, $g(\cdot)$ denotes a scalar functional, $\mathbf{L}$ is a generic linear operator, and $\lambda$ is a regularizing parameter.
One can easily verify that the pseudo-inversion method is a special case, providing the closed form solution given in eq. \eqref{eq:pinv} for the above minimization problem when $\lambda=0$.

Moreover, eq.~\eqref{eq:regularized_techniques_ridge_regression} for \glsentrylong{rr} is a particular application of the Bayesian framework with the function $g(\cdot)$ expressed as the squared $\ell_2$ norm and $\mathbf{L}$ set as the identity operator.
Effectively, the introduction of the regularization function $g(\mathbf{L}\vect{x})$ can be seen as a way to impose the well-posedness to the solution, similarly to how it was done for the regularized singular values. In the framework of Bayesian inference, this can be interpreted as the prior information on the spectra that are expected to be captured by the device.

When obtaining a closed-form solution is not feasible, e.g. when $g(\cdot)$ is not differentiable, \textit{iterative algorithms} become prominent for solving eq. \eqref{eq:bayes}. Approaches such as those based on proximal splitting \cite{BoydV04:book} are commonly employed in such cases. These methods begin with an initial guess, iteratively update the inferred solution in successive steps, and continue this process until convergence is reached.

\subsection{Proposed Reconstruction Solution}
\label{subsec:loris_verhoeven}

In this section, we propose to specify the Bayesian framework with sparsity-inducing priors \cite{gurel2020compressive}. This approach, known as LASSO \cite{tibshirani1996regression}, assumes that the cost function from  eq.~\eqref{eq:bayes} is rewritten as:
\begin{equation}
    \hat{\mathbf{x}}= \argmin{\vect{x}} \frac{1}{2} \|\matr{A} \vect{x} - \vect{y}\|^2_2 +
    \lambda \, \|\matr{L}\vect{x}\|_1
    \,,
    \label{eq:lv}
\end{equation}
where $\|\,\cdot\,\|_1$ denotes the $\ell_1$-norm.

In these approaches, $\mathbf{L}$ is chosen to define a transformation to a sparse domain. We specifically consider two scenarios:

% This formulation is flexible enough to allow the choice of priors that are particularly well suited for the characteristics of the spectrum to reconstruct; we showcase this flexibility in this work using two case scenarios:
% \begin{itemize}
% 	\item Imposing sparsity on the Fourier domain of the spectrum, that is the domain of \gls{opd}, in the case of noisy interferograms.
	
% 	\item Imposing sparsity on the spectrum in the case of sparse spectra, such as in measurements of monochromatic light or light composed of specific wavenumbers.
% \end{itemize}
% In both cases, the key is to define the linear operator $\matr{L}$.

\subsubsection{Sparsity on the Fourier domain of the spectrum}
\label{subsubsec:lv_sparsity_on_the_fourier_domain}

We wish to impose a sparsity-inducing regularizer on the Fourier domain of the spectrum, i.e., the \gls{dct} in this case, where low-amplitude high frequency oscillations can be softly discarded. Said oscillations may occur due to noisy components in nature.
For that, we define $\matr{L}\in \RR{K}{K}$ as the orthogonal version of the Type-II \gls{dct}, whose elements are:
\begin{equation}
	l_{ij} = \sqrt{\frac{2}{J}}
	\cos\left[\frac{\pi}{J} \left(j+\frac{1}{2}\right) i\right]
	_{\forall i, j \in \{0, \dots, K-1\}} ,
	\label{eq:dct_prior}
\end{equation}
with $l_{0j}$ divided by $\sqrt{2}$ $\forall j$.
This transformation allows to express the spectrum in a complementary Fourier domain and is related to the space of the interferogram.
%\andres{Why loosely related? If correctly defined, the inverse DCT of the spectrum is the interferogram of the 2-wave model.}

\subsubsection{Sparsity on the spectrum itself}
\label{subsubsec:lv_sparsity_on_the_spectrum}

In the case of interferograms acquired from light with monochromatic or specific wavenumbers, we know already that the spectra that we wish to reconstruct are sparse. We wish to impose sparsity on the spectrum itself, with LASSO acting as feature selection of the wavenumbers of interest.
Accordingly, we define $\matr{L}$ as the identity operator $\matr{I}_K$, resulting in the following cost function:
\begin{equation}
	\hat{\vect{x}} = \argmin{\vect{x}}{
		\frac{1}{2} \|\vect{y} - \matr{A} \vect{x}\|^2_2
		+
		\lambda \; \|\vect{x}\|_1
	}
\,.
	\label{eq:lv_sparse_spectrum_fista}
\end{equation}

We propose to employ the \gls{lv} optimizer \cite{LoriV11:ip} for the solution of eq. \eqref{eq:lv}. The \gls{lv} optimizer is an iterative algorithm applicable to any inversion problem with a quadratic data term and a non-differentiable convex function $g(\cdot)$. As this solver does not require any matrix inversion and provides easy-to-follow guidelines to set the convergence parameters, it is more suitable for our problem than its competitors \cite{CondKCH23:siam}. In particular, eq. \eqref{eq:lv_sparse_spectrum_fista} is traditionally solved with algorithms based on \gls{ista}, such as \glsentryshort{fista} \cite{BeckT09:jis}, however, the proposed solver is able to generalize to both sparsity scenarios that we consider in our work.

Algorithm \ref{algo:lv_updates} describes the iterative updates of \gls{lv} for the estimation of $\vect{x}$ and of the dual parameter $\vect{u}$.
$\|\matr{A}\|_{\textrm{op}}$ and $\|\matr{L}\|_{\textrm{op}}$ represent the operator norms of $\matr{A}$ and $\matr{L}$ respectively~\cite{CondKCH23:siam}.
$\eta$ and $\tau$ are convergence parameters such that $\eta\tau \leq 1 / \|\matr{L}\|^2_{\textrm{op}}$, and $1\le\rho\le 2$ is the over-relaxation parameter; their specified values were chosen according to the relevant literature \cite{CondKCH23:siam}.

In the third step of the update, $\textrm{prox}_{\lambda\,, g^{\star}} (\vect{u})$ denotes the proximal operator associated to the Fenchel conjugate of $g(\cdot)$ \cite{PariB14:fto}. This is equal to a hard-thresholding operator when $g$ is chosen as the $\ell_1$ norm. That is:
% $\forall\, k \in \{1, \dots, K\}$:
%\begin{subequations}
%	\begin{equation}
%		\textrm{prox}_{\lambda,\, g^{\star}} (u_k) =
%		\begin{cases}
%			-\lambda & \textrm{if } u_k < -\lambda
%			\\
%			u_k & \textrm{if } |u_k| < \lambda
%			\\
%			\lambda & \textrm{if } u_k > \lambda
%		\end{cases}
%		\label{eq:prox_g}
%	\end{equation}
	\begin{equation}
		%\iff
		\textrm{prox}_{\lambda,\, g^{\star}} (\vect{u}) =
		\min(\max(\vect{u}, -\lambda), \lambda)\,,
		\label{eq:prox_g_minmax}
	\end{equation}
%\end{subequations}
where $\min(\cdot, \cdot)$ and $\max(\cdot, \cdot)$ respectively denote the minimum and maximum operator, applied on every element of the first argument.

\begin{algorithm}[t]
	\caption{
		Proposed method using the \glsfmtshort{lv} algorithm \cite{LoriV11:ip}
	}
	\begin{algorithmic}
		\REQUIRE $\matr{A}$, $\matr{L}$, $N_{\textrm{iters}}$
		
		\STATE
		\textbf{Initialize} $\vect{x}^{(0)} = \matr{A}^{\T} \vect{y}$, $\vect{u}^{(0)} = \matr{L} \vect{x}^{(0)}$
		\STATE
		\textbf{Initialize} $\tau = 0.99 / \|\matr{A}\|^2_{op}$, $\eta = 1 / (\tau \|\matr{L}\|^2_{op})$, and $\rho = 1.9$
		\STATE
		\textbf{Define} $\textrm{prox}_{\lambda\, g^{\star}} (\vect{u}) =
		\min(\max(\vect{u}, -\lambda), \lambda)$
		
%		\FOR{$q = 0, \dots, N_{\textrm{iters}-1}$}
		\FOR {$q = 0$ to $N_{iters}-1$}
		\STATE
		$
		\vect{e}^{(q)} = 
		\matr{A}^{\T} (
			\matr{A} \vect{x}^{(q)} - \vect{y}
		)
		$
		\STATE
		$
		\vect{x}^{(q+\frac{1}{2})} = 
		\vect{x}^{(q)} - \tau \left(
			\vect{e}^{(q)}
			+
			\mathbf{L}^{\T} \vect{u}^{(q)}
		\right)
		$
		\STATE
		$
		\vect{u}^{(q+\frac{1}{2})} = 
		\textrm{prox}_{\lambda, g^{\star}} \left(
			\vect{u}^{(q)}
			+
			\eta \matr{L} \vect{x}^{(q+\frac{1}{2})}
		\right)
		$
		\STATE
		$
		\vect{x}^{(q+1)} = 
		\vect{x}^{(q)} - \rho\tau \left(
			\vect{e}^{(q)}
			+
			\mathbf{L}^{\T} \vect{u}^{(q+\frac{1}{2})}
		\right)
		$
		\STATE
		$
		\vect{u}^{(q+1)} = 
		\vect{u}^{(q)} + \rho \left(
			\vect{u}^{(q+\frac{1}{2})} - \vect{u}^{(q)}
		\right)
		$
		\ENDFOR
		
%		\STATE
%		$\hat{\matr{x}} = \matr{x}^{(N_{\textrm{iters}})}$
		
		\RETURN $\hat{\matr{x}} = \matr{x}^{(N_{\textrm{iters}})}$  % , $\vect{e}$
	\end{algorithmic}
	\label{algo:lv_updates}
\end{algorithm}

%%%%%%%%%%%%%%%%%%%%%%%%%%%%%%%%%%%%%%%%%%
\section{Experiments on Simulated Data}
\label{sec:results}

%In this section, we show the results of the spectral reconstruction from simulated and real interferograms.
%The datasets are described in Section \ref{subsec:available_datasets}.
% In Sections \ref{subsec:experiments_reflectivity}, \ref{subsec:experiments_irregular_sampling}, and \ref{subsec:experiments_noise}, we observe the influence of three different categories with simulated interferograms: reflectivity levels, irregular sampling of the \gls{opd}, and noise corruption.
%
% In Section \ref{subsec:real_data_experiments}, we invert real data measured with a real \gls{fp}-based interferometric imaging spectrometer, using monochromatic spectra as input, where the three categories can naturally co-exist.
%
% We note that no noise is added to the simulations in Sections \ref{subsec:experiments_reflectivity} and \ref{subsec:experiments_irregular_sampling}, so the Bayesian framework boils down to a \glsfmtlong{pinv}. In these two sections, we replace \gls{tsvd}, \gls{rr}, and \gls{lv} by a simple \glsfmtlong{pinv}.

In this section, we aim to show the limitations of the spectrum reconstruction algorithms in terms of applicability to various case studies.
To this end  we go through the methods following the order described in Section \ref{sec:inversion_algorithms} and showing that the reconstruction results become less and less accurate the further we move from a baseline case, while taking the chance to discuss on the robustness of the methods.

In particular, we simulate three acquisition scenarios of non-idealities from the textbook models: reflectivity regimes (Section \ref{subsec:experiments_reflectivity}), irregular sampling of the \glspl{opd} (Section \ref{subsec:experiments_irregular_sampling}), and noise corruption (\ref{subsec:experiments_noise}).

\subsection{Dataset Description}
\label{subsec:available_datasets}

% In the following we separate the datasets to be used in our experiments. In the first class, labeled \textit{simulated datasets}, only the spectra are available and the corresponding interferograms are simulated under specific assumptions that are detailed in the relevant sections.

The simulated datasets include two collection of spectra, labeled Solar
and Specim.
In particular, Solar contains $M=22$ solar spectra acquired at different times of the day.
Its spectral support is $\Omega = [1.000,\;2.850]$ \si{\um}$^{-1}$.
Specim contains the spectra of the central pixel of each of the $M=24$ color boxes of a \glsfmtlong{cc}, acquired with
the \textit{Specim IQ} hyperspectral camera.
% \cite{SpecimIQ}.
Its spectral support is $\Omega = [0.996,\;2.517]$ \si{\um}$^{-1}$.
When simulating interferograms from such spectra, we consider $L = 319$ \gls{opd} samples, to match the specifications of the interferometric spectrometer \gls{imspoc} UV 2.
The simulation is carried out using the transfer matrix obtained by sampling eq. \eqref{eq:transmittance_response_fabry_perot_infty} for the \gls{mbi} and applying the model of eq. \eqref{eq:direct_model} to simulate the interferogram.

Unless otherwise stated, no noise is added, and the \gls{opd} support is regularly sampled with a step size $\Delta \delta = 0.175$ \si{\um}, i.e., $\delta \in [0,\,55.65]$ \si{\um}. We also set a reflectivity $\mathcal{R}=0.2$ and transmissivity $\mathcal{T}=1$.

% \textcolor{red}{Put interferograms here}

%\input{floating_tex/dataset_information_simulated.tex}

% These interferograms are then measured using the \gls{fp}-based \gls{imspoc} UV 2 device. The instrument is composed of an array of $319$ \glsentrylong{fp} etalons with increasing thickness of step size $\Delta d = 87.5 \, \si{\nm}$. 
% The characteristics of the two datasets are summarized in \tablename\;\ref{tab:real_dataset_information}. 
\subsection{Expermental Setup}

We perform a qualitative analysis of the reconstruction by varying the parameters from the baseline acquisition model described in the previous section. 
% The choice of these parameters is detailed in the appropriate section.
We compare the \gls{idct} method from eq. \eqref{eq:idct}, Haar method \cite{al-saeed-2016-fourier-trans}, and the \glsentryfull{pinv} of eq. \eqref{eq:pinv}. When noise is present, we also compare with \gls{tsvd} \cite{GoluHO99:jmaa} and \gls{rr} \cite{Hans90:jssc}  that we detailed in Section \ref{subsec:svd_based_approaches}, as well as our proposed method.
Those were ignored in the noiseless case, as the optimal parametric choice makes them equivalent to the pseudo-inversion.

In terms of notation, we denote by $\matr{Y} \in \RR{L}{M}$ the set of observed interferograms, by $\matr{X} \in \RR{K}{M}$ the reference spectra, and by $\hat{\matr{X}} \in \RR{K}{M}$ the corresponding reconstructed spectra.

The results are assessed quantitatively using the \gls{rmse} quality index:
\begin{equation}
	\textrm{RMSE} = \frac{
		\|\matr{X} - \hat{\matr{X}}\|_F^2
	}{
		\|\matr{X}\|_F^2
	}
	\,.
\end{equation}

\subsection{Reflectivity}
\label{subsec:experiments_reflectivity}

In this section we assume that the transfer matrix $\mathbf{A}$ follows the \gls{mbi} formulation with constant reflectivity values in the set $\mathcal{R} = \{0.2, 0.4, 0.7\}$, as well as a case where the reflectivity varies with respect to $\sigma$, extracted from the characterization of a real device \cite{PicoGDFL23:oe}. \figurename\,\ref{fig:reflectivity_levels} shows these reflectivity regimes.
% For simplicity, no noise is added, and the \gls{opd} support is regularly sampled with a step size $\Delta \delta = 0.175~\si{\um}$, i.e., $\delta \in [0,\,55.65]~\si{\um}$.
The Haar method is not applicable to the case of varying reflectivity, but we still apply it assuming that the reflectivity is equal to its mean value over the spectral range of reconstruction.

\def \x {0.85}

\begin{figure}[t]
	\centering
	\includegraphics[width=\x\linewidth]{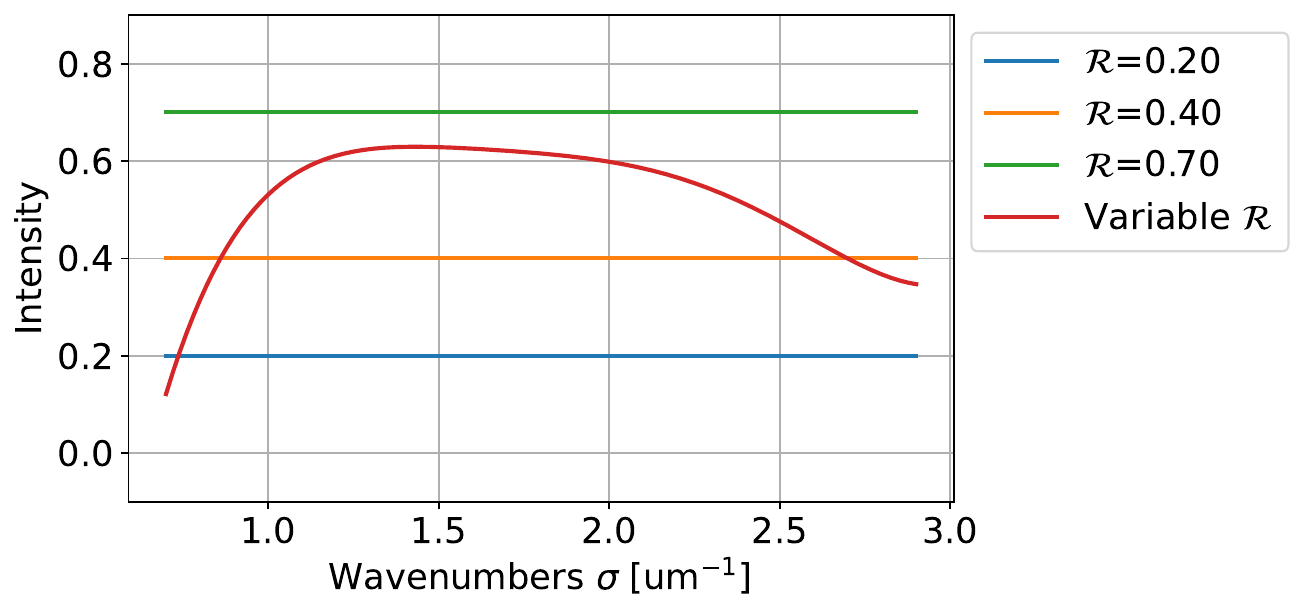}
	\caption{
		Different reflectivity regimes.
	}
	\label{fig:reflectivity_levels}
\end{figure}

\figurename\,\ref{fig:reconstruction_reflectivity_levels} shows the results of inverting simulated Solar and Specim interferograms in each case. As expected, the \gls{idct} performs better for low values of reflectivity, as the model is more similar to the \gls{tbi}, as especially evident for the Specim case. The Haar method is more robust to the increase of reflectivity, but the reconstruction results tend to fail for high reflectivity regimes, possibly since such regimes are susceptible to aliasing, and are inaccurate for varying reflectivity. The pseudo-inversion generates basically perfect matches, as the noiseless case requires no regularization and we exploit the perfect knowledge of the system through $\mathbf{A}$.

\def \x {0.23}
\def \y {0.016}
\def \z {1.6}

\begin{figure*}[t]
	\centering
	
	% \subfloat{%
		\raisebox{\z cm}{\rotatebox[origin=c]{90}{\normalsize{\texttt{\textbf{Solar}}}}}
	% }
	\subfloat[$\mathcal{R}=0.2$]{\label{subfig:solar/fp_0_low_r}%
		\includegraphics[width=\x\linewidth]{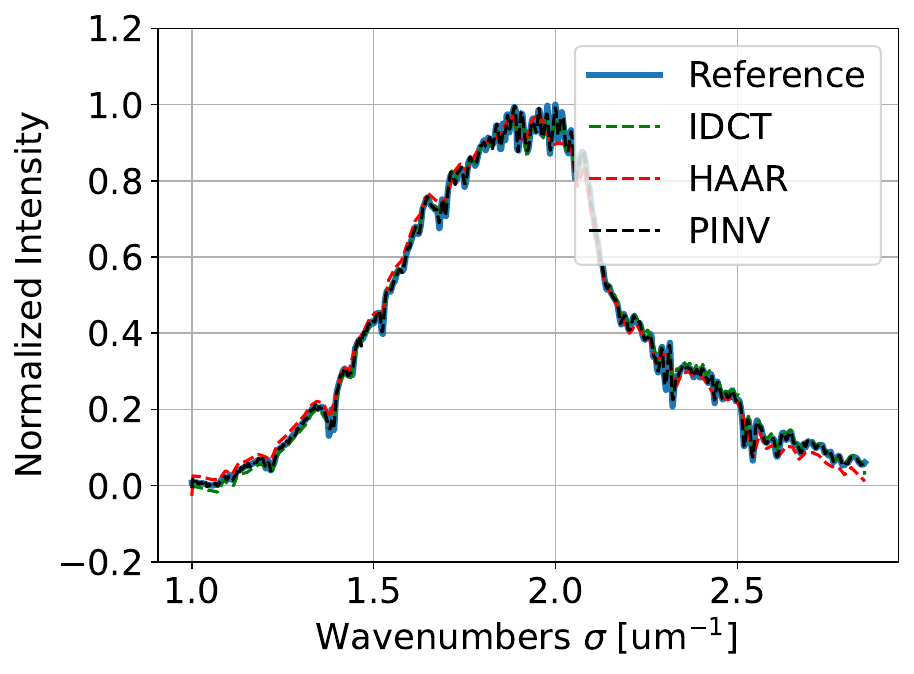}
	}
	\subfloat[$\mathcal{R}=0.4$]{\label{subfig:solar/fp_0_med_r}%
		\includegraphics[width=\x\linewidth]{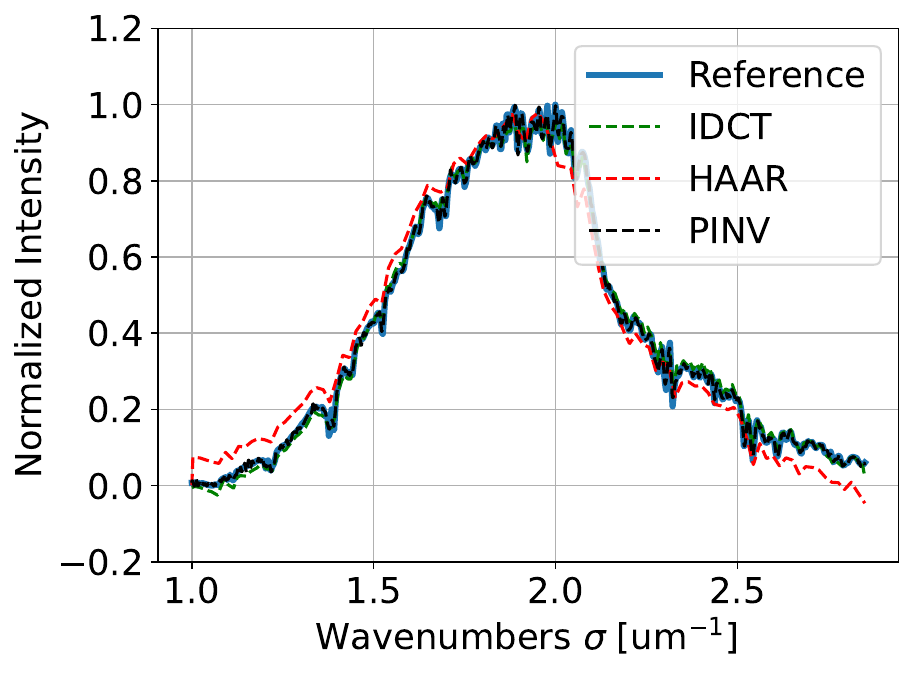}
	}
	\subfloat[$\mathcal{R}=0.7$]{\label{subfig:solar/fp_0_hig_r}%
		\includegraphics[width=\x\linewidth]{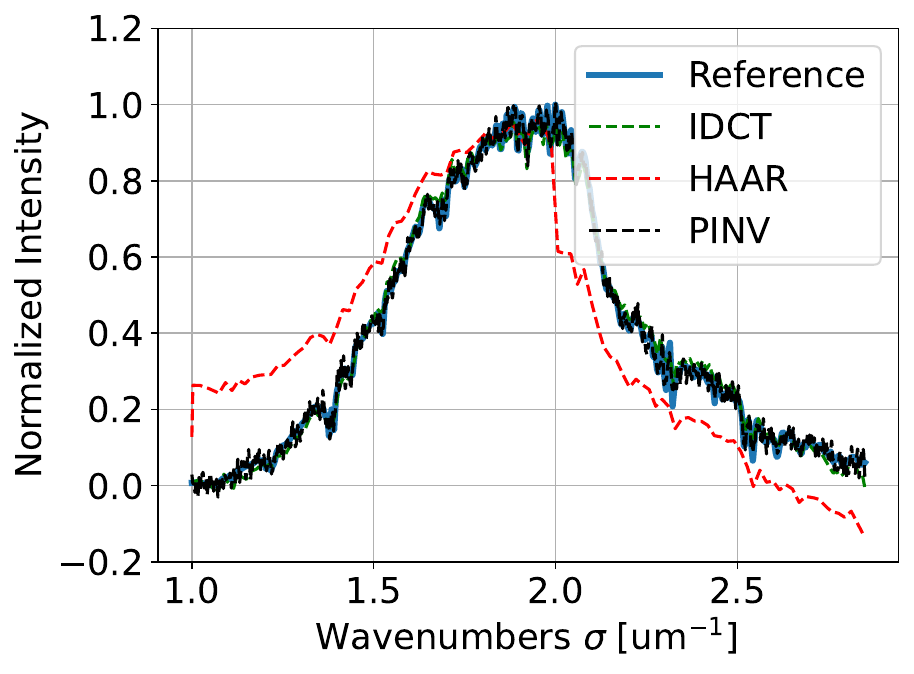}
	}
	\subfloat[Varying $\mathcal{R}$]{\label{subfig:solar/fp_0_var_r}%
		\includegraphics[width=\x\linewidth]{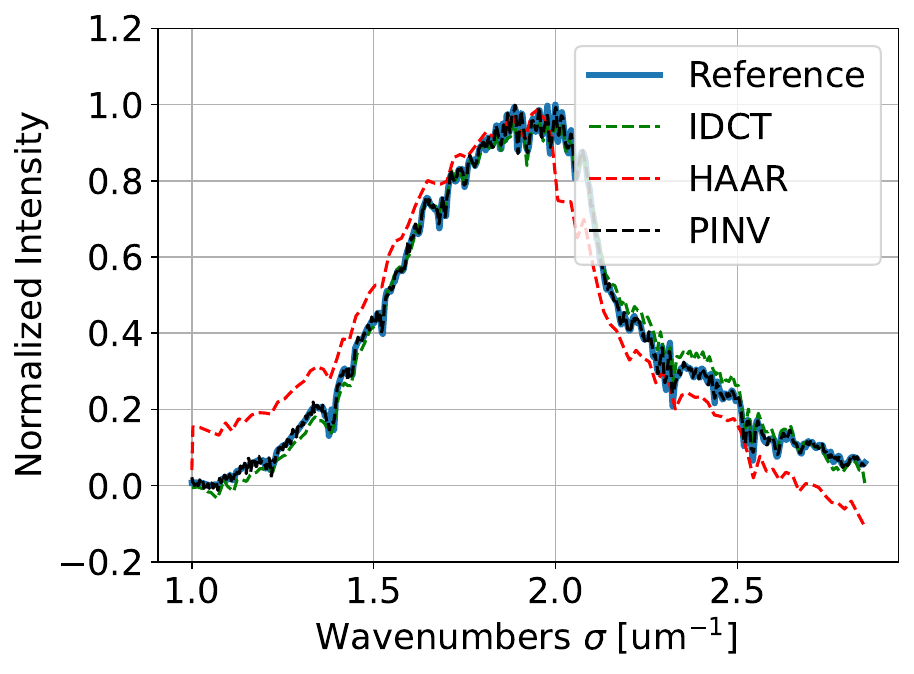}
	}
	
	% \subfloat{%
		\raisebox{\z cm}{\rotatebox[origin=c]{90}{\normalsize{\texttt{\textbf{Specim}}}}}
	% }
	\subfloat[$\mathcal{R}=0.2$]{\label{subfig:specim/fp_0_low_r}%
		\includegraphics[width=\x\linewidth]{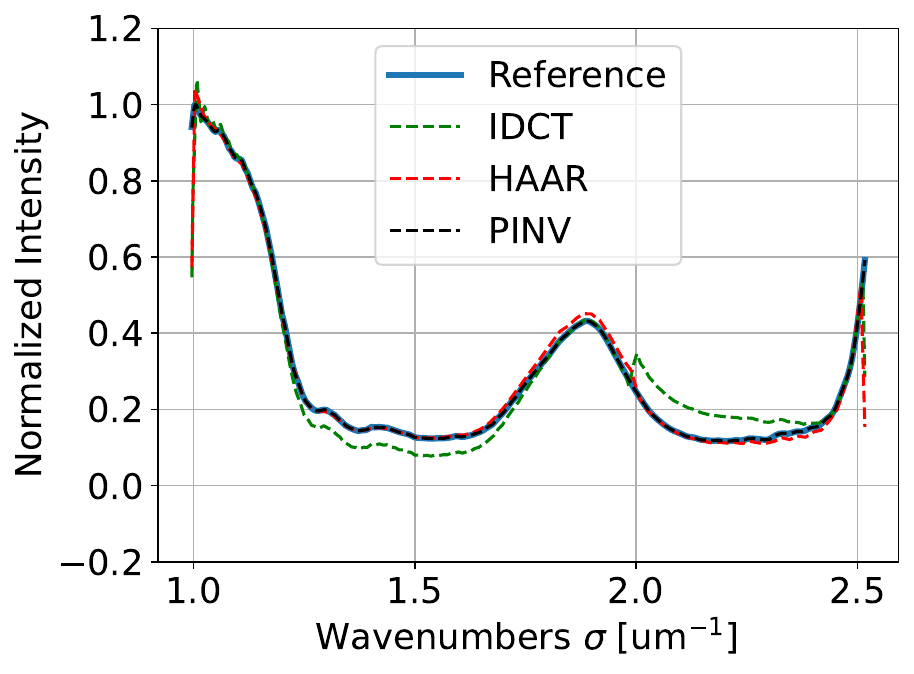}
	}
	\subfloat[$\mathcal{R}=0.4$]{\label{subfig:specim/fp_0_med_r}%
		\includegraphics[width=\x\linewidth]{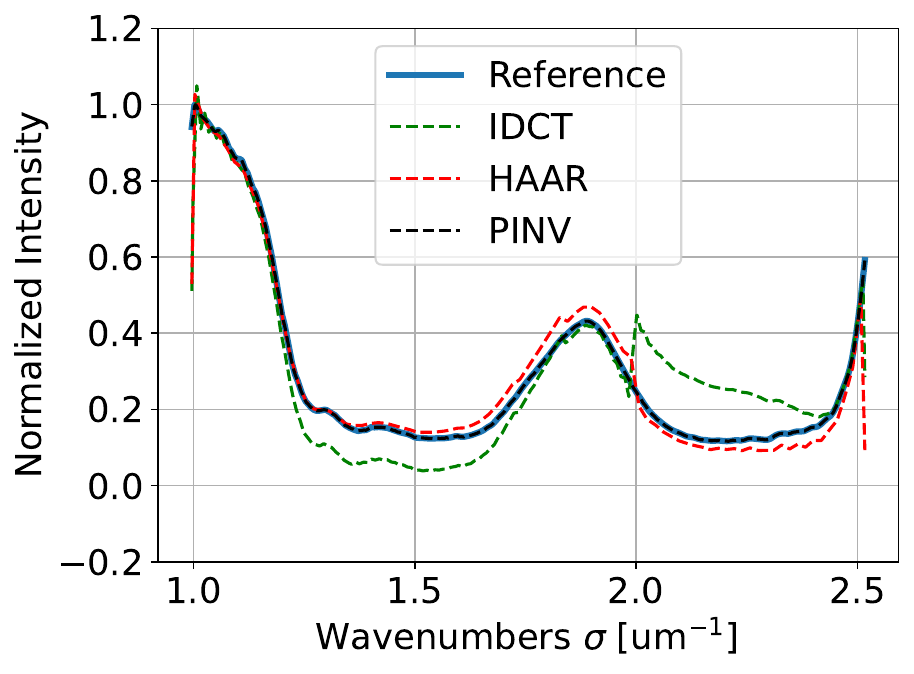}
	}
	\subfloat[$\mathcal{R}=0.7$]{\label{subfig:specim/fp_0_hig_r}%
		\includegraphics[width=\x\linewidth]{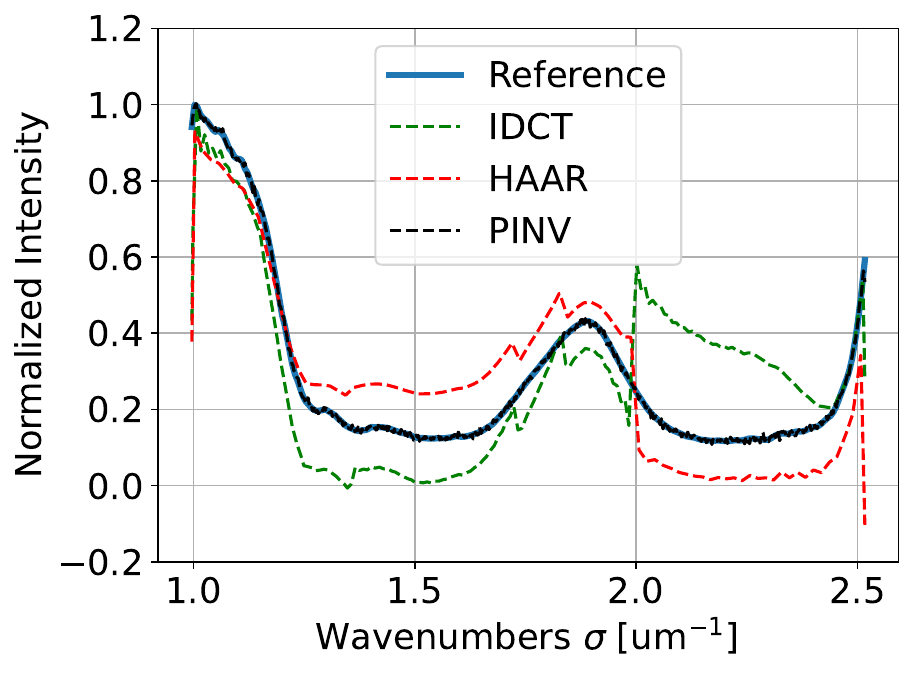}
	}
	\subfloat[Varying $\mathcal{R}$]{\label{subfig:specim/fp_0_var_r}%
		\includegraphics[width=\x\linewidth]{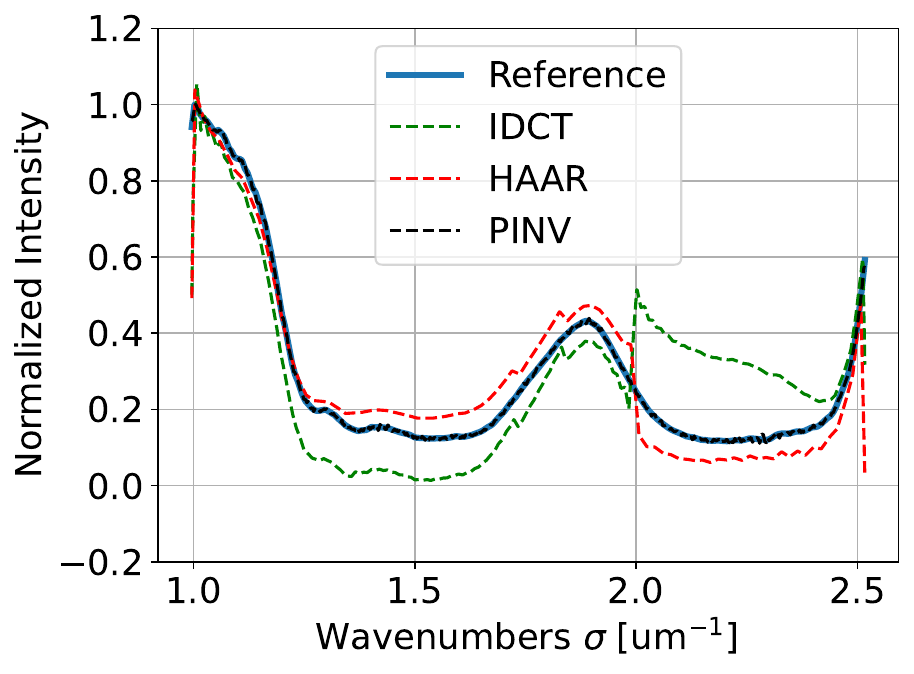}
	}
	
	\caption{
		Reconstruction of spectra with different values of reflectivity. The two rows refer respectively to the Solar and Specim datasets.
	}
	\label{fig:reconstruction_reflectivity_levels}
\end{figure*}

\def \x {0.46}

\begin{figure}[t]
	\centering
	
	\subfloat[Solar]{\label{subfig:irregular_sampling/solar}%
		\includegraphics[width=\x\linewidth]{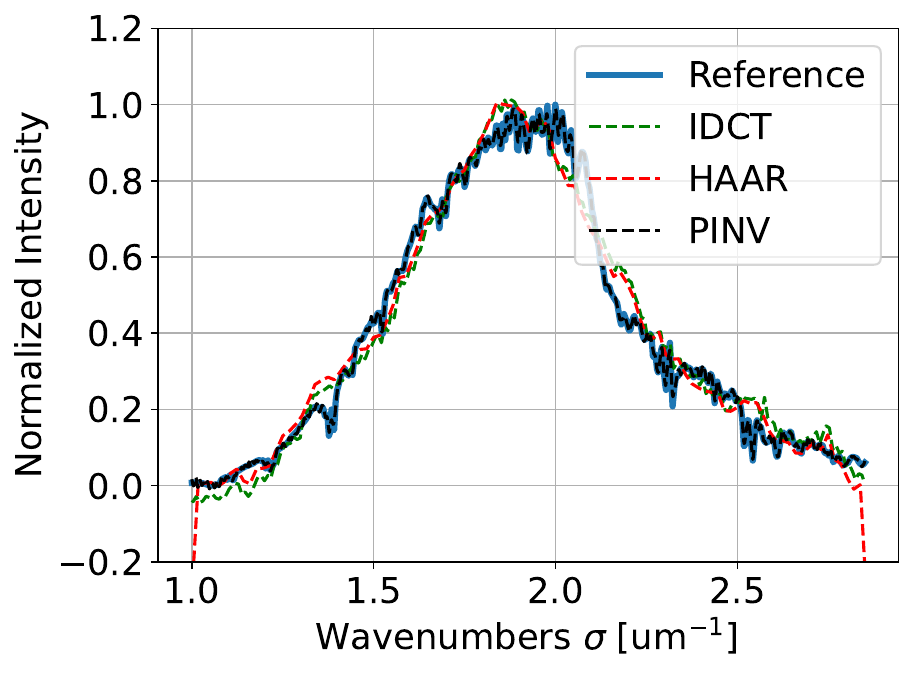}
	}
	\subfloat[Specim]{\label{subfig:irregular_sampling/specim}%
		\includegraphics[width=\x\linewidth]{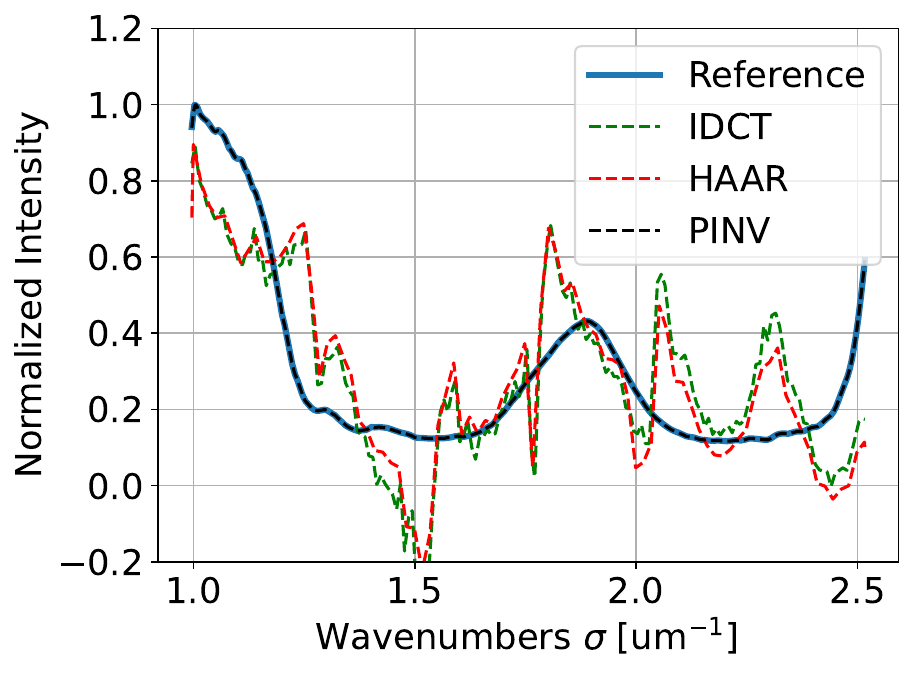}
	}
	
	\caption{
		Reconstruction with irregular \gls{opd} sampling.
	}
	\label{fig:reconstruction_irregular_sampling}
\end{figure}

%\begin{figure}[t]
%	\centering
%	
%	\begin{minipage}[b]{\linewidth}
%		\centering
%		\begin{subfigure}[b]{\x\linewidth}
%			\centering
%			\includegraphics[width=\linewidth]{figures/simulated/irregular_sampling/solar/fp_0_low_r/spectrum_comparison/acquisition_000.pdf}
%			\caption{
%				Solar
%			}
%			\label{subfig:irregular_sampling/solar}
%		\end{subfigure}
%		\begin{subfigure}[b]{\x\linewidth}
%			\centering
%			\includegraphics[width=\textwidth]{figures/simulated/irregular_sampling/specim/fp_0_low_r/spectrum_comparison/acquisition_000.pdf}
%			\caption{
%				SPECIM
%			}
%			\label{subfig:irregular_sampling/specim}
%		\end{subfigure}
%	\end{minipage}
%	
%	\caption{
%		Reconstruction with irregular sampling.
%	}
%	\label{fig:reconstruction_irregular_sampling}
%\end{figure}

\def \x {0.46}
\def \z {1.6}

\begin{figure}[t]
\centering

% \subfloat{%
\raisebox{\z cm}{\rotatebox[origin=c]{90}{\normalsize{\texttt{\textbf{Solar}}}}}
% }
\subfloat[$\text{SNR}=20$ \si{\decibel}]{\label{subfig:solar_reconstruction_snr_60}%
\includegraphics[width=\x\linewidth]{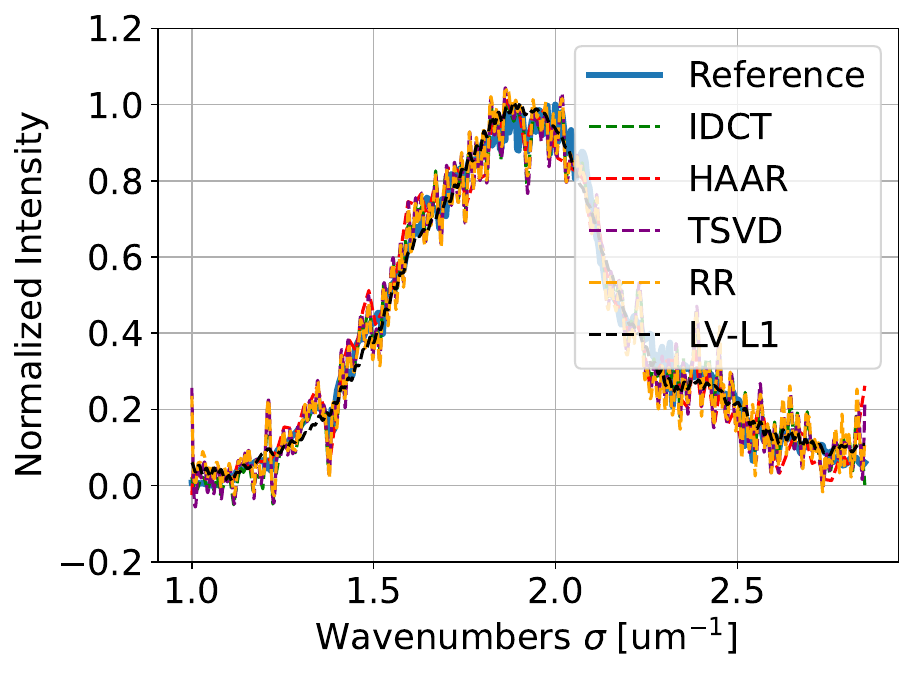}
}
\subfloat[$\text{SNR}=15$ \si{\decibel}]{\label{subfig:solar_reconstruction_snr_50}%
\includegraphics[width=\x\linewidth]{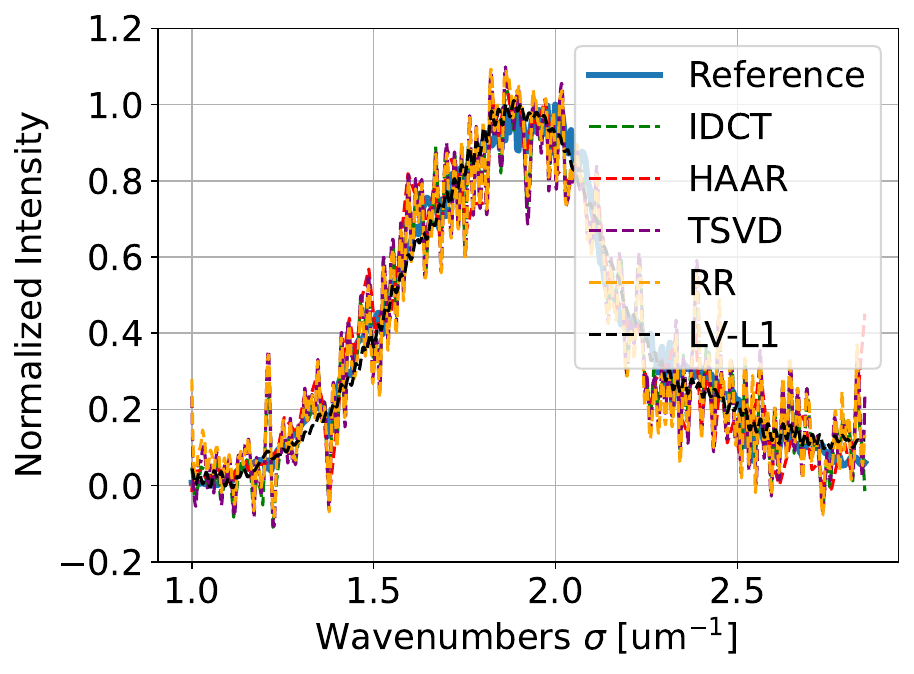}
}

% \subfloat{%
\raisebox{\z cm}{\rotatebox[origin=c]{90}{\normalsize{\texttt{\textbf{Specim}}}}}
% }
\subfloat[$\text{SNR}=20$ \si{\decibel}]{\label{subfig:specim_reconstruction_snr_60}%
\includegraphics[width=\x\linewidth]{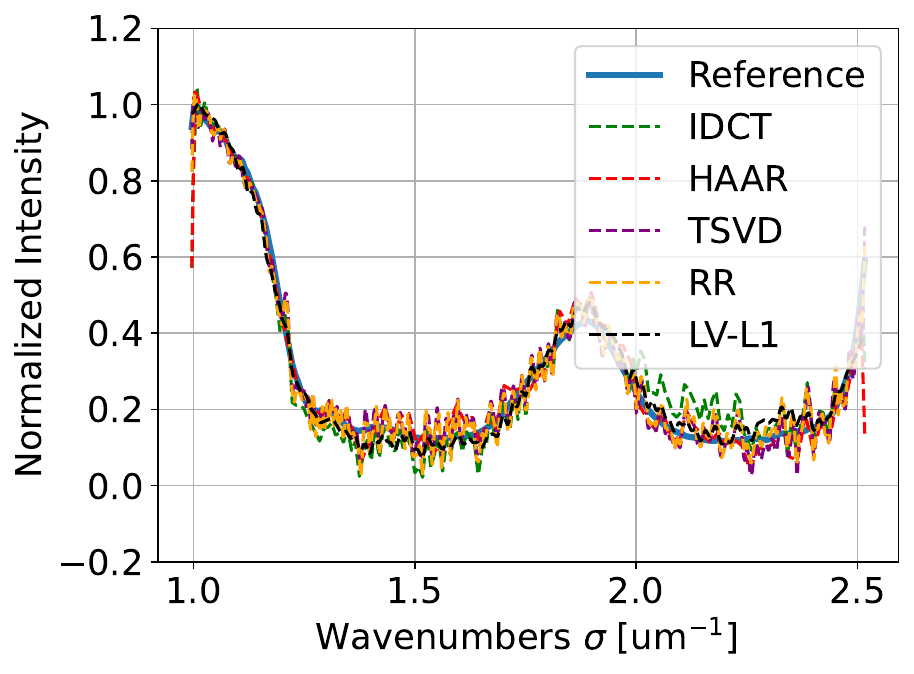}
}
\subfloat[$\text{SNR}=15$ \si{\decibel}]{\label{subfig:specim_reconstruction_snr_50}%
\includegraphics[width=\x\linewidth]{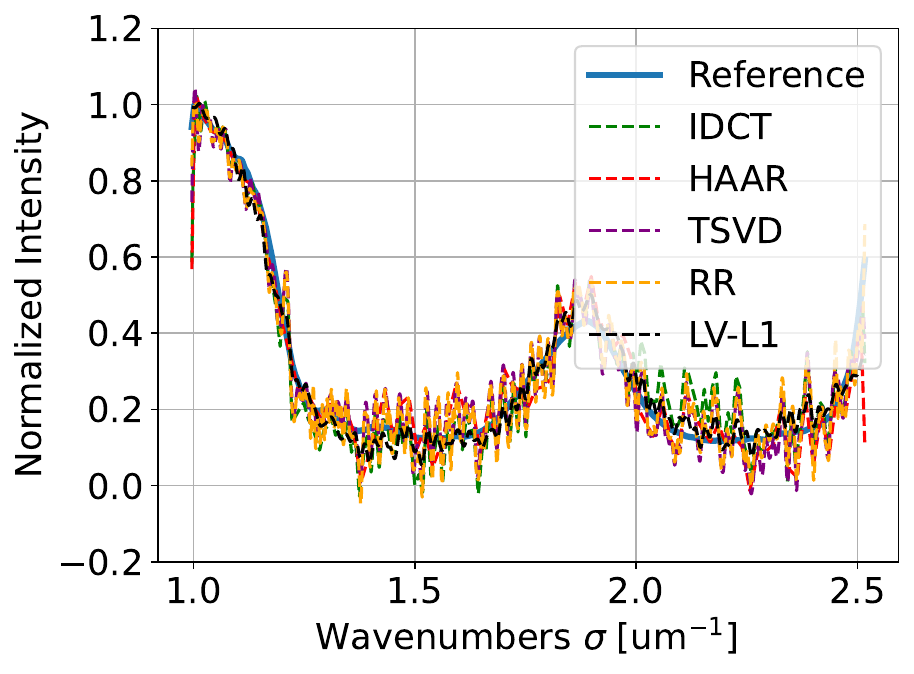}
}

\caption{
Reconstruction of spectra from interferograms with noise corruption. The two rows refer respectively to the Solar and Specim datasets.
}
\label{fig:simulated_data_experiments}
\end{figure}

\subsection{Irregular Sampling}
\label{subsec:experiments_irregular_sampling}

Compared to the baseline setup, in this experiment we simulate the interferograms assuming that the \gls{opd} domain is irregularly sampled, or in other words that $\Delta\delta$ is not constant. In particular, we use the \gls{opd} values that were characterized in the study of \cite{PicoGDFL23:oe}, whose mean and standard deviation on $\Delta\delta$ are $0.17$ and $0.15$ \si{\um}, respectively.

\figurename\,\ref{fig:reconstruction_irregular_sampling} shows the results of inverting simulated Solar and Specim interferograms.
As expected, the Fourier-based approaches, \gls{idct} and Haar, exhibit many mismatches as they require the \gls{opd} support to be regularly sampled.
% A preprocessing step is possibly needed to fix the support.
The pseudo-inverse however shows perfect matches as this non-ideality is already embedded in the matrix $\matr{A}$.

\subsection{Noise Corruption}
\label{subsec:experiments_noise}

%\subsubsection{Experimental setup}
%\label{subsec:simulated_data_experimental_setup}

In this experiment we aim to show the robustness of the proposed algorithm when the interferograms are corrupted by noise. In particular, we corrupt the simulated interferograms by adding Gaussian noise with SNR of $20$ and $15$ \si{\decibel}.
% For the simulations, we choose the following parameters inspired by the \gls{imspoc} UV 2 device:
%\begin{itemize}
% 	\item We consider $L = 319$ \gls{opd} samples with a step size $\Delta \delta = 0.175~\si{\um}$ where $\delta \in [0,\,55.65]~\si{\um}$.
	
% 	\item We consider $\mathcal{R} = 0.2$ and $\mathcal{T} = 1$ constant $\forall \sigma_k$.
% \end{itemize}

% After simulating the interferograms, denoted by $\vect{Y}$, we corrupt them with Gaussian noise with SNR of $20$ and $15$ \si{\decibel}.
% Then, we reconstruct the set of spectra, denoted by $\hat{\matr{X}}$, using the inversion algorithms \gls{idct}, Haar \cite{al-saeed-2016-fourier-trans}, \gls{tsvd} \cite{GoluHO99:jmaa}, \gls{rr} \cite{Hans90:jssc}, and \gls{lv} with sparsity on the Fourier domain of the spectrum.
%\subsubsection{Results and discussion}
%\label{subsec:simulated_data_results_discussion}

\begin{table}[t]
	\begin{center}
		\caption{Results of the simulated experiments showing the \gls{rmse} and the optimal regularizing parameter $\lambda_{\mathrm{opt}}$.
%			 The best results are marked in \textbf{bold}.
		 }
		\label{tab:simulated_data_experiments}
	    \begin{tabular}{l|c||c|c"c|c}
			% \multicolumn{2}{c||}{ } &
			% \multicolumn{4}{c}{\textbf{\glsfmtlong{fp} $\infty$-wave model}}
			% \\
	        \multicolumn{2}{c||}{ } &
	        \multicolumn{2}{c"}{\textbf{SNR} $=20$~\si{\decibel}} &
	        \multicolumn{2}{c}{\textbf{SNR} $=15$~\si{\decibel}}
	        \\
	        \hline
	        \textbf{Dataset} &
	        \textbf{Method} &
	        $\lambda_{\textrm{opt}}$ &
	        \gls{rmse} &
	        $\lambda_{\textrm{opt}}$ &
	        \gls{rmse}
	        \\
	        \hline\hline
	        \multirow{5}{*}{\textbf{Solar}}
& \gls{idct} & - & 0.087 & - & 0.145 \\
& Haar \cite{al-saeed-2016-fourier-trans} & - & 0.099 & - & 0.161 \\
& \gls{tsvd} \cite{Hans90:jssc} & 0.648 & 0.110 & 0.648 & 0.186 \\
& \gls{rr} \cite{GoluHO99:jmaa} & 7.565 & 0.113 & 10.723 & 0.186 \\
& Ours & 16.768 & \textbf{0.071} & 29.471 & \textbf{0.079} \\
			\thickhline
			\multirow{5}{*}{\textbf{Specim}}
& \gls{idct} & - & 0.142 & - & 0.180 \\
& Haar \cite{al-saeed-2016-fourier-trans} & - & 0.112 & - & 0.160 \\
& \gls{tsvd} \cite{Hans90:jssc} & 0.539 & 0.098 & 0.533 & 0.166 \\
& \gls{rr} \cite{GoluHO99:jmaa} & 6.136 & 0.102 & 7.055 & 0.173 \\
& Ours & 7.197 & \textbf{0.084} & 12.649 & \textbf{0.109} \\
% 	        \thickhline
% 	        \multirow{5}{*}{\texttt{\textbf{SHINE}}}
% & \gls{idct} & - & 0.139 & - & 0.202 \\
% & Haar \cite{al-saeed-2016-fourier-trans} & - & 0.138 & - & 0.187 \\
% & \gls{tsvd} \cite{GoluHO99:jmaa} & 0.648 & 0.105 & 0.648 & 0.183 \\
% & \gls{rr} \cite{Hans90:jssc} & 5.337 & 0.113 & 6.136 & 0.196 \\
% & Ours & 11.514 & \textbf{0.063} & 22.230 & \textbf{0.072} \\
	    \end{tabular}
	\end{center}
\end{table}

For each method, the optimal regularization parameter $\lambda_{opt}$ is chosen with a grid search to minimize the \gls{rmse}.
\tablename\;\ref{tab:simulated_data_experiments} shows a summary of the results.
\figurename\,\ref{fig:simulated_data_experiments} shows a selected spectrum from each of the Solar and Specim datasets.
The proposed method based on \gls{lv} outperforms the other approaches in terms of \gls{rmse}, which are greatly affected by the noise.
We also give some insights on the regularizing parameter and robustness to noise.

In \tablename\;\ref{tab:simulated_data_experiments}, first, the \gls{idct} and the Haar methods fail to attenuate the noise.
Second, as the SNR decreases, the values of $\lambda_{\textrm{opt}}$ in \gls{rr} and our proposed method increase. 

In fact, for low SNRs, the higher-order harmonics get confused with the noise energy. This limits the resolving power of the data term in the cost function, demanding for a larger penalization factor.
% that applying higher penalization for the favor of induced sparsity attenuates high oscillation components.
%, which is a trade-off.
%
On the other hand, the values of $\lambda_{\textrm{opt}}$ in the case of \gls{tsvd} remain unchanged.
%For instance, in the Solar case, $\lambda_{\min}$ for \gls{tsvd} is always around $0.65$,
%which means that the penalization on the singular values of the transfer matrix fails to attenuate any noise.

\figurename\,\ref{fig:simulated_data_experiments} confirms this analysis.
The reconstruction based on induced sparsity fits better with the reference, noting that some high oscillation components are lost in low SNR regimes \cite{cook1995multiplex}.
For the other approaches, even if the reconstruction loosely follows the shape of the references, the spectra remain noisy.
% While the other approaches follow the shape of the reference, the reconstruction remains noisy.

%This shows a limitation for methods that penalize on the singular values of the transfer matrix and further supports the reliance on methods that reduce the noise without directly affecting important information in the model.

\section{Experiments on Real Data}
\label{sec:real_data_experiments}

In this section, we perform spectral reconstructions for real interferograms measured using a multi-aperture \gls{fp}-based device, comparing the accuracy the results among the methods that were discussed in Section~\ref{sec:inversion_algorithms}. In particular we aim to reconstruct monochromatic sources with the same optical energy and verify if the inferred reconstruction stays flat across the wavenumber range.

\def \x {0.38}
\def \y {0.57}

\begin{figure}[t]
	\centering
	
	\subfloat[Camera prototype]{
		\includegraphics[width=\x\linewidth]{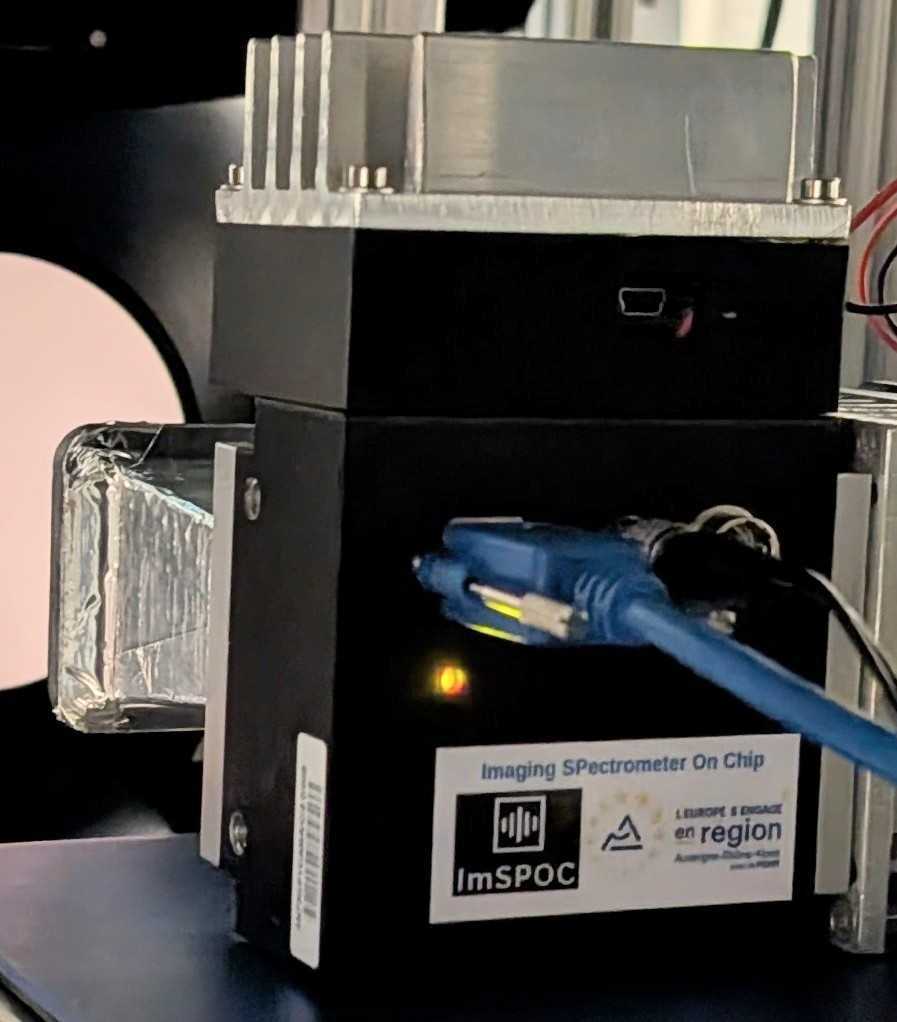}
		\label{fig:imspoc_prototype}
	}
	\subfloat[Array of \glspl{fpi}]{
		\includegraphics[width=\y\linewidth]{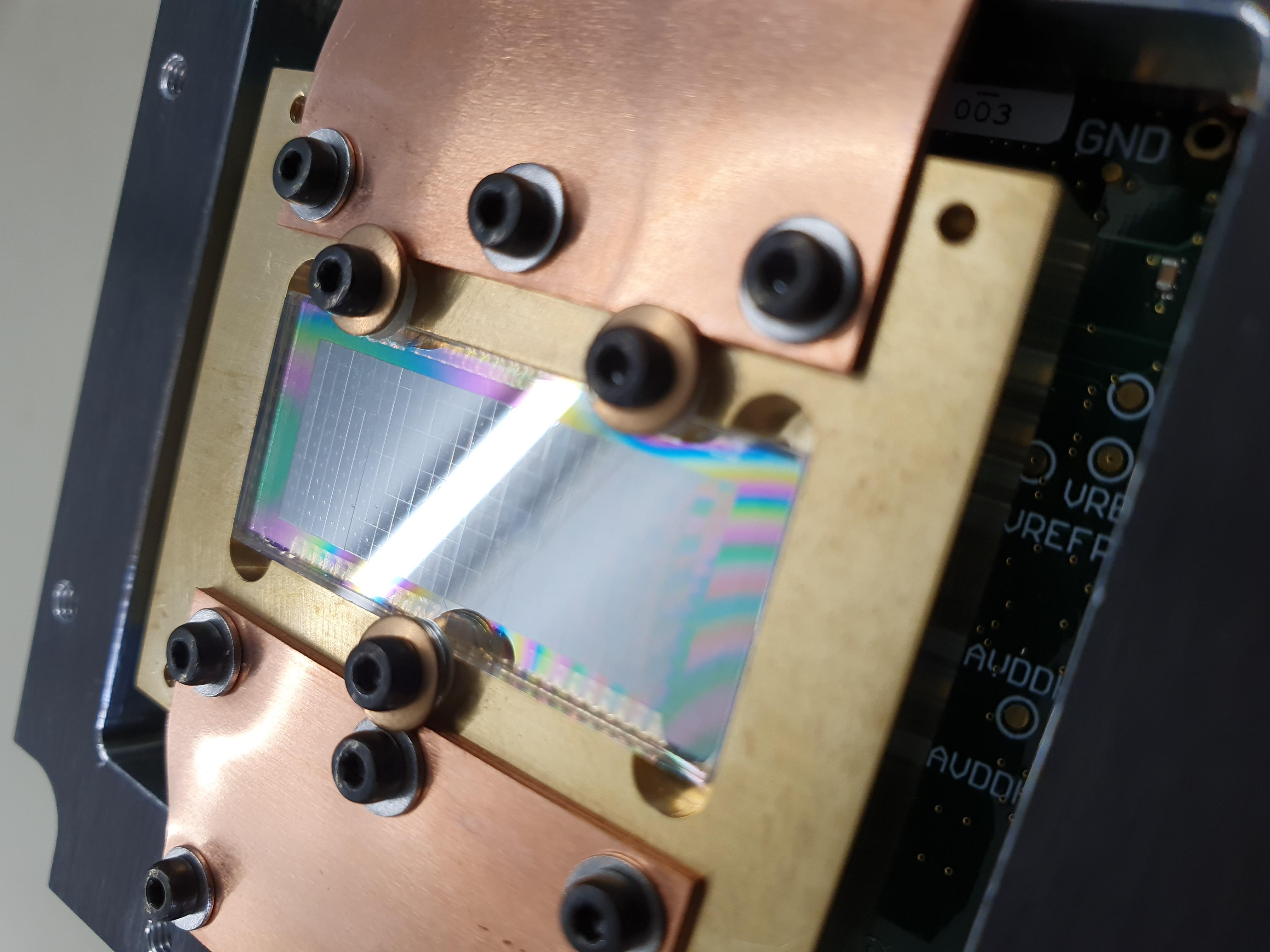}
		\label{fig:imspoc_array}
	}
	
	\caption{Left: The \gls{imspoc} prototype used to acquire the real datasets. Right: The array of $11 \times 29$ \glsentrylongpl{fp} etalons that composes the prototype.}
	\label{fig:imspoc}
	
\end{figure}

\begin{table}[t]
	\caption{
		Characteristics of  the real interferometric datasets captured with \gls{imspoc} UV 2 \cite{GuerLFD18:imspoc}.
		The range of central wavenumbers and that of \glspl{opd} \cite{PicoGDFL23:oe} are reported.
	}
	\label{tab:real_dataset_information}
	\centering
		\begin{tabular}{l"c"c|c"c}
			\begin{tabular}{@{}c@{}}
				\textbf{Dataset}
			\end{tabular} &
			\begin{tabular}{@{}c@{}}
				\textbf{No. of} \\
				\textbf{acq.s},
				$M$
			\end{tabular} &
			\begin{tabular}{@{}c@{}}
				\textbf{No. of} \\
				\textbf{\glspl{opd}},
				$L$
			\end{tabular} &
			\begin{tabular}{@{}c@{}}
			\textbf{\gls{opd}} \\
			\textbf{range} $\delta$ (\si{\um})
			\end{tabular} &
	%		\begin{tabular}{@{}c@{}}
	%			\textbf{No. of} \\
	%			\textbf{channels},
	%			$K$
	%		\end{tabular} &
			\begin{tabular}{@{}c@{}}
				\textbf{Central wn.} \\
%				\textbf{wavenumber} \\
				\textbf{range} (\si{\um}$^{-1}$)
			\end{tabular}
	%		&
	%		\begin{tabular}{@{}c@{}}
	%			$\Delta \sigma$, \\
	%			(\si{\mm}$^{-1}$)
	%		\end{tabular}
			\\
			\thickhline
			\textbf{MC-451} & $451$ &
			$319$ & $[1.79, \; 55.87]$ &
			$[1.00, \; 2.85]$
	%		& $4.1111$
			\\
			\textbf{MC-651} & $651$ &
			$319$ & $[1.75, \; 55.78]$ &
			$[1.10, \; 2.85]$
	%		& $2.6923$
			\\
		\end{tabular}
\end{table}

\def \x {0.48}
\def \y {0.48}

\begin{figure}[t]
	\centering
	
	\subfloat[MC-451 dataset]{
		\includegraphics[width=\x\linewidth]{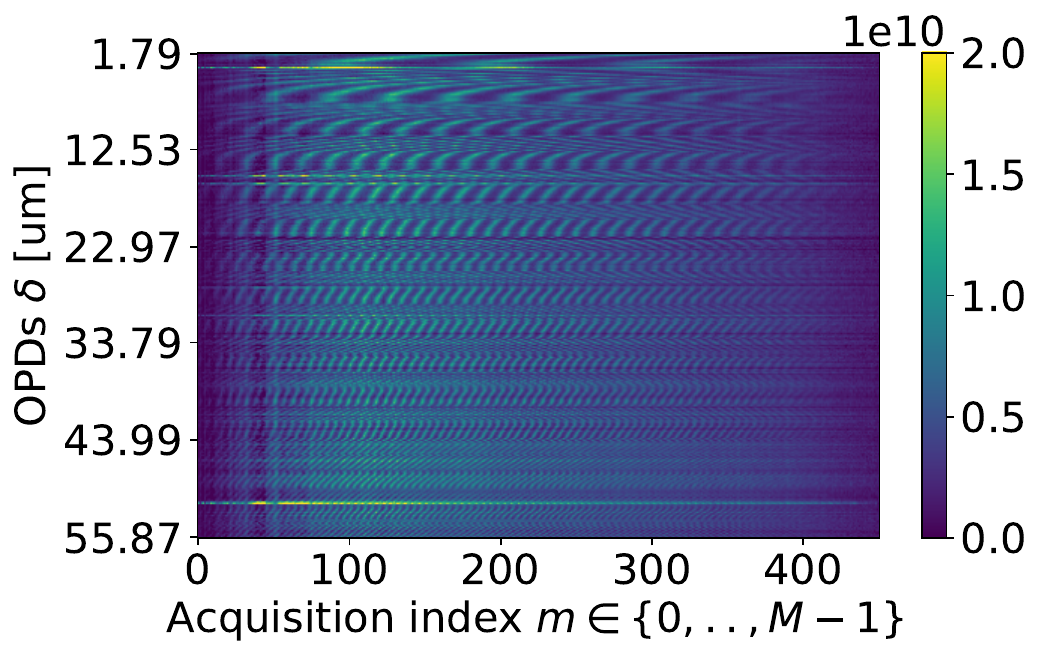}
		\label{fig:mc451_dataset}
	}
	\subfloat[MC-651 dataset]{
		\includegraphics[width=\y\linewidth]{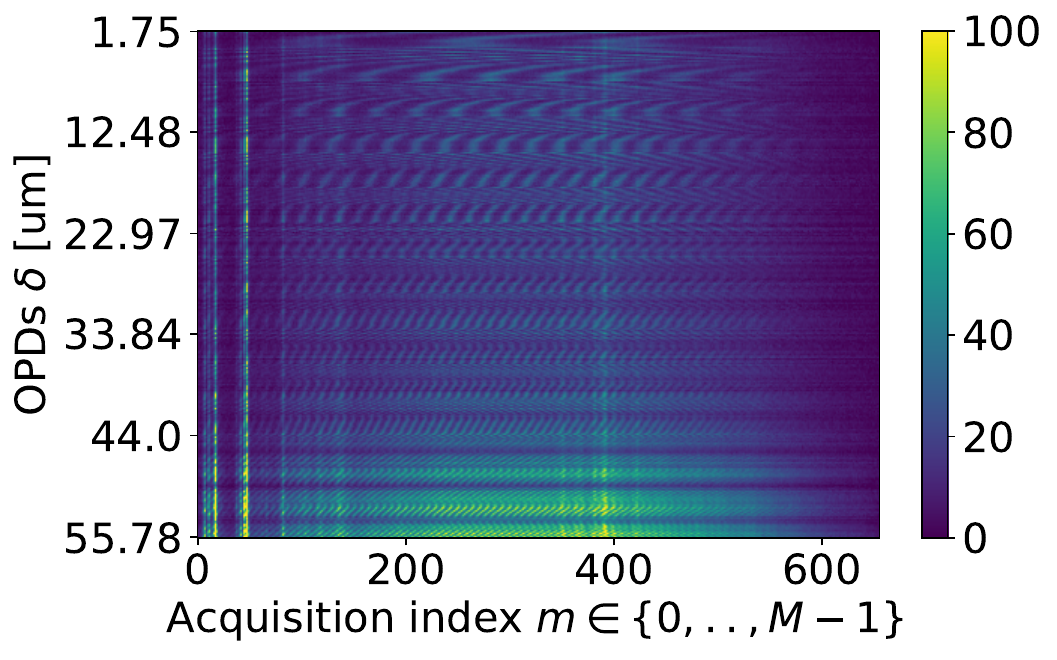}
		\label{fig:mc651_dataset}
	}
	
	\caption{Visualization of the real datasets. Each column represents an interferogram acquired from a monochromatic source.}
	\label{fig:real_datasets}
	
\end{figure}

%\begin{figure}[h]
%	\centering
%	
%	\begin{subfigure}[b]{0.495\linewidth}
%		\centering
%		\includegraphics[width=\textwidth]{figures/real/invert_mc451/imspoc_uv_2_mc451/dataset/dataset.pdf}
%		\caption{\texttt{MC-451} dataset}
%		\label{fig:mc451_dataset}
%	\end{subfigure}
%	\begin{subfigure}[b]{0.485\linewidth}
%		\centering
%		\includegraphics[width=\textwidth]{figures/real/invert_mc651/imspoc_uv_2_mc451/dataset/dataset.pdf}
%		\caption{\texttt{MC-651} dataset}
%		\label{fig:mc651_dataset}
%	\end{subfigure}
%	
%	\caption{Visualization of the real datasets, where each column is an interferogram acquisition of a monochromatic source.}
%	\label{fig:real_datasets}
%	
%\end{figure}

% In the second class, labeled \textit{real datasets}, the interferograms are measured with a interferometric spectrometer, under a controlled scenario that allowing to know the paired input spectra.

\subsection{Dataset Description}

% We acquire two real interferometric datasets.
The real datasets include two collections of interferograms, labeled MC-451 and MC-651, whose characteristics are summarized in \tablename\;\ref{tab:real_dataset_information}.
These interferograms were measured using the \gls{imspoc} UV 2 device, shown in \figurename \ref{fig:imspoc_prototype}. This instrument is composed of an array of $319$ \glsentrylong{fp} etalons with nominal increasing thickness of step size $\Delta d = 87.5$ \si{\nm}, shown in \figurename \ref{fig:imspoc_array}.

The acquisitions were performed in a controlled environment, with the input spectra being obtained by modulating a known light source through a tunable diffraction grating, using the setup described in \cite{PicoGDFL23:oe}.
%In other words, each acquired interferogram represents the response of the system to an impulse spectrum centered at a given wavenumber.
Given the monochromatic nature of the inputs, we assume in this study that the columns of the reference spectra $\mathbf{X}\in\R{K \times M}$ are Dirac pulses centered at the nominal central wavenumber of the diffraction grating. In particular, we have $M\!=\!451$ and $M\!=\!651$ central wavenumbers for the MC-451 and the MC-651 datasets, respectively.

\subsection{Experimental Setup}
\label{subsec:real_data_experimental_setup}

% For the experiments, we employ the same methods described in Section \ref{subsec:simulated_data_experimental_setup}, except the \gls{idct}, because not applicable to this scenario, and the \glsfmtlong{pinv}, since .

For the reconstruction, we assume that the transfer matrix $\mathbf{A}$ is once again obtained as a sampled version of eq. \eqref{eq:transmittance_response_fabry_perot_infty} for the \gls{mbi}. However, compared to the previous set of experiments, we use here a 5-th order polynomial function to express the dependency of $\mathcal{R}(\sigma)$ and $\mathcal{T}(\sigma)$ on the wavenumbers $\sigma$. Moreover, the \gls{opd} support is not regularly sampled and starts from $\delta_{min} \approx 1.75$, missing around 10 samples in average.
In the experiments, the \gls{opd} support is extrapolated from $\delta=0$, and the transfer matrix can then be generated for any range of values of wavenumbers and \glspl{opd}.

% The set of parameters $\Theta$ of the $\infty$-wave model $\matr{A}_{\Theta}$ are estimated from the calibration data (instrumental response) of the \gls{imspoc} UV 2 prototype, with the \gls{irca} developed by the second author~\cite{PicoGDFL23:oe}. In this case, $\mathcal{R}(\sigma_k)$ and $\mathcal{T}(\sigma_k)$ are assumed to be samples of a 5th-degree polynomial function. Moreover, the \gls{opd} support is not regularly sampled and starts from $\delta_{min} \neq 0$, missing around 10 samples in average.

The expression of reflectivity, transmissivity,
% phase shift,
and \gls{opd} that we plug in the transfer matrix $\mathbf{A}$ follows the characterization of the instrument performed by the second author in \cite{PicoGDFL23:oe}. The characterization makes use only of the MC-451 dataset.

On the other hand, the interferograms of MC-651, which were acquired at a different time, were never used in the calibration process. This makes the MC-651 dataset particularly valuable for testing, as it provides an independent set of data that was not seen during training (i.e., calibration).

%and was acquired at a different date than the one that we aim to reconstruct, MC-651, in order to make the interferogram we aim to invert as independent as possible from the model. Moreover, the MC-651 dataset has a finer wavenumber resolution $\Delta\sigma$.

\figurename\,\ref{fig:transfer_matrix_real_mc451} shows the coefficients of  $\mathbf{A}$ and its singular values. The condition number is high, showing that the problem is ill-conditioned and cannot be solved without some sort of regularization. Consequently, compared to the previous testbed, we do not provide results for the \glsfmtlong{pinv} method. We also exclude the \gls{idct} and Haar methods, as they are not applicable to this scenario. For our proposed method, we impose sparsity in the spectral domain, following eq. \eqref{eq:lv_sparse_spectrum_fista}.

\def \y {0.48}
\def \z {0.48}

\begin{figure}[t]
    \centering
    
    \subfloat[Transfer matrix]{\label{subfig:mc451_transfer_matrix}%
        \includegraphics[width=\y\linewidth]{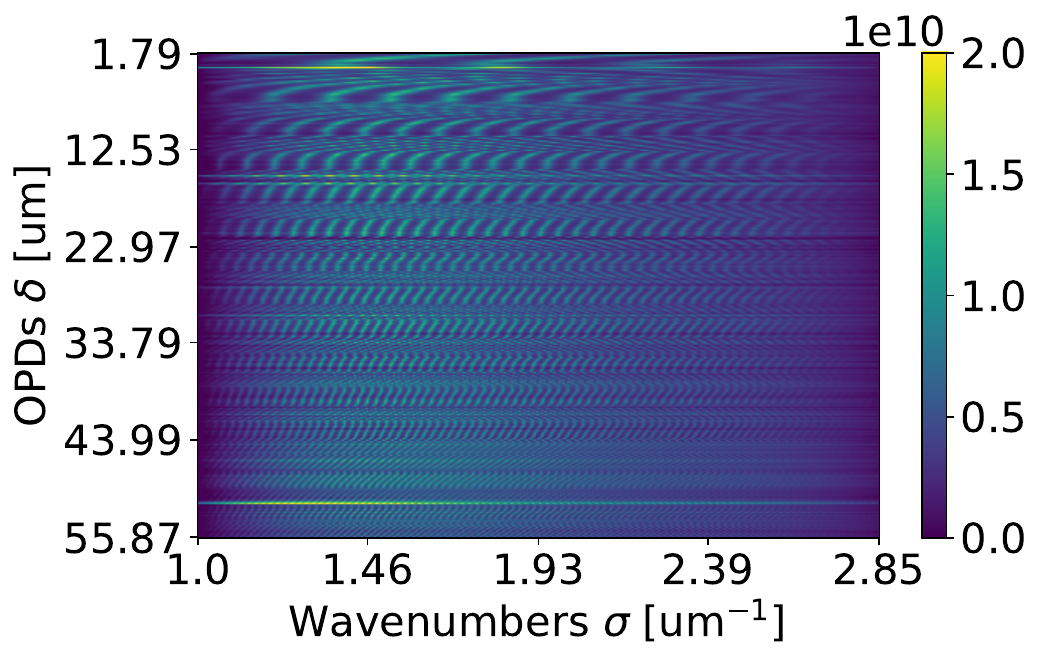}
    }
    \subfloat[Singular values]{\label{subfig:mc451_singular_values}%
        \includegraphics[width=\z\linewidth]{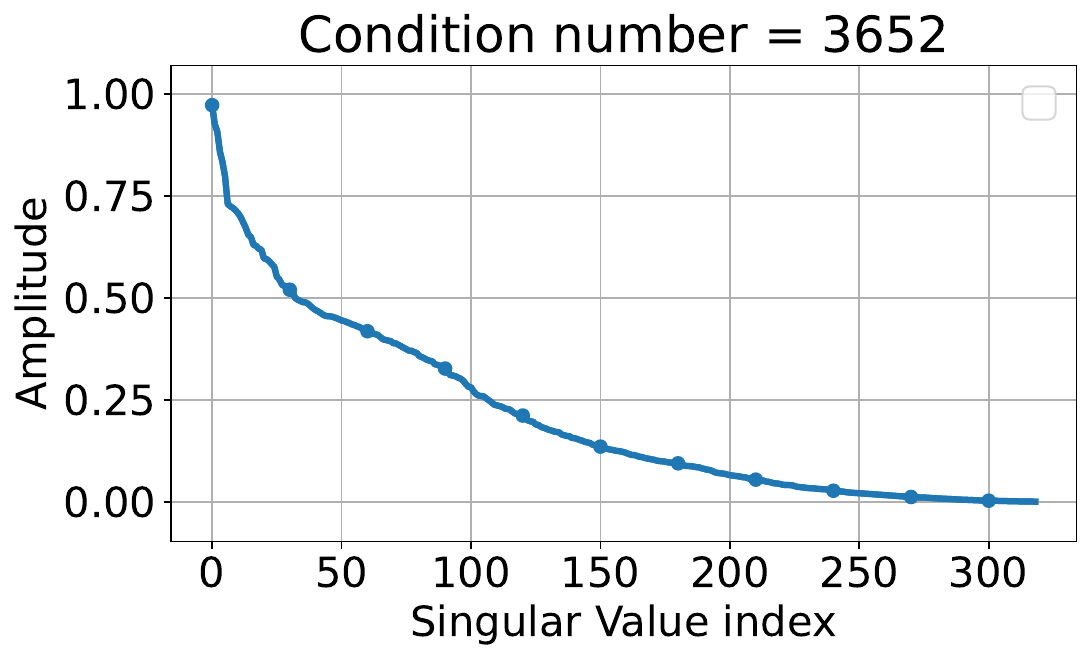}
    }
    
    \caption{Transfer matrix $\matr{A}$, parametrized from the dataset MC-451.}
    \label{fig:transfer_matrix_real_mc451}
\end{figure}

%\begin{figure*}[t]
%    
%    \centering
%
%    \begin{minipage}[b]{\x\linewidth}
%    	\centering
%    	
%        \begin{subfigure}[b]{\y\linewidth}
%            \centering
%            \includegraphics[width=\textwidth]{figures/real/invert_mc451/imspoc_uv_2_mc451/transfer_matrix/transfer_matrix.pdf}
%            \caption{
%                Transfer matrix
%            }
%            \label{subfig:mc451_transfer_matrix}
%        \end{subfigure}
%        \begin{subfigure}[b]{\z\linewidth}
%            \centering
%            \includegraphics[width=\textwidth]{figures/real/invert_mc451/imspoc_uv_2_mc451/transfer_matrix/singular_values.pdf}
%            \caption{
%            	Singular values
%            }
%            \label{subfig:mc451_singular_values}
%        \end{subfigure}
%        \begin{subfigure}[b]{\z\linewidth}
%            \centering
%            \includegraphics[width=\textwidth]{figures/real/invert_mc451/imspoc_uv_2_mc451/transfer_matrix/opd_response_idx_20.pdf}
%            \caption{
%                \gls{opd} response
%            }
%            \label{subfig:mc451_opd_response}
%        \end{subfigure}
%    \end{minipage}
%
%
%    \caption{
%		Transfer matrix $\matr{A}$, parametrized from the dataset \texttt{MC-451}, and used to invert both \texttt{MC-451} and \texttt{MC-651}.
%	}
%	\label{fig:transfer_matrix_real_mc451}
%\end{figure*}

First, we assess the results based on the \gls{rmse} metric, with the identity matrix $\matr{I}_{K}$ as reference.
Second, we introduce the \textit{number of \gls{mcw}} as a quality metric; this metric defines the number of times the maxima of the reconstructed spectra match with the nominal central wavenumbers of the corresponding interferograms. For instance, denoting by $\hat{\vect{x}}_m$ the $m$-th column of $\hat{\matr{X}}$ and $\hat{k}=\argmax{k}(\hat{\vect{x}}_m)$:
\small
\begin{equation}
	\text{\gls{mcw}}(\hat{\matr{X}})
	=
	\sum_{m=0}^{M-1} \mathbbm{1}_{\{m, \hat{k}\}}
	\text{ with }
	\mathbbm{1}_{\{m, k\}} =
	\begin{cases}
		1, \;\; \text{if } k = m. \\
		0, \;\; \text{otherwise}.
	\end{cases}
\end{equation}
\normalsize

For each method, we explored the range of regularization parameters with a grid search within reasonable intervals.

\subsection{Reconstruction Results}
\label{subsec:real_data_results_discussion}

\tablename\;\ref{tab:real_experiment_results} shows a summary of the results.
Overall, the proposed solution outperforms the other techniques in terms of \gls{rmse}, even at a lower number of iterations, and is at least competitive in terms of the number of \gls{mcw} with respect to \gls{tsvd} and \gls{rr}.

\begin{table}[t]
		\caption{
			Results of the real experiments showing the \glspl{rmse} and the number of \gls{mcw}.
			The best results are marked in \textbf{bold}.
            }
		\label{tab:real_experiment_results}
		\centering
	    \begin{tabular}{l|c||c|c|c|c}
			\multicolumn{6}{c}{\textbf{\gls{imspoc} UV 2: Array of \glsfmtlong{fp} $\infty$-wave model}}
	        \\
	        \hline
	        \textbf{Dataset} &
	        \textbf{Method} &
	        $\lambda_{\textrm{opt}}$ &
	        \begin{tabular}{@{}c@{}}
	        	Diagonal \\
	        	\glsfmtshort{rmse}
	        \end{tabular} &
		    \begin{tabular}{@{}c@{}}
			    Full \\
			    \glsfmtshort{rmse}
			\end{tabular} &
			\begin{tabular}{@{}c@{}}
				No. of \\
				\glsfmtshort{mcw}
			\end{tabular}
	        \\
	        \hline\hline
	        \multirow{4}{*}{\textbf{MC-451}}
& \gls{tsvd} \cite{Hans90:jssc} & 0.600 & 0.499 & 0.963 & 408 \\
& \gls{rr} \cite{GoluHO99:jmaa} & 0.316 & 0.472 & 0.937 & \underline{423} \\
		\cline{2-6}
%& Ours (1k iters.) & 0.185 & 0.398 & 0.810 & 401 \\
& Ours (1k iters.) & 0.215 & 0.388 & 0.800 & 403 \\
& Ours (50k iters.) & 0.054 & \textbf{0.221} & \textbf{0.463} & \textbf{429} \\
	        \thickhline
	        \multirow{4}{*}{\textbf{MC-651}}
& \gls{tsvd} \cite{Hans90:jssc} & 0.320 & 0.784 & 1.244 & \textbf{444} \\
& \gls{rr} \cite{GoluHO99:jmaa} & 2.512 & 0.781 & 1.242 & 429 \\
		\cline{2-6}
%& Ours (1k iters.) & 0.858 & 0.545 & 0.959 & 428 \\
& Ours (1k iters.) & 1.000 & 0.547 & 0.959 & 426 \\
& Ours (50k iters.) & 0.341 & \textbf{0.542} & \textbf{0.879} & \underline{440} \\
	    \end{tabular}
\end{table}

\def \x {0.238}
\def \w {0.016}
\def \z {1.8}

\begin{figure*}[ht]
	\centering
	
	% \subfloat{\label{subfig:invert_mc451_label}%
		\raisebox{\z cm}{\rotatebox[origin=c]{90}{\texttt{\textbf{MC-451}}}}
	% }
	\subfloat[Reference]{\label{subfig:invert_mc451/imspoc_uv_2_mc451/spectrum_matrices/reference}%
		\includegraphics[width=\x\linewidth]{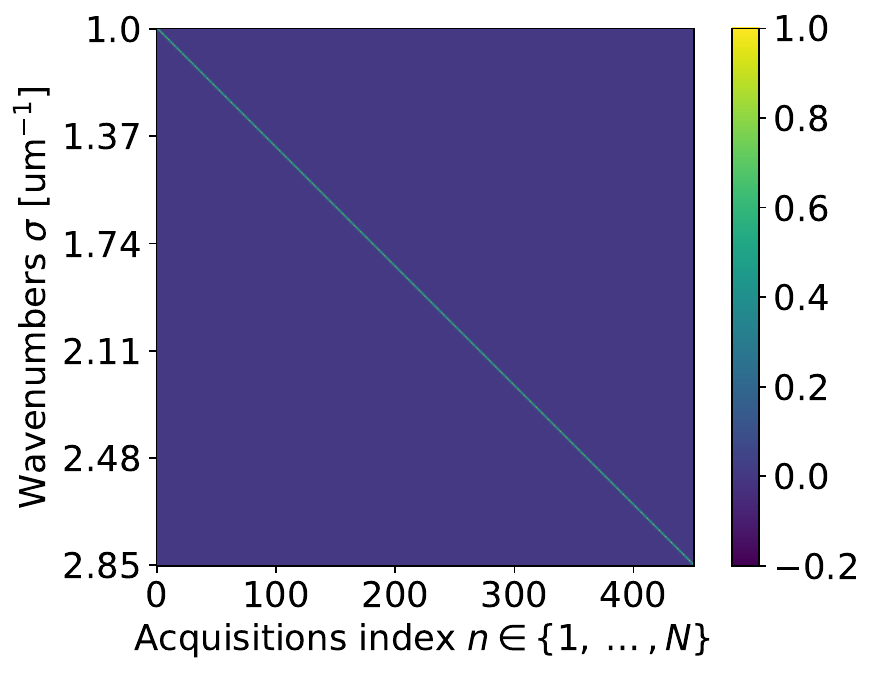}
	}
	\subfloat[\gls{tsvd} \cite{Hans90:jssc}]{\label{subfig:invert_mc451/imspoc_uv_2_mc451/spectrum_matrices/tsvd}%
		\includegraphics[width=\x\linewidth]{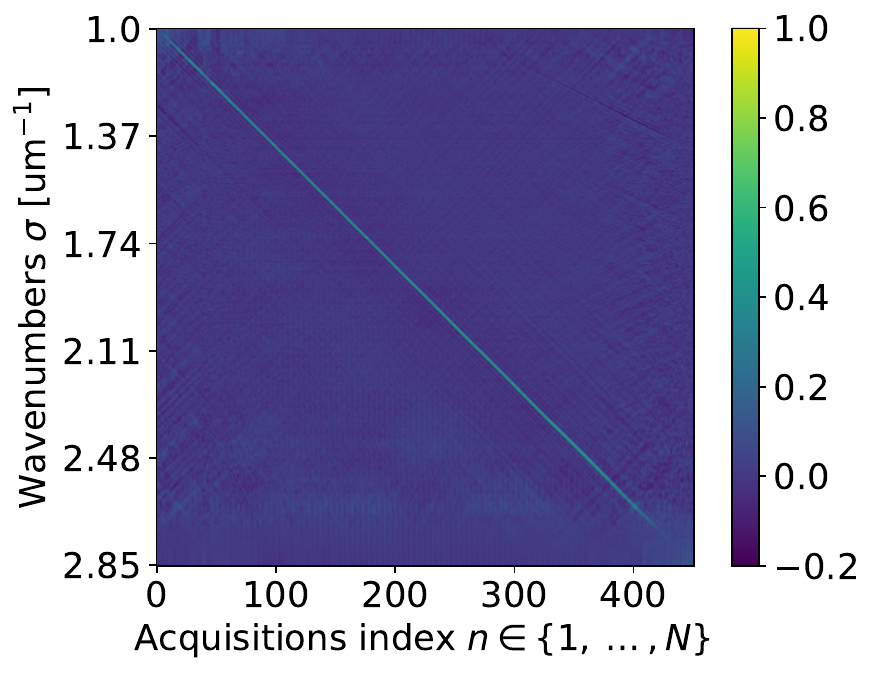}
	}
	\subfloat[\gls{rr} \cite{GoluHO99:jmaa}]{\label{subfig:invert_mc451/imspoc_uv_2_mc451/spectrum_matrices/rr}%
		\includegraphics[width=\x\linewidth]{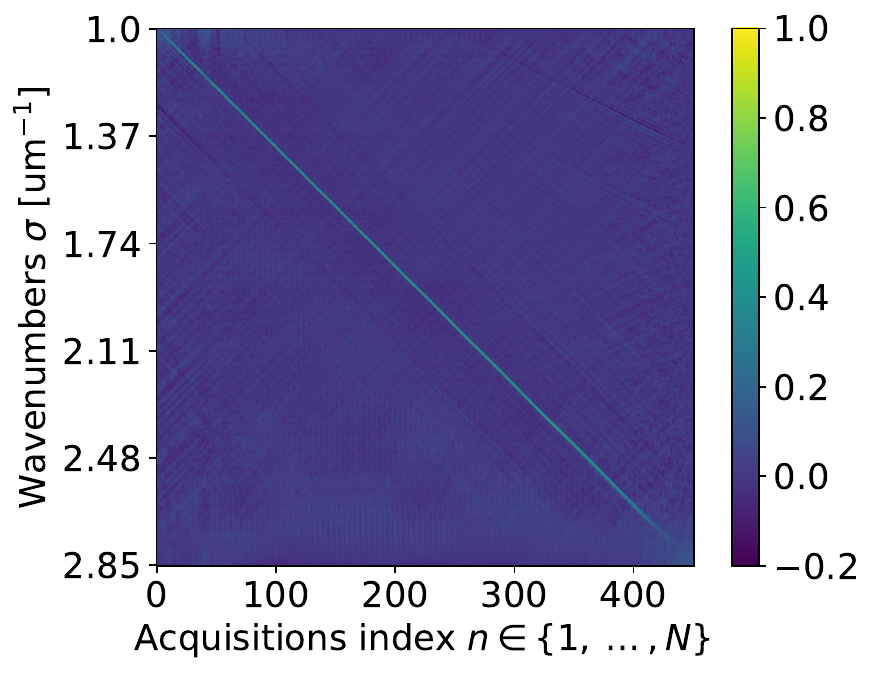}
	}
	\subfloat[Ours]{\label{subfig:invert_mc451/imspoc_uv_2_mc451/spectrum_matrices/lv}%
		\includegraphics[width=\x\linewidth]{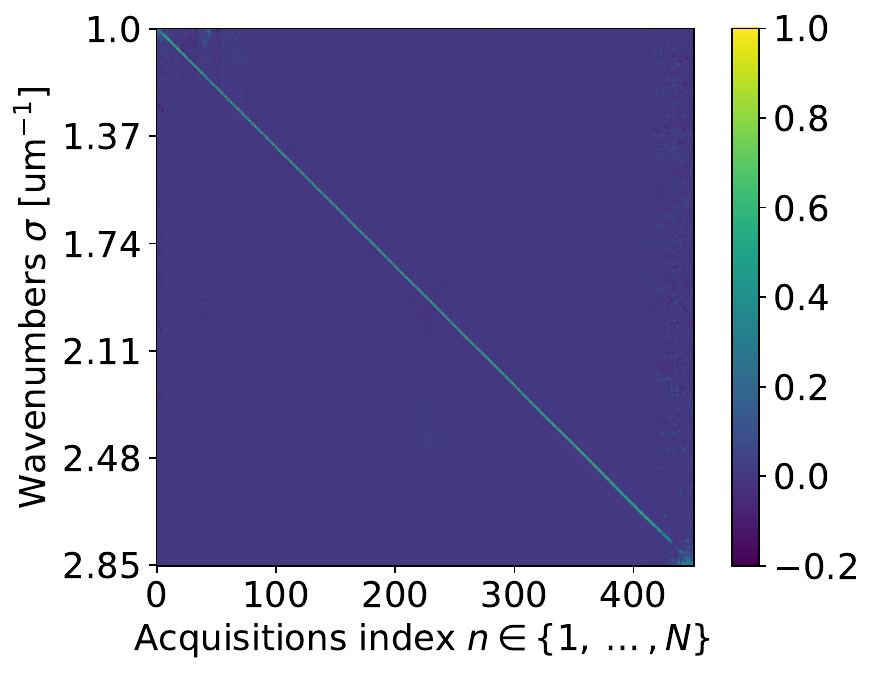}
	}
	
	% \subfloat{%
		\raisebox{\z cm}{\rotatebox[origin=c]{90}{\texttt{\textbf{MC-651}}}}
	% }
	\subfloat[Reference]{\label{subfig:invert_mc651/imspoc_uv_2_mc451/spectrum_matrices/reference}%
		\includegraphics[width=\x\linewidth]{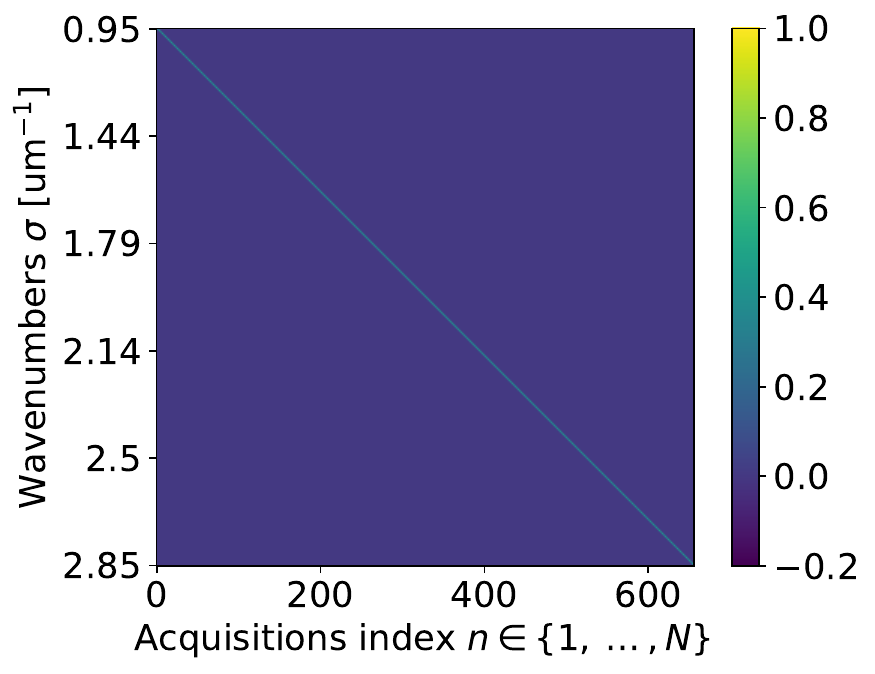}
	}
	\subfloat[\gls{tsvd} \cite{Hans90:jssc}]{\label{subfig:invert_mc651/imspoc_uv_2_mc451/spectrum_matrices/tsvd}%
		\includegraphics[width=\x\linewidth]{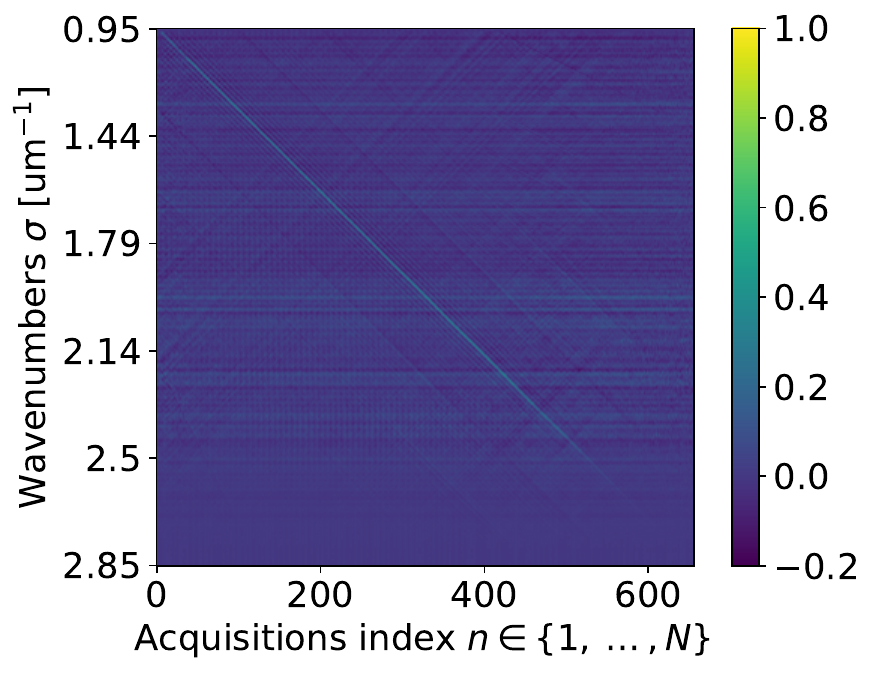}
	}
	\subfloat[\gls{rr} \cite{GoluHO99:jmaa}]{\label{subfig:invert_mc651/imspoc_uv_2_mc451/spectrum_matrices/rr}%
		\includegraphics[width=\x\linewidth]{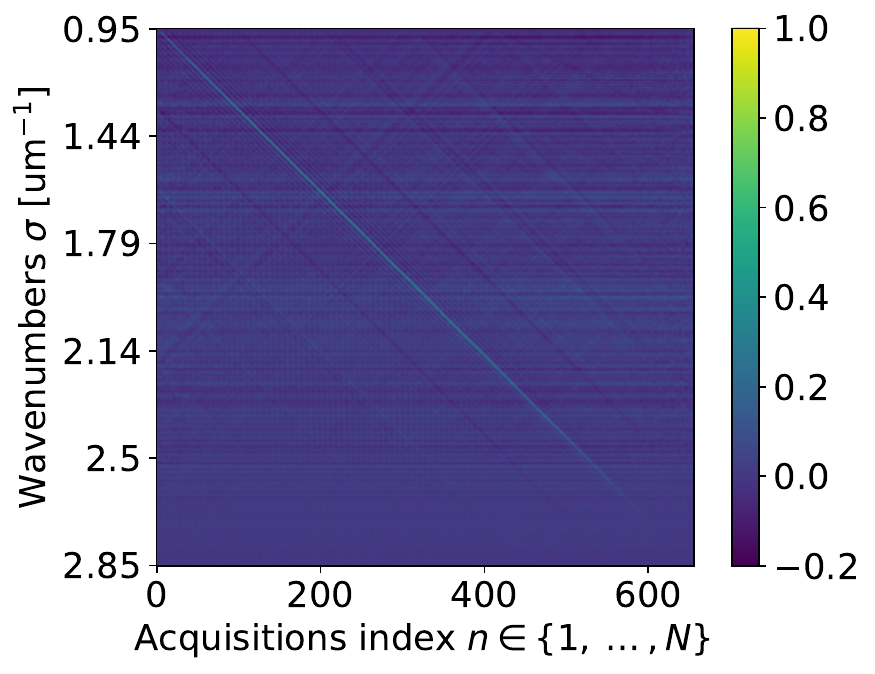}
	}
	\subfloat[Ours]{\label{subfig:invert_mc651/imspoc_uv_2_mc451/spectrum_matrices/lv}%
		\includegraphics[width=\x\linewidth]{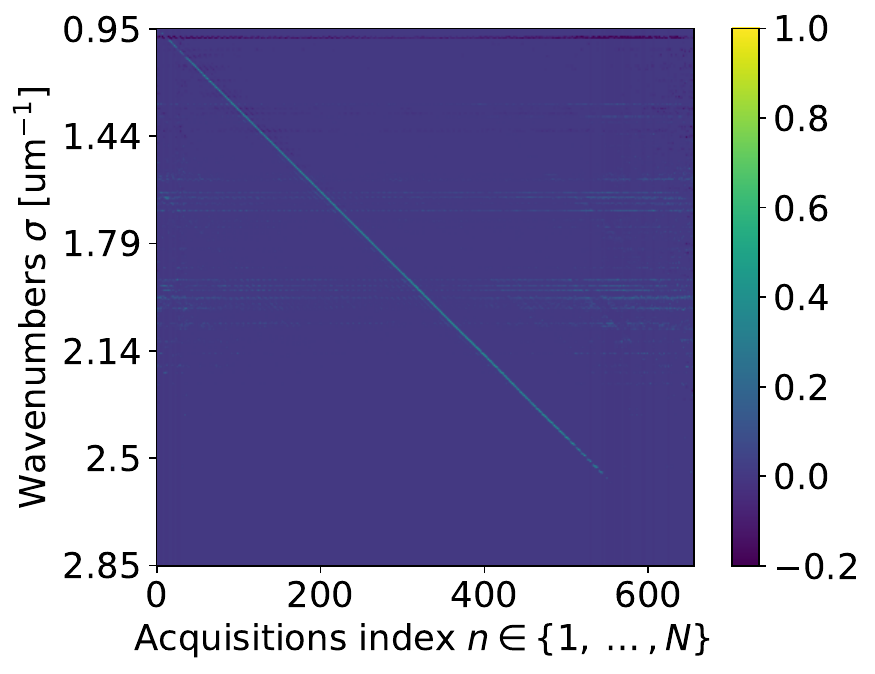}
	}
	
	\caption{
            Visualization of the reconstructed monochromatic datasets in $\hat{\matr{X}}$. Each column of $\hat{\matr{X}}$ represents a reconstructed monochromatic spectrum.
            The two rows refer to the \texttt{MC-451} and \texttt{MC-651} datasets, respectively.
	}
	\label{fig:spectral_matrix_reconstruction}
\end{figure*}

\figurename\,\ref{fig:spectral_matrix_reconstruction} shows the reconstructed spectral matrix $\hat{\matr{X}}$ for the MC-451 and the MC-651 datasets. In this case, the corresponding reference is an identity matrix.
%%\figurename\,\ref{fig:mc651_acquisition_compare} shows a selected spectrum reconstruction from each dataset.
Looking at the figures, for the two datasets, there is a significant improvement with the sparse prior compared to the \gls{tsvd} approach and the $\ell_2$-norm in \gls{rr}. 
Our proposed approach provides the least noisy reconstructions, especially around the peaks.

%\subsubsection{Matching central wavenumbers}

\figurename\,\ref{fig:matching_maxima_intensity} plots the maxima of the reconstructed spectra with respect to the associated nominal central wavenumbers, for the two datasets.
Moreover, the points in the plots are color-coded in orange if the maximum value of the reconstructed spectra matches with the central wavenumber of the reference.
Compared to conventional techniques, our algorithm produces a plot of the maxima of the reconstructed spectra that is overall flatter.
This is more evident for the MC-451, as the transfer matrix $\mathbf{A}$ closely matches the acquisition conditions. However, decent results are also obtained for MC-651 dataset.

\def \x {0.3}
\def \y {0.02}
\def \z {2.1}

\begin{figure*}[ht!]
	\centering
	
	% \subfloat{\label{subfig:invert_mc451_label}%
		\raisebox{\z cm}{\rotatebox[origin=c]{90}{\texttt{\textbf{MC-451}}}}
	% }
	\subfloat[\gls{tsvd} \cite{Hans90:jssc}]{\label{subfig:invert_mc451/imspoc_uv_2_mc451/matching_maxima_intensity/tsvd}%
		\includegraphics[width=\x\linewidth]{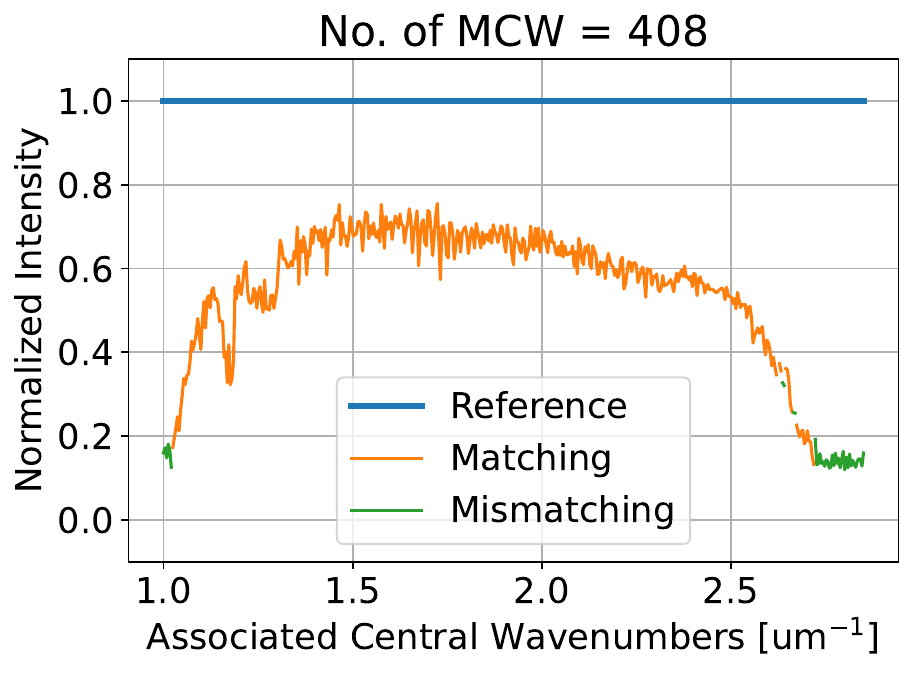}
	}
	\subfloat[\gls{rr} \cite{GoluHO99:jmaa}]{\label{subfig:invert_mc451/imspoc_uv_2_mc451/matching_maxima_intensity/rr}%
		\includegraphics[width=\x\linewidth]{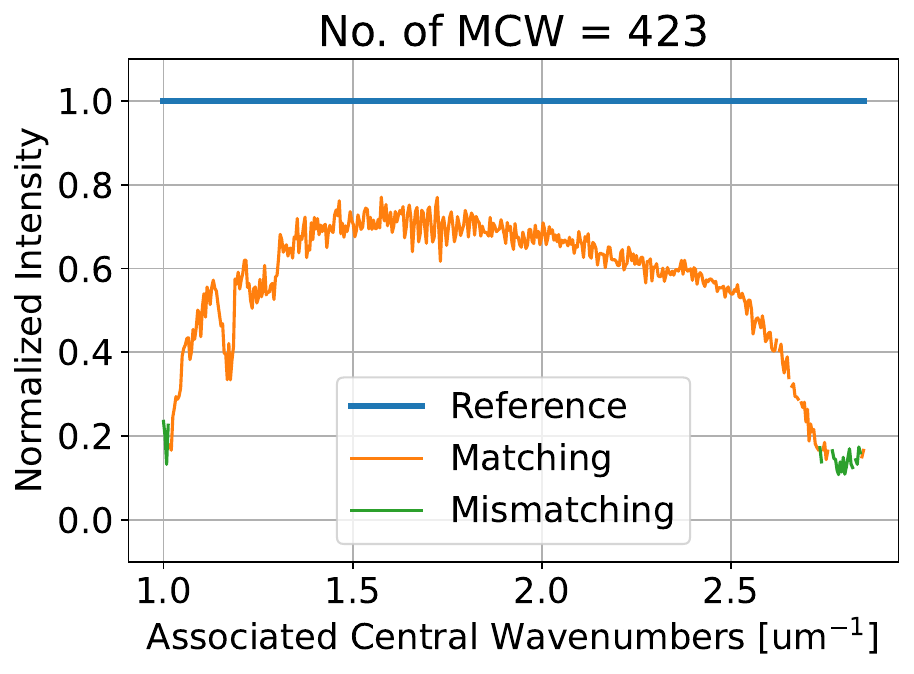}
	}
	\subfloat[Ours]{\label{subfig:invert_mc451/imspoc_uv_2_mc451/matching_maxima_intensity/lv}%
		\includegraphics[width=\x\linewidth]{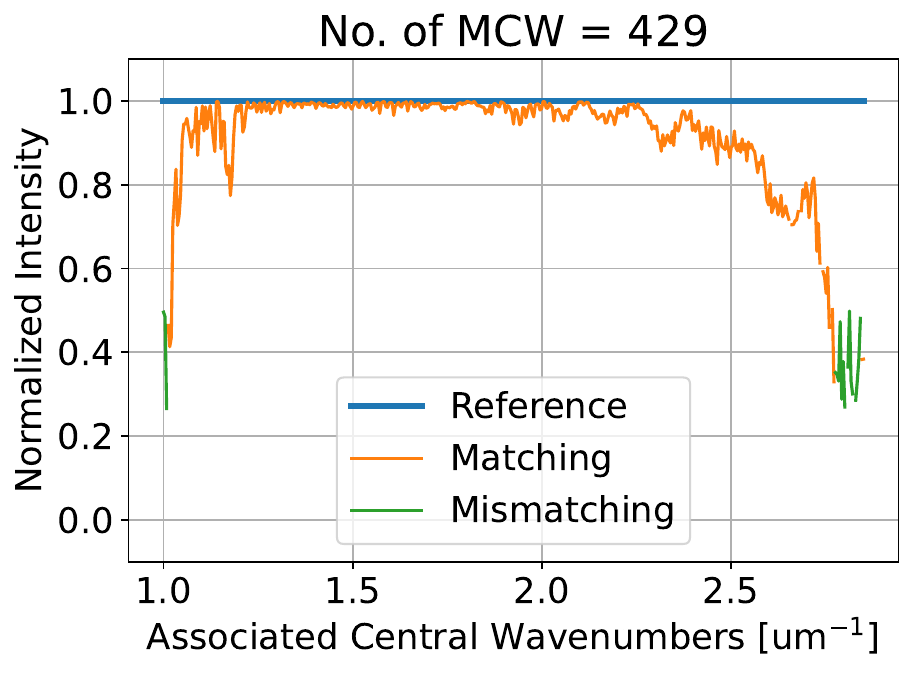}
	}
	
	% \subfloat{\label{subfig:invert_mc651_label}%
		\raisebox{\z cm}{\rotatebox[origin=c]{90}{\texttt{\textbf{MC-651}}}}
	% }
	\subfloat[\gls{tsvd} \cite{Hans90:jssc}]{\label{subfig:invert_mc651/imspoc_uv_2_mc451/matching_maxima_intensity/tsvd}%
		\includegraphics[width=\x\linewidth]{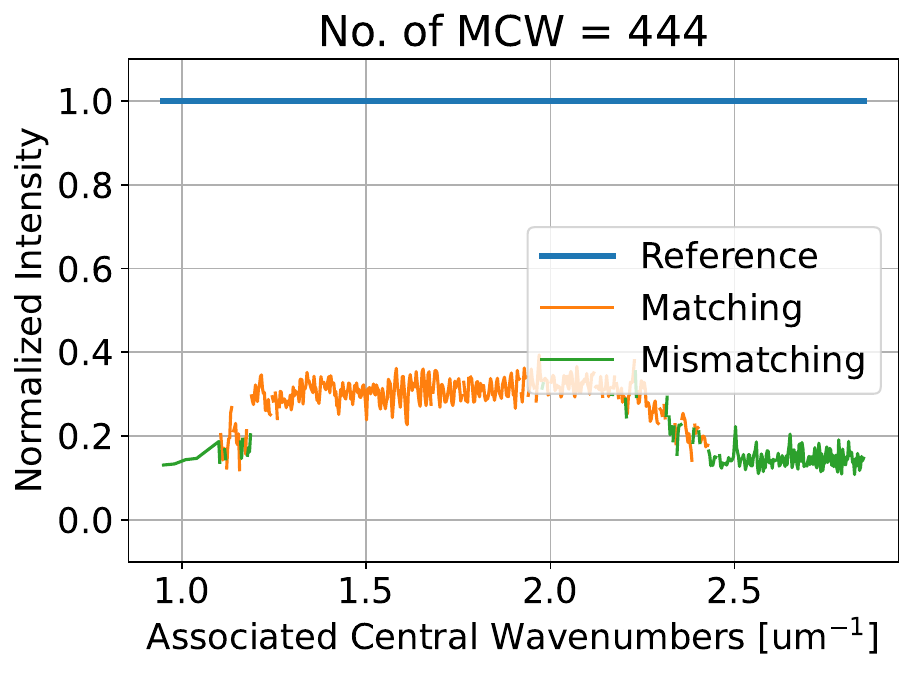}
	}
	\subfloat[\gls{rr} \cite{GoluHO99:jmaa}]{\label{invert_mc651/imspoc_uv_2_mc451/matching_maxima_intensity/rr}%
		\includegraphics[width=\x\linewidth]{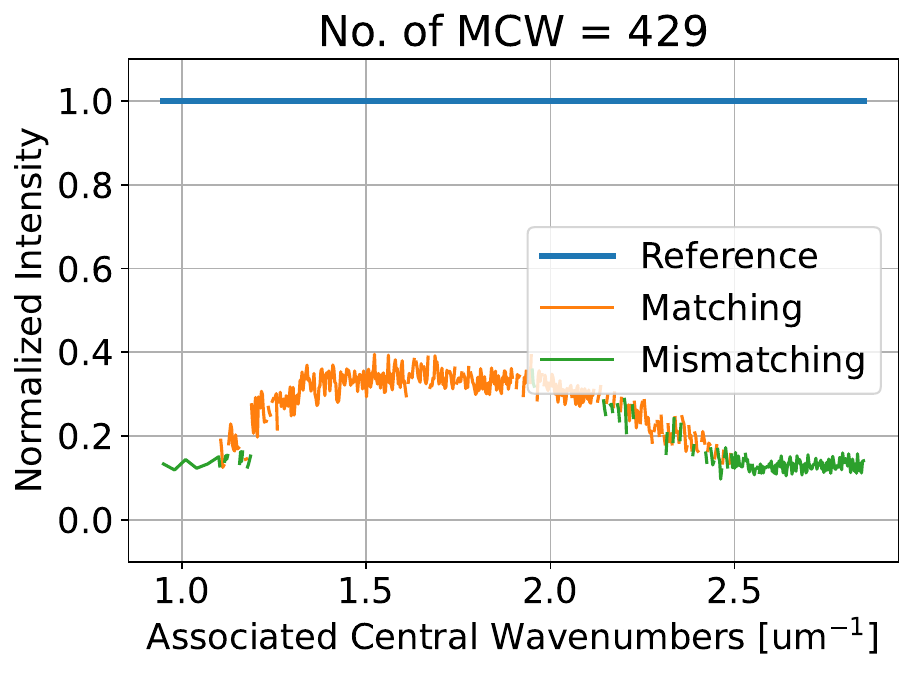}
	}
	\subfloat[Ours]{\label{subfig:invert_mc651/imspoc_uv_2_mc451/matching_maxima_intensity/lv}%
		\includegraphics[width=\x\linewidth]{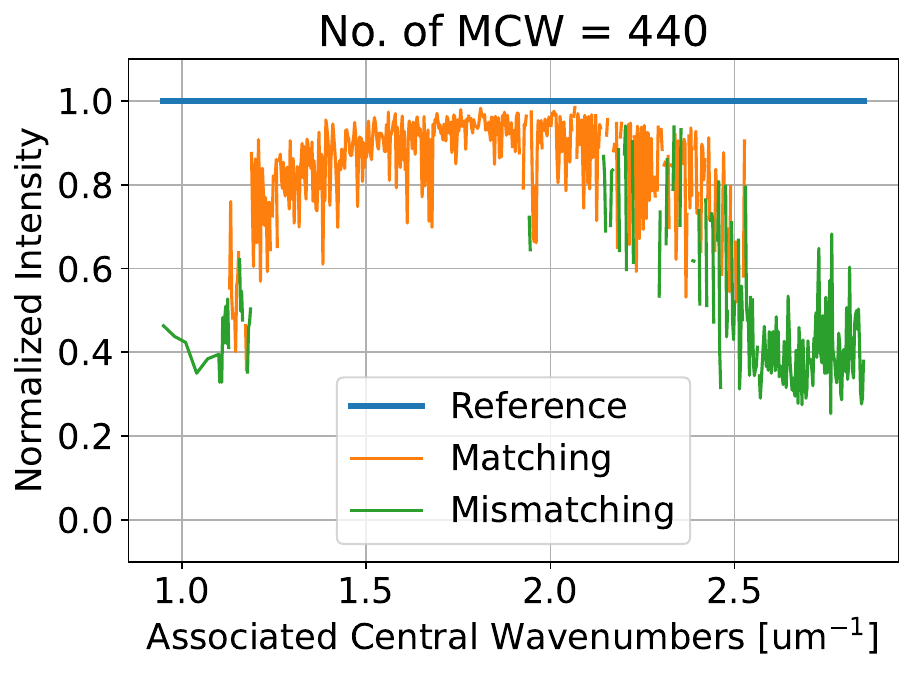}
	}
	
	\caption{
		Visualization of the maximum of the reconstructed spectra in $\hat{\matr{X}}$, classified into those that match with the corresponding nominal central wavenumbers. The two rows refer to the \texttt{MC-451} and \texttt{MC-651} datasets, respectively.
	}
	\label{fig:matching_maxima_intensity}
\end{figure*}

\section{Conclusion}
\label{sec:conclusions}

In this review, we tackle the problem of \gls{mbi} spectroscopy, as we propose a unified framework for the analysis and the spectrum reconstruction in both \gls{tbi} and \gls{mbi}.
At its core, the framework relies on the representation of the continuous physical models by the discretized transfer matrix and the Bayesian formulation of the reconstruction problem.
First, we perform a numerical analysis on the transfer matrices in \gls{tbi} and \gls{mbi} under the textbook formulation and formalize their limitations in terms of spectral resolution and the condition number.
Second, we extend the range of spectrum reconstruction techniques while trying to tackle real-world non-idealities, as we showcase an increasingly generic framework based on the Bayesian inversion.
The comparisons are validated on simulated interferograms by considering one non-ideality at a time (reflectivity levels, irregular sampling, and noise measurements), and on real interferograms acquired from monochromatic spectral sources.
%
% In the future, we plan to expand the reconstruction to hyperspectral cubes by taking into account spatial information, and extend our database with more acquisitions for deep learning approaches.

\section*{Acknowledgments}

%The authors would like to
We thank Dr. Bernard Shmidt for assistance with acquiring
%and calibrating
the \texttt{SHINE} and \texttt{SPECIM} datasets, and Dr. Silvère Gousset for help with the \texttt{MC-451} and \texttt{MC-651} datasets.

\bibliographystyle{IEEEtran}
% \small
\bibliography{refs}


\begin{IEEEbiography}[
	{
		\includegraphics[width=1in,height=1.25in,clip,keepaspectratio]{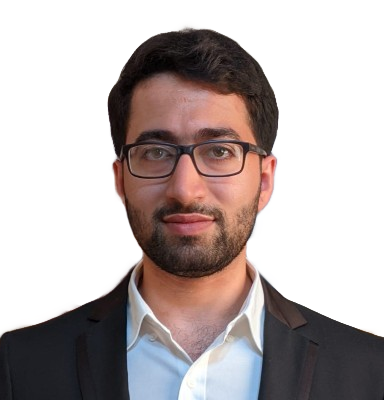}
	}
]{Mohamad Jouni}
	(Member, IEEE) received the B.Eng. degree in Computer and Communications Engineering from the Lebanese University, Beirut, Lebanon, in 2016, and the M.Sc. and Ph.D. degrees in Signal and Image Processing from Grenoble Institute of Technology and the University of Grenoble Alpes, Grenoble, France, in 2017 and 2021 respectively.
	Since 2021, he has been a Postdoctoral Researcher at Grenoble Institute of Technology, Grenoble, France.
	In 2019 and 2023, he was a visiting researcher for 10 and 6 weeks, respectively, at Tokyo Institute of Technology, Tokyo, Japan.
	His interests include computational imaging, tensor algebra, hybrid AI methods, applications of multimodal and hyperspectral data analysis, and technology transfer.
\end{IEEEbiography}

\begin{IEEEbiography}[
	{
		\includegraphics[width=1in,height=1.25in,clip,keepaspectratio]{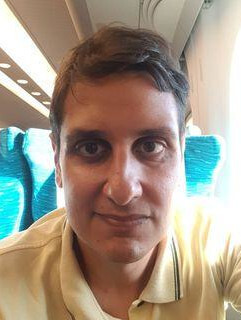}
	}
]{Daniele Picone}
	(Member, IEEE) received the B.Sc. and M.Sc. degrees in electronic engineering from the University of Salerno, Salerno, Italy, in 2008 and 2016, respectively, and the Ph.D degree in Signal and Image processing form the University of Grenoble Alpes, Grenoble, France, in 2021.
	In 2016, he was a Research fellow for 4 months at University of Salerno. In 2019, he was a visiting researcher at the Tokyo Insitute of Technology, Japan for three months. Since December 2021, he was a Postdoctoral fellow for one year with the University of Grenoble Alpes, Grenoble, France, and then with the Grenoble Institute of Technology (Grenoble-INP), Grenoble, France, starting from January 2023.
	He is conducting his research at the Grenoble Images Speech Signals and Automatics Laboratory (GIPSA-Lab).
	His main research activities are in the fields of image processing and remote sensing, with applications mainly involving computational imaging, optimization, data fusion, and hyperspectral data processing.
\end{IEEEbiography}

\begin{IEEEbiography}[
	{
		\includegraphics[width=1in,height=1.25in,clip,keepaspectratio]{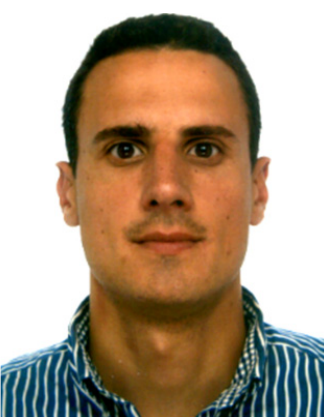}
	}
]{Mauro Dalla~Mura}
	(Senior Member, IEEE) received the B.Sc. and M.Sc. degrees in Telecommunication Engineering from the University of Trento, Italy in 2005 and 2007, respectively.
	He obtained in 2011 a joint Ph.D. degree in Information and Communication Technologies (Telecommunications Area) from the University of Trento, Italy and in Electrical and Computer Engineering from the University of Iceland, Iceland.
	In 2011 he was a Research fellow at Fondazione Bruno Kessler, Trento, Italy, conducting research on computer vision.
	He is currently an Assistant Professor at Grenoble Institute of Technology (Grenoble INP), France since 2012. He is conducting his research at the Grenoble Images Speech Signals and Automatics Laboratory (GIPSA-Lab).  He is a Junior member of the Institut Universitaire de France (2021-2026).
	Dr. Dalla~Mura has been appointed "Specially Appointed Associate Professor" at the School of Computing, Tokyo Institute of Technology, Japan for 2019-2022.
	His main research activities are in the fields of remote sensing, computational imaging, image and signal processing.
	Dr. Dalla~Mura was the recipient of the IEEE GRSS Second Prize in the Student Paper Competition of the 2011 IEEE IGARSS 2011 and co-recipient of the Best Paper Award of the International Journal of Image and Data Fusion for the year 2012-2013 and the Symposium Paper Award for IEEE IGARSS 2014.
	Dr. Dalla~Mura was the IEEE GRSS Chapter's Committee Chair for 2020-2021. He was President of the IEEE GRSS French Chapter 2016-2020 (he previously served as Secretary 2013-2016). In 2017 the IEEE GRSS French Chapter was the recipient of the IEEE GRSS Chapter Award and the ``Chapter of the year 2017'' from the IEEE French Section.
	He is on the Editorial Board of the IEEE Journal of Selected Topics in Applied Earth Observations and Remote Sensing (J-STARS) since 2016.
\end{IEEEbiography}

\end{document}